\documentclass{jfp1}

\makeatletter
\@ifundefined{lhs2tex.lhs2tex.sty.read}%
  {\@namedef{lhs2tex.lhs2tex.sty.read}{}%
   \newcommand\SkipToFmtEnd{}%
   \newcommand\EndFmtInput{}%
   \long\def\SkipToFmtEnd#1\EndFmtInput{}%
  }\SkipToFmtEnd

\newcommand\ReadOnlyOnce[1]{\@ifundefined{#1}{\@namedef{#1}{}}\SkipToFmtEnd}
\usepackage{amstext}
\usepackage{amssymb}
\usepackage{stmaryrd}
\DeclareFontFamily{OT1}{cmtex}{}
\DeclareFontShape{OT1}{cmtex}{m}{n}
  {<5><6><7><8>cmtex8
   <9>cmtex9
   <10><10.95><12><14.4><17.28><20.74><24.88>cmtex10}{}
\DeclareFontShape{OT1}{cmtex}{m}{it}
  {<-> ssub * cmtt/m/it}{}

\DeclareFontShape{OT1}{cmtt}{bx}{n}
  {<5><6><7><8>cmtt8
   <9>cmbtt9
   <10><10.95><12><14.4><17.28><20.74><24.88>cmbtt10}{}
\DeclareFontShape{OT1}{cmtex}{bx}{n}
  {<-> ssub * cmtt/bx/n}{}

\newcommand{\Conid}[1]{\mathit{#1}}
\newcommand{\Varid}[1]{\mathit{#1}}
\newcommand{\anonymous}{\kern0.06em \vbox{\hrule\@width.5em}}
\newcommand{\plus}{\mathbin{+\!\!\!+}}
\newcommand{\bind}{\mathbin{>\!\!\!>\mkern-6.7mu=}}

\usepackage{polytable}

\@ifundefined{mathindent}%
  {\newdimen\mathindent\mathindent\leftmargini}%
  {}%

\def\resethooks{%
  \global\let\SaveRestoreHook\empty
  \global\let\ColumnHook\empty}
\newcommand*{\savecolumns}[1][default]%
  {\g@addto@macro\SaveRestoreHook{\savecolumns[#1]}}
\newcommand*{\restorecolumns}[1][default]%
  {\g@addto@macro\SaveRestoreHook{\restorecolumns[#1]}}
\newcommand*{\aligncolumn}[2]%
  {\g@addto@macro\ColumnHook{\column{#1}{#2}}}

\resethooks

\newcommand{\onelinecommentchars}{\quad-{}- }
\newcommand{\commentbeginchars}{\enskip\{-}
\newcommand{\commentendchars}{-\}\enskip}

\newcommand{\visiblecomments}{%
  \let\onelinecomment=\onelinecommentchars
  \let\commentbegin=\commentbeginchars
  \let\commentend=\commentendchars}

\newcommand{\invisiblecomments}{%
  \let\onelinecomment=\empty
  \let\commentbegin=\empty
  \let\commentend=\empty}

\visiblecomments

\newlength{\blanklineskip}
\newcommand{\hsindent}[1]{\quad}
\let\hspre\empty
\let\hspost\empty

\EndFmtInput
\makeatother
\ReadOnlyOnce{polycode.fmt}%
\makeatletter

\newcommand{\hsnewpar}[1]%
  {{\parskip=0pt\parindent=0pt\par\vskip #1\noindent}}

\newcommand{\hscodestyle}{}

\newcommand{\sethscode}[1]%
  {\expandafter\let\expandafter\hscode\csname #1\endcsname
   \expandafter\let\expandafter\endhscode\csname end#1\endcsname}

  {\par\noindent
   \advance\leftskip\mathindent
   \hscodestyle
   \let\\=\@normalcr
   \let\hspre\(\let\hspost\)%
   \pboxed}%
  {\endpboxed\)%
   \par\noindent
   \ignorespacesafterend}

  {\hsnewpar\abovedisplayskip
   \advance\leftskip\mathindent
   \hscodestyle
   \let\hspre\(\let\hspost\)%
   \pboxed}%
  {\endpboxed%
   \hsnewpar\belowdisplayskip
   \ignorespacesafterend}

  {\hsnewpar\abovedisplayskip
   \advance\leftskip\mathindent
   \hscodestyle
   \let\\=\@normalcr
   \(\pboxed}%
  {\endpboxed\)%
   \hsnewpar\belowdisplayskip
   \ignorespacesafterend}

\newcommand{\plainhs}{\sethscode{plainhscode}}

\plainhs

  {\hsnewpar\abovedisplayskip
   \advance\leftskip\mathindent
   \hscodestyle
   \let\\=\@normalcr
   \(\parray}%
  {\endparray\)%
   \hsnewpar\belowdisplayskip
   \ignorespacesafterend}

  {\parray}{\endparray}

  {\(\parray}{\endparray\)}

\def\codeframewidth{\arrayrulewidth}
\RequirePackage{calc}

  {\parskip=\abovedisplayskip\par\noindent
   \hscodestyle
   \arrayrulewidth=\codeframewidth
   \tabular{@{}|p{\linewidth-2\arraycolsep-2\arrayrulewidth-2pt}|@{}}%
   \hline\framedhslinecorrect\\{-1.5ex}%
   \let\endoflinesave=\\
   \let\\=\@normalcr
   \(\pboxed}%
  {\endpboxed\)%
   \framedhslinecorrect\endoflinesave{.5ex}\hline
   \endtabular
   \parskip=\belowdisplayskip\par\noindent
   \ignorespacesafterend}

\newcommand{\framedhslinecorrect}[2]%
  {#1[#2]}

  {\(\def\column##1##2{}%
   \let\>\undefined\let\<\undefined\let\\\undefined
   \newcommand\>[1][]{}\newcommand\<[1][]{}\newcommand\\[1][]{}%
   \def\fromto##1##2##3{##3}%
   }{\) }%

  {\let\orighscode=\hscode
   \let\origendhscode=\endhscode
   \def\endhscode{\def\hscode{\endgroup\def\@currenvir{hscode}\\}\begingroup}
   \orighscode\def\hscode{\endgroup\def\@currenvir{hscode}}}%
  {\origendhscode
   \global\let\hscode=\orighscode
   \global\let\endhscode=\origendhscode}%

\makeatother
\EndFmtInput
\jdate{January 2018}
\pubyear{2018}
\pagerange{\pageref{firstpage}--\pageref{lastpage}}
\doi{...}

\usepackage{tikz}

\usepackage{filecontents}
\usepackage{pgfplots, pgfplotstable}
\usepgfplotslibrary{statistics}

\usepackage{makecell}

\usepackage[T1]{fontenc}
\usepackage[utf8]{inputenc}

\usepackage{hyperref}

\usepackage[round,sectionbib]{natbib}
\renewcommand\refname{References}

\usepackage{csquotes}

\usepackage{microtype}
\DisableLigatures[>,<]{encoding = T1,family=tt*} 

\renewcommand{\cite}[1]{\citep{#1}}

\usepackage{xifthen}

\newboolean{anonymous}
\setboolean{anonymous}{False}

\usepackage[british]{babel}

\usepackage{siunitx}
\usepackage{array}
\newcolumntype{L}[1]{>{\raggedright\let\newline\\\arraybackslash\hspace{0pt}}m{#1}}
\newcolumntype{C}[1]{>{\centering\let\newline\\\arraybackslash\hspace{0pt}}m{#1}}
\newcolumntype{R}[1]{>{\raggedleft\let\newline\\\arraybackslash\hspace{0pt}}m{#1}}

\usepackage{xspace}
\usepackage{xcolor}

\newcommand{\comm}[2]{}
\newcommand{\olcomment}[1]{\comm{OL}{#1}}
\newcommand{\mbcomment}[1]{\comm{MB}{#1}}
\newcommand{\ptcomment}[1]{\comm{Phil}{#1}}

\DeclareRobustCommand{\xth}{\textsuperscript{th}\xspace}
\DeclareRobustCommand{\st}{\textsuperscript{st}\xspace}
\DeclareRobustCommand{\nd}{\textsuperscript{nd}\xspace}
\DeclareRobustCommand{\xrd}{\textsuperscript{rd}\xspace}

\newcommand{\done}{\textcolor{green}{\bfseries done!}\xspace}
\DeclareRobustCommand{\hairspn}{\hspace{1pt}\nolinebreak}

\DeclareRobustCommand{\ie}{{i.\hairspn{}e.~}}
\DeclareRobustCommand{\eg}{{e.\hairspn{}g.~}}

\DeclareRobustCommand{\xth}{\textsuperscript{th}\xspace}
\DeclareRobustCommand{\st}{\textsuperscript{st}\xspace}
\DeclareRobustCommand{\nd}{\textsuperscript{nd}\xspace}
\DeclareRobustCommand{\xrd}{\textsuperscript{rd}\xspace}

\usepackage{booktabs}
\usepackage{multirow}

\usepackage{csvsimple}

\usepackage{tablefootnote}

\microtypecontext{spacing=nonfrench} 

\usepackage{subfig}

\title{Arrows for Parallel Computation}
\ifthenelse{\boolean{anonymous}}{%
\author{Submission ID xxxxxx}
}{%
 \author[M. Braun, O. Lobachev, and P. Trinder]%
        {\textls*{MARTIN BRAUN}\\
         University Bayreuth, 95440 Bayreuth, Germany\\
		 \textls*{OLEG LOBACHEV}\\
		 University Bayreuth, 95440 Bayreuth, Germany\\
		 \and\ \textls*{PHIL TRINDER}\\
		 Glasgow University, Glasgow, G12 8QQ, Scotland}
}

\begin{document}

\label{firstpage}

\def\SymbReg{\textsuperscript{\textregistered}}

\maketitle

\begin{abstract}
Arrows are a general interface for computation and an alternative to Monads for API design. In contrast to Monad-based parallelism, we explore the use of Arrows for specifying generalised parallelism. Specifically, we define an Arrow-based language and implement it using multiple parallel Haskells.

As each parallel computation is an Arrow, such parallel Arrows (PArrows) can be readily composed and transformed as such.
To allow for more sophisticated communication schemes between computation nodes in distributed systems, we utilise the concept of Futures to wrap direct communication.

To show that PArrows have similar expressive power as existing parallel languages, we implement several algorithmic skeletons and four benchmarks. 
Benchmarks show that our framework does not induce any notable performance overhead. We conclude that Arrows have considerable potential for composing parallel programs and for producing programs that can execute on multiple parallel language implementations. 
%
%
\end{abstract}

\tableofcontents

	%
	%

\section{Introduction}
\label{sec:introduction}

Parallel functional languages have a long history of being used for experimenting with novel parallel programming paradigms. Haskell, which we focus on in this paper, has  several mature implementations. We regard here in-depth
Glasgow parallel Haskell or short GpH (its Multicore SMP implementation, in particular), the
\ensuremath{\Conid{Par}} Monad, and Eden, a distributed memory parallel Haskell. These
languages represent orthogonal approaches. Some use a Monad, even if
only for the internal representation. Some introduce additional
language constructs. Section \ref{sec:parallelHaskells} gives a short overview over these languages.

A key novelty in this paper is to use Arrows to represent parallel computations. They seem a natural fit as they can be thought of as a more general function arrow (\ensuremath{\to }) and serve as general interface to computations while not being as restrictive as Monads \citep{HughesArrows}. Section \ref{sec:arrows} gives a short introduction to Arrows.

We provide an Arrows-based type class and implementations for the three above mentioned parallel Haskells.
Instead of 
introducing a new low-level parallel backend to implement our
Arrows-based interface, we define a shallow-embedded DSL for Arrows. This DSL
is defined as a common interface with varying implementations in
the existing parallel Haskells. Thus, we not only define a parallel programming interface in a
novel manner -- we tame the zoo of parallel Haskells. We provide a
common, very low-penalty programming interface that allows to switch
the parallel implementations at will. The induced penalty was in the single-digit percent range, with means typically under 2\% overhead in our measurements over the varying cores configuration (Section~\ref{sec:benchmarks}). Further implementations, based on HdpH or a Frege implementation (on the Java Virtual Machine), are viable, too.

\paragraph{Contributions.}
%
%
We propose an Arrow-based encoding for parallelism based on a new Arrow combinator \ensuremath{\Varid{parEvalN}\mathbin{::}[\mskip1.5mu \Varid{arr}\;\Varid{a}\;\Varid{b}\mskip1.5mu]\to \Varid{arr}\;[\mskip1.5mu \Varid{a}\mskip1.5mu]\;[\mskip1.5mu \Varid{b}\mskip1.5mu]}. A parallel Arrow is still an Arrow, hence the resulting parallel Arrow can still be used in the same way as a potential sequential version. In this paper we evaluate the expressive power of such a formalism in the context of parallel programming.

\begin{itemize}
\item We introduce a parallel evaluation formalism using Arrows. One big advantage of our specific approach is that we do not have to introduce any new types, facilitating composability (Section~\ref{sec:parallel-arrows}).
\item We show that PArrow programs can readily exploit multiple parallel language implementations. We demonstrate the use of GpH, a \ensuremath{\Conid{Par}} Monad, and Eden. We do not re-implement all the parallel internals, as we host this functionality in the \ensuremath{\Conid{ArrowParallel}} type class, which abstracts all parallel implementation logic. The implementations can easily be swapped, so we are not bound to any specific one.

This has many practical advantages. For example, during development we can run the program in a simple GHC-compiled variant using GpH and afterwards deploy it on a cluster by converting it into an Eden program, by just replacing the \ensuremath{\Conid{ArrowParallel}} instance and compiling with Eden's GHC variant (Section~\ref{sec:parallel-arrows}).
\item We extend the PArrows formalism with \ensuremath{\Conid{Future}}s to enable direct communication of data between nodes in a distributed memory setting similar to Eden's Remote Data \citep[\ensuremath{\Conid{RD}},][]{Dieterle2010}. Direct communication is useful in a distributed memory setting because it allows for inter-node communication without blocking the master-node. (Section~\ref{sec:futures})
\item We demonstrate the expressiveness of PArrows by using them to define common algorithmic skeletons (Section~\ref{sec:skeletons}), and by using these skeletons to implement four benchmarks (Section~\ref{sec:benchmarks}).
\item We practically demonstrate that Arrow parallelism has a low performance overhead compared with existing approaches, \eg the mean over all cores of relative mean overhead was less than $3.5\%$ and less than $0.8\%$ for all benchmarks with GpH and Eden, respectively. As for \ensuremath{\Conid{Par}} Monad, the mean of mean overheads was in our favour in all benchmarks (Section~\ref{sec:benchmarks}).
\end{itemize}

PArrows are open source and are available from \url{https://github.com/s4ke/Parrows}.
	\section{Related Work}
\label{sec:related-work}

\ptcomment{The non-strict semantics of Haskell, and the fact that reduction encapsulates computations as closures, makes it relatively easy to define alternate parallelisations. A range of approaches have been explored, including data parallelism (Chakravarty et al., 2007; Keller et al.,
2010), GPU-based approaches (Mainland & Morrisett, 2010; Svensson, 2011), software
transactional memory (Harris et al., 2005; Perfumo et al., 2008). The Haskell–GPU bridge
Accelerate (Chakravarty et al., 2011; Clifton-Everest et al., 2014; McDonell et al., 2015)
deserves a special mention. <<Why does it deserve special mention?>>  Accelerate is completely orthogonal to our approach. A good survey of parallel Haskells can be found in ~\cite{Marlow2013}.

Our PArrow implementation uses three task parallel languages as backends: the
GpH (Trinder et al., 1996, 1998) parallel Haskell dialect and its multicore version (Marlow
et al., 2009), the \ensuremath{\Conid{Par}} Monad (Marlow et al., 2011; Foltzer et al., 2012), and Eden (Loogen
et al., 2005; Loogen, 2012). These languages are under active development, for example a combined shared and distributed memory implementation of GpH is available (Aljabri et al., 2014,
2015). Research on Eden includes a low-level  implementation (Berthold, 2008; Berthold et al., 2016), work on skeleton
composition (Dieterle et al., 2016), communication (Dieterle et al., 2010a), and generation
of process networks (Horstmeyer & Loogen, 2013). The definitions of new Eden skeletons is a specific focus (Hammond et al., 2003; Berthold
& Loogen, 2006; Berthold et al., 2009b,c; Dieterle et al., 2010b; de la Encina et al., 2011;
Dieterle et al., 2013; Janjic et al., 2013).

Other task parallel Haskells, related to Eden, GpH, and the \ensuremath{\Conid{Par}} Monad, include the following.  
HdpH (Maier et al., 2014; Stewart et al., 2016) is an extension of \ensuremath{\Conid{Par}} Monad to
heterogeneous clusters. LVish (Kuper et al., 2014) is a communication-centred extension
of \ensuremath{\Conid{Par}} Monad.}

\paragraph{Parallel Haskells.}
The non-strict semantics of Haskell, and the fact that reduction encapsulates computations as closures, makes it relatively easy to define alternate parallelisations. A range of approaches have been explored, including data parallelism \cite{Chakravarty2007,Keller:2010:RSP:1932681.1863582}, GPU-based approaches \cite{Mainland:2010:NEC:2088456.1863533,obsidian-phd}, software transactional memory \cite{Harris:2005:CMT:1065944.1065952,Perfumo:2008:LST:1366230.1366241}.
The Haskell--GPU bridge Accelerate
\cite{Chakravarty:2011:AHA:1926354.1926358,CMCK14,McDonell:2015:TRC:2887747.2804313}
is completely orthogonal to our approach.
A~good survey of parallel Haskells can be found in~\citet{marlow2013parallel}.

Our PArrow implementation uses three task parallel languages as backends: the GpH \cite{Trinder1996,Trinder1998a} parallel Haskell dialect and its multicore version \cite{Marlow2009}, the \ensuremath{\Conid{Par}} Monad \cite{par-monad,Foltzer:2012:MPC:2398856.2364562}, and Eden \cite{eden,Loogen2012}. These languages are under active development, for example a combined shared and distributed memory implementation of GpH is available \cite{Aljabri:2013:DIG:2620678.2620682,Aljabri2015}.
Research on Eden includes low-level  implementation
\cite{JostThesis,berthold_loidl_hammond_2016}, skeleton composition
\cite{dieterle_horstmeyer_loogen_berthold_2016}, communication \cite{Dieterle2010}, and generation of process networks \cite{Horstmeyer2013}. The definitions of new Eden skeletons is a specific focus \cite{doi:10.1142/S0129626403001380,Eden:PARCO05,Berthold2009-mr,Berthold2009-fft,dieterle2010skeleton,delaEncina2011,Dieterle2013,janjic2013space}.

Other task parallel Haskells related to Eden, GpH, and the \ensuremath{\Conid{Par}} Monad include the following.  
HdpH \cite{Maier:2014:HDS:2775050.2633363,stewart_maier_trinder_2016} is an extension of \ensuremath{\Conid{Par}} Monad to
heterogeneous clusters. LVish \cite{Kuper:2014:TPE:2666356.2594312} is a communication-centred extension
of \ensuremath{\Conid{Par}} Monad.

\olcomment{OLD:
\paragraph{Parallel Haskells.}
Of course, the three parallel Haskell flavours we use as backends: the GpH \cite{Trinder1996,Trinder1998a} parallel Haskell dialect and its multicore version \cite{Marlow2009}, the \ensuremath{\Conid{Par}} Monad \cite{par-monad,Foltzer:2012:MPC:2398856.2364562}, and Eden \cite{eden,Loogen2012} are related to this work. We use these languages as backends: our DSL can switch from one to another at user's command.

HdpH \cite{Maier:2014:HDS:2775050.2633363,stewart_maier_trinder_2016} is an extension of \ensuremath{\Conid{Par}} Monad to heterogeneous clusters. LVish \cite{Kuper:2014:TPE:2666356.2594312} is a communication-centred extension of \ensuremath{\Conid{Par}} Monad.
Further parallel Haskell approaches include pH \cite{ph-book}, research work done on distributed variants of GpH \cite{Trinder1996,Aljabri:2013:DIG:2620678.2620682,Aljabri2015}, and low-level Eden implementation \cite{JostThesis,berthold_loidl_hammond_2016}. Skeleton composition \cite{dieterle_horstmeyer_loogen_berthold_2016}, communication \cite{Dieterle2010}, and generation of process networks \cite{Horstmeyer2013} are recent in-focus research topics in Eden. This also includes the definitions of new skeletons \cite{doi:10.1142/S0129626403001380,Eden:PARCO05,Berthold2009-mr,Berthold2009-fft,dieterle2010skeleton,delaEncina2011,Dieterle2013,janjic2013space}.

Alternative approaches include data parallelism \cite{Chakravarty2007,Keller:2010:RSP:1932681.1863582}, GPU-based approaches \cite{Mainland:2010:NEC:2088456.1863533,obsidian-phd}, software transactional memory \cite{Harris:2005:CMT:1065944.1065952,Perfumo:2008:LST:1366230.1366241}.
The Haskell--GPU bridge Accelerate
\cite{Chakravarty:2011:AHA:1926354.1926358,CMCK14,McDonell:2015:TRC:2887747.2804313}
deserves a special mention. Accelerate is completely orthogonal to our
approach. \citeauthor{marlow2013parallel} authored a recent book in
\citeyear{marlow2013parallel} on parallel Haskells.
}

\paragraph{Algorithmic skeletons.}
Algorithmic skeletons were introduced by \citet{Cole1989}.
Early publications on this topic include \cite{DANELUTTO1992205,darlington1993parallel,botorog1996efficient,Lengauer1997,Gorlatch1998}. \citet{SkeletonBook} consolidated early reports on high-level programming approaches.
Types of algorithmic skeletons include \ensuremath{\Varid{map}}-, \ensuremath{\Varid{fold}}-, and \ensuremath{\Varid{scan}}-based parallel programming patterns, special applications such as divide-and-conquer or topological skeletons.

The \ensuremath{\Varid{farm}} skeleton \citep{Hey1990185,Eden:PPDP01,Kuchen05} is a statically task-balanced parallel \ensuremath{\Varid{map}}. When tasks' durations cannot be foreseen, a dynamic load balancing (\ensuremath{\Varid{workpool}}) brings a lot of improvement \citep{Rudolph:1991:SLB:113379.113401,doi:10.1142/S0129626403001380,Hippold2006,PADL08HMWS,Marlow2009}. For special tasks \ensuremath{\Varid{workpool}} skeletons can be extended with dynamic task creation \cite{WPEuropar06,Dinan:2009:SWS:1654059.1654113,brown2010ever}. Efficient load-balancing schemes for \ensuremath{\Varid{workpool}}s are subject of research \cite{Blumofe:1999:SMC:324133.324234,Acar:2000:DLW:341800.341801,vanNieuwpoort:2001:ELB:568014.379563,Chase:2005:DCW:1073970.1073974,4625841,Michael:2009:IWS:1594835.1504186}.
The \ensuremath{\Varid{fold}} (or \ensuremath{\Varid{reduce}}) skeleton was implemented in various skeleton libraries \cite{Kuchen2002,5361825,BUONO20102095,Dastgeer:2011:ASM:1984693.1984697}, as also its inverse, \ensuremath{\Varid{scan}} \cite{Bischof2002,harris2007parallel}.
Google \ensuremath{\Varid{map}}--\ensuremath{\Varid{reduce}} \cite{Dean:2008:MSD:1327452.1327492,Dean:2010:MFD:1629175.1629198} is more special than just a composition of the two skeletons \cite{LAMMEL20081,Berthold2009-mr}.

The effort is ongoing, including topological skeletons \cite{Eden:PARCO05}, special-purpose skeletons for computer algebra \cite{Berthold2009-fft,lobachev-phd,Lobachev2012,janjic2013space}, iteration skeletons \cite{Dieterle2013}. The idea of \citet{scscp} is to use a parallel Haskell to orchestrate further software systems to run in parallel. \citet{dieterle_horstmeyer_loogen_berthold_2016} compare the composition of skeletons to stable process networks.

\paragraph{Arrows.}
Arrows were introduced by \citet{HughesArrows} as a less restrictive alternative to Monads, in essence they are a generalised function arrow~\ensuremath{\to }. \citet{Hughes2005} presents a tutorial on Arrows. \citet{jacobs_heunen_hasuo_2009,LINDLEY201197,ATKEY201119} develop theoretical background of Arrows. \citet{Paterson:2001:NNA:507669.507664} introduced a new notation for Arrows. Arrows have applications in information flow research \cite{1648705,LI20101974,Russo:2008:LLI:1411286.1411289}, invertible programming \cite{Alimarine:2005:BAA:1088348.1088357}, and quantum computer simulation \cite{vizzotto_altenkirch_sabry_2006}. But probably most prominent application of Arrows is Arrow-based functional reactive programming, AFRP \cite{Nilsson:2002:FRP:581690.581695,Hudak2003,Czaplicki:2013:AFR:2499370.2462161}.
\citet{Liu:2009:CCA:1631687.1596559} formally define a more special kind of Arrows that capsule the computation more than regular Arrows do and thus enable optimisations. Their approach would allow parallel composition, as their special Arrows would not interfere with each other in concurrent execution. In contrast, we capture a whole parallel computation as a single entity: our main instantiation function \ensuremath{\Varid{parEvalN}} makes a single (parallel) Arrow out of list of Arrows. \citet{Huang2007} utilise Arrows for parallelism, but strikingly different from our approach. They use Arrows to orchestrate several tasks in robotics. We, however, propose a general interface for parallel programming, while remaining completely in Haskell.

\paragraph{Arrows in other languages.}
Although this work is centred on Haskell implementation of Arrows, it is applicable to any functional programming language where parallel evaluation and Arrows can be defined. Basic definitions of PArrows are possible in the Frege language\footnote{GitHub project page at \url{https://github.com/Frege/frege}} (which is basically Haskell on the JVM). However, they are beyond the scope of this work, as are similar experiments with the Eta language\footnote{Eta project page at \url{http://eta-lang.org}}, 
a new approach to Haskell on the JVM.

\citet{achten2004arrows,achten2007arrow} use an Arrow implementation in Clean for better handling of typical GUI tasks. \citet{Dagand:2009:ORD:1481861.1481870} used Arrows in OCaml in the implementation of a distributed system.

        \section{Background}
	\label{sec:background}
	This section gives a short overview of Arrows
        (Section~\ref{sec:arrows}) and of GpH, the \ensuremath{\Conid{Par}} Monad, and
        Eden, the three parallel Haskells which we base our DSL on
        (Section~\ref{sec:parallelHaskells}).
	\subsection{Arrows}
\label{sec:arrows}
Arrows were introduced by \citet{HughesArrows} as a general interface for computation and a less restrictive generalisation of Monads. \citeauthor{HughesArrows} motivates the broader interface of Arrows with the example of a parser with added static meta-information that can not satisfy the monadic bind operator \ensuremath{(\bind )\mathbin{::}\Varid{m}\;\Varid{a}\to (\Varid{a}\to \Varid{m}\;\Varid{b})\to \Varid{m}\;\Varid{b}} (with \ensuremath{\Varid{m}} being a Monad)\footnote{In the example a parser of the type \ensuremath{\Conid{Parser}\;\Varid{s}\;\Varid{a}} with static meta information \ensuremath{\Varid{s}} and result \ensuremath{\Varid{a}} is shown to not be able to use the static information \ensuremath{\Varid{s}} without applying the monadic function \ensuremath{\Varid{a}\to \Varid{m}\;\Varid{b}}. With Arrows this is possible.}.

\begin{figure}[t]
\centering
\parbox[c][17em]{0.49\linewidth}{%
\vfill
\centering
	\includegraphics{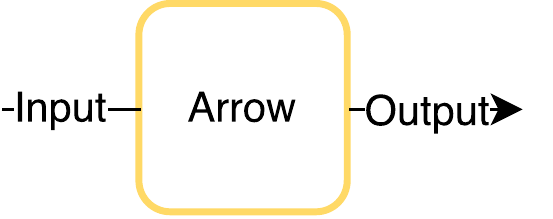}
\vfill
}
\parbox[c][17em]{0.49\linewidth}{%
\vfill
\centering
	{\includegraphics[scale=0.6]{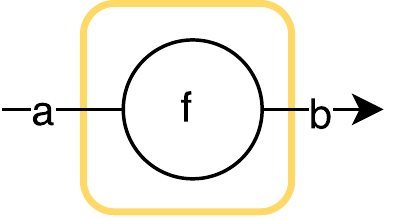}}
	{\includegraphics[scale=0.6]{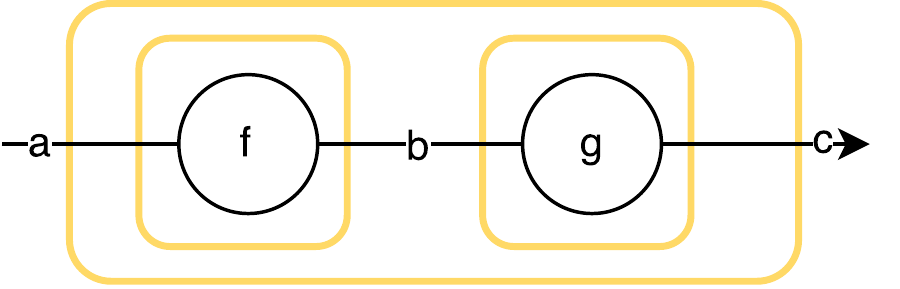}}
	{\includegraphics[scale=0.6]{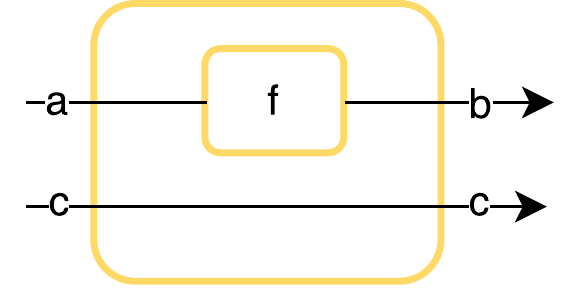}}
\vfill
}
\caption{Schematic depiction of  an Arrow (left) and its basic
  combinators \ensuremath{\Varid{arr}}, \ensuremath{\mathbin{>\!\!>\!\!>}} and \ensuremath{\Varid{first}} (right).}
\label{fig:arrow-sch}
\label{fig:arrows-viz}
\end{figure}

An Arrow \ensuremath{\Varid{arr}\;\Varid{a}\;\Varid{b}} represents a computation that converts an input \ensuremath{\Varid{a}} to an output \ensuremath{\Varid{b}}. This is defined in the \ensuremath{\Conid{Arrow}} type class shown in Fig.~\ref{fig:ArrowDefinition}.
To lift an ordinary function to an Arrow, \ensuremath{\Varid{arr}} is used, analogous to the monadic \ensuremath{\Varid{return}}. Similarly, the composition operator \ensuremath{\mathbin{>\!\!>\!\!>}} is analogous to the monadic composition \ensuremath{\bind } and combines two Arrows \ensuremath{\Varid{arr}\;\Varid{a}\;\Varid{b}} and \ensuremath{\Varid{arr}\;\Varid{b}\;\Varid{c}} by \enquote{wiring} the outputs of the first to the inputs to the second to get a new Arrow \ensuremath{\Varid{arr}\;\Varid{a}\;\Varid{c}}. Lastly, the \ensuremath{\Varid{first}} operator takes the input Arrow \ensuremath{\Varid{arr}\;\Varid{a}\;\Varid{b}} and converts it into an Arrow on pairs \ensuremath{\Varid{arr}\;(\Varid{a},\Varid{c})\;(\Varid{b},\Varid{c})} that leaves the second argument untouched. It allows us to to save input across Arrows. Figure~\ref{fig:arrows-viz} shows a graphical representation of these basic Arrow combinators.
The most prominent instances of this interface are regular functions \ensuremath{(\to )}
and the Kleisli type (Fig.~\ref{fig:ArrowDefinition}), which wraps monadic functions, e.g.  \ensuremath{\Varid{a}\to \Varid{m}\;\Varid{b}}.

\begin{figure}[t]
\centering
\begin{hscode}\SaveRestoreHook
\column{B}{@{}>{\hspre}l<{\hspost}@{}}%
\column{3}{@{}>{\hspre}l<{\hspost}@{}}%
\column{9}{@{}>{\hspre}l<{\hspost}@{}}%
\column{E}{@{}>{\hspre}l<{\hspost}@{}}%
\>[B]{}\mathbf{class}\;\Conid{Arrow}\;\Varid{arr}\;\mathbf{where}{}\<[E]%
\\
\>[B]{}\hsindent{3}{}\<[3]%
\>[3]{}\Varid{arr}\mathbin{::}(\Varid{a}\to \Varid{b})\to \Varid{arr}\;\Varid{a}\;\Varid{b}{}\<[E]%
\\
\>[B]{}\hsindent{3}{}\<[3]%
\>[3]{}(\mathbin{>\!\!>\!\!>})\mathbin{::}\Varid{arr}\;\Varid{a}\;\Varid{b}\to \Varid{arr}\;\Varid{b}\;\Varid{c}\to \Varid{arr}\;\Varid{a}\;\Varid{c}{}\<[E]%
\\
\>[B]{}\hsindent{3}{}\<[3]%
\>[3]{}\Varid{first}\mathbin{::}\Varid{arr}\;\Varid{a}\;\Varid{b}\to \Varid{arr}\;(\Varid{a},\Varid{c})\;(\Varid{b},\Varid{c}){}\<[E]%
\\[\blanklineskip]%
\>[B]{}\mathbf{instance}\;\Conid{Arrow}\;(\to )\;\mathbf{where}{}\<[E]%
\\
\>[B]{}\hsindent{9}{}\<[9]%
\>[9]{}\Varid{arr}\;\Varid{f}\mathrel{=}\Varid{f}{}\<[E]%
\\
\>[B]{}\hsindent{9}{}\<[9]%
\>[9]{}\Varid{f}\mathbin{>\!\!>\!\!>}\Varid{g}\mathrel{=}\Varid{g}\mathbin{\circ}\Varid{f}{}\<[E]%
\\
\>[B]{}\hsindent{9}{}\<[9]%
\>[9]{}\Varid{first}\;\Varid{f}\mathrel{=}\lambda (\Varid{a},\Varid{c})\to (\Varid{f}\;\Varid{a},\Varid{c}){}\<[E]%
\\[\blanklineskip]%
\>[B]{}\mathbf{data}\;\Conid{Kleisli}\;\Varid{m}\;\Varid{a}\;\Varid{b}\mathrel{=}\Conid{Kleisli}\;\{\mskip1.5mu \Varid{run}\mathbin{::}\Varid{a}\to \Varid{m}\;\Varid{b}\mskip1.5mu\}{}\<[E]%
\\[\blanklineskip]%
\>[B]{}\mathbf{instance}\;\Conid{Monad}\;\Varid{m}\Rightarrow \Conid{Arrow}\;(\Conid{Kleisli}\;\Varid{m})\;\mathbf{where}{}\<[E]%
\\
\>[B]{}\hsindent{9}{}\<[9]%
\>[9]{}\Varid{arr}\;\Varid{f}\mathrel{=}\Conid{Kleisli}\;(\Varid{return}\mathbin{\circ}\Varid{f}){}\<[E]%
\\
\>[B]{}\hsindent{9}{}\<[9]%
\>[9]{}\Varid{f}\mathbin{>\!\!>\!\!>}\Varid{g}\mathrel{=}\Conid{Kleisli}\;(\lambda \Varid{a}\to \Varid{f}\;\Varid{a}\bind \Varid{g}){}\<[E]%
\\
\>[B]{}\hsindent{9}{}\<[9]%
\>[9]{}\Varid{first}\;\Varid{f}\mathrel{=}\Conid{Kleisli}\;(\lambda (\Varid{a},\Varid{c})\to \Varid{f}\;\Varid{a}\bind \lambda \Varid{b}\to \Varid{return}\;(\Varid{b},\Varid{c})){}\<[E]%
\ColumnHook
\end{hscode}\resethooks
\vfill
\caption{The \ensuremath{\Conid{Arrow}} type class and its two most typical instances.}
\label{fig:ArrowDefinition}
\end{figure}

\begin{figure}[h]
	\centering
	\begin{tabular}{cc}
{\label{t1}}{\includegraphics[width = 1.5in]{first}} &
{\label{fig:secondImg}}{\includegraphics[width = 1.5in]{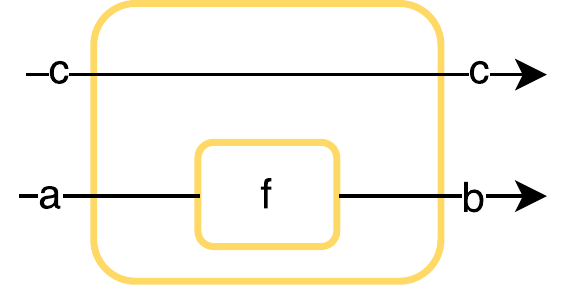}} \\
\ensuremath{\Varid{first}} & \ensuremath{\Varid{second}} \\
\midrule
{}{\includegraphics[width = 1.5in]{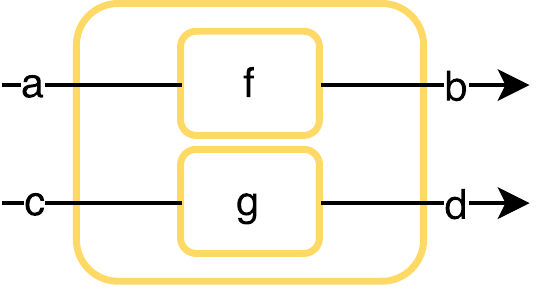}} &
{}{\includegraphics[width = 1.5in]{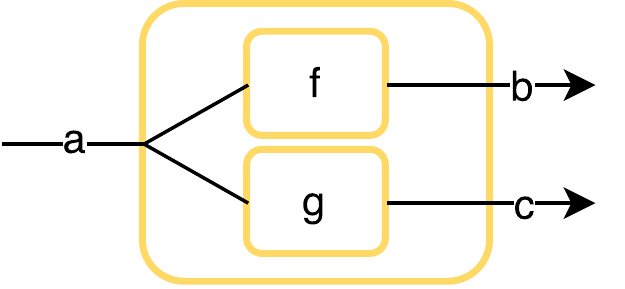}}\\
\ensuremath{(\mathbin{*\!*\!*})}\label{fig:***Img} & \ensuremath{(\mathbin{\&\!\&\!\&})} \label{fig:&&&Img} \\
	\end{tabular}
	\caption{Visual depiction of syntactic sugar for Arrows.}
	\label{fig:syntacticSugarArrows}
\end{figure}
Hughes also defined some syntactic sugar (Fig.~\ref{fig:syntacticSugarArrows}): \ensuremath{\Varid{second}}, \ensuremath{\mathbin{*\!*\!*}} and \ensuremath{\mathbin{\&\!\&\!\&}}. 
\ensuremath{\Varid{second}} is the mirrored version of \ensuremath{\Varid{first}} (Appendix~\ref{utilfns}).
The \ensuremath{\mathbin{*\!*\!*}} function combines \ensuremath{\Varid{first}} and \ensuremath{\Varid{second}} to handle two inputs in one arrow, and is defined as follows:
\begin{hscode}\SaveRestoreHook
\column{B}{@{}>{\hspre}l<{\hspost}@{}}%
\column{E}{@{}>{\hspre}l<{\hspost}@{}}%
\>[B]{}(\mathbin{*\!*\!*})\mathbin{::}\Conid{Arrow}\;\Varid{arr}\Rightarrow \Varid{arr}\;\Varid{a}\;\Varid{b}\to \Varid{arr}\;\Varid{c}\;\Varid{d}\to \Varid{arr}\;(\Varid{a},\Varid{c})\;(\Varid{b},\Varid{d}){}\<[E]%
\\
\>[B]{}\Varid{f}\mathbin{*\!*\!*}\Varid{g}\mathrel{=}\Varid{first}\;\Varid{f}\mathbin{>\!\!>\!\!>}\Varid{second}\;\Varid{g}{}\<[E]%
\ColumnHook
\end{hscode}\resethooks
The \ensuremath{\mathbin{\&\!\&\!\&}} combinator, which constructs an Arrow that outputs two different values like \ensuremath{\mathbin{*\!*\!*}}, but takes only one input, is:
\begin{hscode}\SaveRestoreHook
\column{B}{@{}>{\hspre}l<{\hspost}@{}}%
\column{E}{@{}>{\hspre}l<{\hspost}@{}}%
\>[B]{}(\mathbin{\&\!\&\!\&})\mathbin{::}\Conid{Arrow}\;\Varid{arr}\Rightarrow \Varid{arr}\;\Varid{a}\;\Varid{b}\to \Varid{arr}\;\Varid{a}\;\Varid{c}\to \Varid{arr}\;\Varid{a}\;(\Varid{b},\Varid{c}){}\<[E]%
\\
\>[B]{}\Varid{f}\mathbin{\&\!\&\!\&}\Varid{g}\mathrel{=}\Varid{arr}\;(\lambda \Varid{a}\to (\Varid{a},\Varid{a}))\mathbin{>\!\!>\!\!>}\Varid{f}\mathbin{*\!*\!*}\Varid{g}{}\<[E]%
\ColumnHook
\end{hscode}\resethooks
A~first short example given by Hughes on how to use Arrows is addition with Arrows:
\begin{hscode}\SaveRestoreHook
\column{B}{@{}>{\hspre}l<{\hspost}@{}}%
\column{E}{@{}>{\hspre}l<{\hspost}@{}}%
\>[B]{}\Varid{add}\mathbin{::}\Conid{Arrow}\;\Varid{arr}\Rightarrow \Varid{arr}\;\Varid{a}\;\Conid{Int}\to \Varid{arr}\;\Varid{a}\;\Conid{Int}\to \Varid{arr}\;\Varid{a}\;\Conid{Int}{}\<[E]%
\\
\>[B]{}\Varid{add}\;\Varid{f}\;\Varid{g}\mathrel{=}\Varid{f}\mathbin{\&\!\&\!\&}\Varid{g}\mathbin{>\!\!>\!\!>}\Varid{arr}\;(\lambda (\Varid{u},\Varid{v})\to \Varid{u}\mathbin{+}\Varid{v}){}\<[E]%
\ColumnHook
\end{hscode}\resethooks

As we can rewrite the monadic bind operation \ensuremath{(\bind )} with only the Kleisli type into \ensuremath{\Varid{m}\;\Varid{a}\to \Conid{Kleisli}\;\Varid{m}\;\Varid{a}\;\Varid{b}\to \Varid{m}\;\Varid{b}}, but not with a general Arrow \ensuremath{\Varid{arr}\;\Varid{a}\;\Varid{b}}, we can intuitively get an idea of why Arrows must be a generalisation of Monads. While this also means that a general Arrow can not express everything a Monad can, \citet{HughesArrows} shows in his parser example that this trade-off is worth it in some cases.

In this paper we will show that parallel computations can be expressed with this more general interface of Arrows without requiring Monads. We also do not restrict the compatible Arrows to ones which have \ensuremath{\Conid{ArrowApply}} instances but instead only require instances for \ensuremath{\Conid{ArrowChoice}} (for if-then-else constructs) and \ensuremath{\Conid{ArrowLoop}} (for looping). Because of this, we have a truly more general interface as compared to a monadic one.

While we could have based our DSL on Profunctors as well, we chose Arrows\mbcomment{cite missing?} for this paper since they they allow for a more direct way of thinking about parallelism than general Profunctors because of their composability. However, they are a promising candidate for future improvements of our DSL. Some Profunctors, especially ones supporting a composition operation, choice, and looping, can already be adapted to our interface as shown in Appendix~\ref{app:profunctorArrows}.
	\subsection{Short introduction to parallel Haskells}
\label{sec:parallelHaskells}
\label{sec:parEvalNIntro}
\begin{figure}[t]
  \centering
	\includegraphics[scale=0.7]{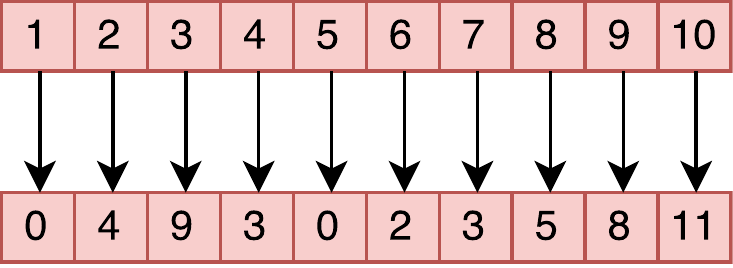}
	\caption{Schematic illustration of \ensuremath{\Varid{parEvalN}}. A list of inputs is transformed by different functions in parallel.}
	\label{fig:parEvalN}
\end{figure}

In its purest form, parallel computation (on functions) can be looked at as the execution of some functions \ensuremath{\Varid{a}\to \Varid{b}} in parallel or \ensuremath{\Varid{parEvalN}\mathbin{::}[\mskip1.5mu \Varid{a}\to \Varid{b}\mskip1.5mu]\to [\mskip1.5mu \Varid{a}\mskip1.5mu]\to [\mskip1.5mu \Varid{b}\mskip1.5mu]}, as also Figure~\ref{fig:parEvalN} symbolically shows.

In this section, we will implement this non-Arrow version which will later be adapted for usage in our Arrow-based parallel Haskell.

There exist several parallel Haskells already. Among the most important are probably GpH \citep[based on \ensuremath{\Varid{par}} and \ensuremath{\Varid{pseq}} \enquote{hints},][]{Trinder1996,Trinder1998a}, the \ensuremath{\Conid{Par}} Monad \citep[a monad for deterministic parallelism,][]{par-monad,Foltzer:2012:MPC:2398856.2364562}, Eden \citep[a parallel Haskell for distributed memory,][]{eden,Loogen2012}, HdpH \citep[a Template Haskell-based parallel Haskell for distributed memory,][]{Maier:2014:HDS:2775050.2633363,stewart_maier_trinder_2016} and LVish \citep[a \ensuremath{\Conid{Par}} extension with focus on communication,][]{Kuper:2014:TPE:2666356.2594312}.

As the goal of this paper is not to re-implement yet another parallel runtime, but to represent parallelism with Arrows, we base our efforts on existing work which we wrap as backends behind a common interface. For this paper we chose GpH for its simplicity, the \ensuremath{\Conid{Par}} Monad to represent a monadic DSL, and Eden as a distributed parallel Haskell.

LVish and HdpH were not chosen as the former does not differ from the original \ensuremath{\Conid{Par}} Monad with regard to how we would have used it in this paper, while the latter (at least in its current form) does not comply with our representation of parallelism due to its heavy reliance on Template Haskell.

We will now go into some detail on GpH, the \ensuremath{\Conid{Par}} Monad and Eden, and also give their respective implementations of the non-Arrow version of \ensuremath{\Varid{parEvalN}}.


\subsubsection{Glasgow parallel Haskell -- GpH}
\label{sec:GpHIntro}
GpH \cite{Marlow2009,Trinder1998a} is one of the simplest ways to do parallel processing found in standard GHC.\footnote{The Multicore implementation of GpH is available on Hackage under \url{https://hackage.haskell.org/package/parallel-3.2.1.0}, compiler support is integrated in the stock GHC.} Besides some basic primitives (\ensuremath{\Varid{par}} and \ensuremath{\Varid{pseq}}), it ships with parallel evaluation strategies for several types which can be applied with \ensuremath{\Varid{using}\mathbin{::}\Varid{a}\to \Conid{Strategy}\;\Varid{a}\to \Varid{a}}, which is exactly what is required for an implementation of \ensuremath{\Varid{parEvalN}}.

\begin{hscode}\SaveRestoreHook
\column{B}{@{}>{\hspre}l<{\hspost}@{}}%
\column{18}{@{}>{\hspre}l<{\hspost}@{}}%
\column{E}{@{}>{\hspre}l<{\hspost}@{}}%
\>[B]{}\Varid{parEvalN}\mathbin{::}(\Conid{NFData}\;\Varid{b})\Rightarrow [\mskip1.5mu \Varid{a}\to \Varid{b}\mskip1.5mu]\to [\mskip1.5mu \Varid{a}\mskip1.5mu]\to [\mskip1.5mu \Varid{b}\mskip1.5mu]{}\<[E]%
\\
\>[B]{}\Varid{parEvalN}\;\Varid{fs}\;\Varid{as}\mathrel{=}\mathbf{let}\;\Varid{bs}\mathrel{=}\Varid{zipWith}\;(\mathbin{\$})\;\Varid{fs}\;\Varid{as}{}\<[E]%
\\
\>[B]{}\hsindent{18}{}\<[18]%
\>[18]{}\mathbf{in}\;\Varid{bs}\mathbin{`\Varid{using}`}\Varid{parList}\;\Varid{rdeepseq}{}\<[E]%
\ColumnHook
\end{hscode}\resethooks

\begin{figure}[t]
	\includegraphics[scale=0.5]{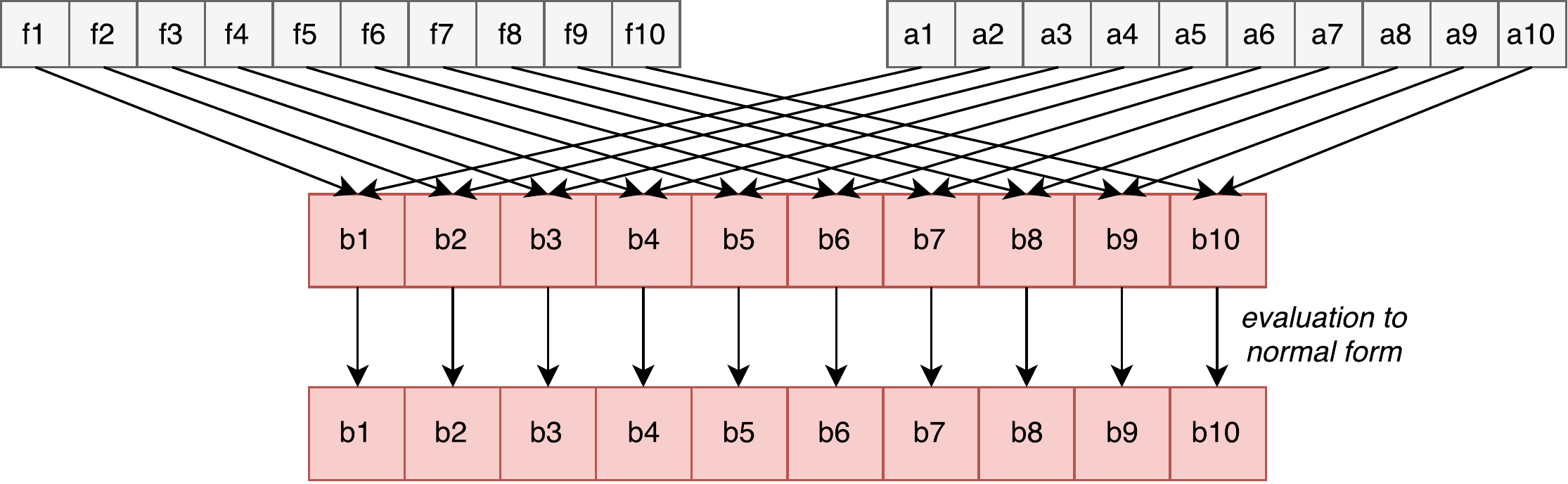}
	\caption{\ensuremath{\Varid{parEvalN}} (GpH).}
	\label{fig:parEvalNMulticoreImg}
\end{figure}

In the above definition of \ensuremath{\Varid{parEvalN}} we just apply the list of functions \ensuremath{[\mskip1.5mu \Varid{a}\to \Varid{b}\mskip1.5mu]} to the list of inputs \ensuremath{[\mskip1.5mu \Varid{a}\mskip1.5mu]} by zipping them with the application operator \ensuremath{\mathbin{\$}}. 
We then evaluate this lazy list \ensuremath{[\mskip1.5mu \Varid{b}\mskip1.5mu]} according to a \ensuremath{\Conid{Strategy}\;[\mskip1.5mu \Varid{b}\mskip1.5mu]} with the \ensuremath{\Varid{using}\mathbin{::}\Varid{a}\to \Conid{Strategy}\;\Varid{a}\to \Varid{a}} operator. We construct this strategy with \ensuremath{\Varid{parList}\mathbin{::}\Conid{Strategy}\;\Varid{a}\to \Conid{Strategy}\;[\mskip1.5mu \Varid{a}\mskip1.5mu]} and \ensuremath{\Varid{rdeepseq}\mathbin{::}\Conid{NFData}\;\Varid{a}\Rightarrow \Conid{Strategy}\;\Varid{a}} where the latter is a strategy which evaluates to normal form. Other strategies like e.g. evaluation to weak head normal form are available as well. It also allows for custom \ensuremath{\Conid{Strategy}} implementations to be used.
Fig.~\ref{fig:parEvalNMulticoreImg} shows a visual representation of this code.


\subsubsection{\ensuremath{\Conid{Par}} Monad}
The \ensuremath{\Conid{Par}} Monad\footnote{The \ensuremath{\Conid{Par}} monad can be found in the \texttt{monad-par} package on Hackage under \url{https://hackage.haskell.org/package/monad-par-0.3.4.8/}.} introduced by \citet{par-monad}, is a Monad designed for composition of parallel programs. Let:

\begin{hscode}\SaveRestoreHook
\column{B}{@{}>{\hspre}l<{\hspost}@{}}%
\column{9}{@{}>{\hspre}l<{\hspost}@{}}%
\column{E}{@{}>{\hspre}l<{\hspost}@{}}%
\>[B]{}\Varid{parEvalN}\mathbin{::}(\Conid{NFData}\;\Varid{b})\Rightarrow [\mskip1.5mu \Varid{a}\to \Varid{b}\mskip1.5mu]\to [\mskip1.5mu \Varid{a}\mskip1.5mu]\to [\mskip1.5mu \Varid{b}\mskip1.5mu]{}\<[E]%
\\
\>[B]{}\Varid{parEvalN}\;\Varid{fs}\;\Varid{as}\mathrel{=}\Varid{runPar}\mathbin{\$}{}\<[E]%
\\
\>[B]{}\hsindent{9}{}\<[9]%
\>[9]{}(\Varid{sequenceA}\;(\Varid{map}\;(\Varid{return}\mathbin{\circ}\Varid{spawn})\;(\Varid{zipWith}\;(\mathbin{\$})\;\Varid{fs}\;\Varid{as})))\bind \Varid{mapM}\;\Varid{get}{}\<[E]%
\ColumnHook
\end{hscode}\resethooks

\begin{figure}[t]
	\includegraphics[scale=0.5]{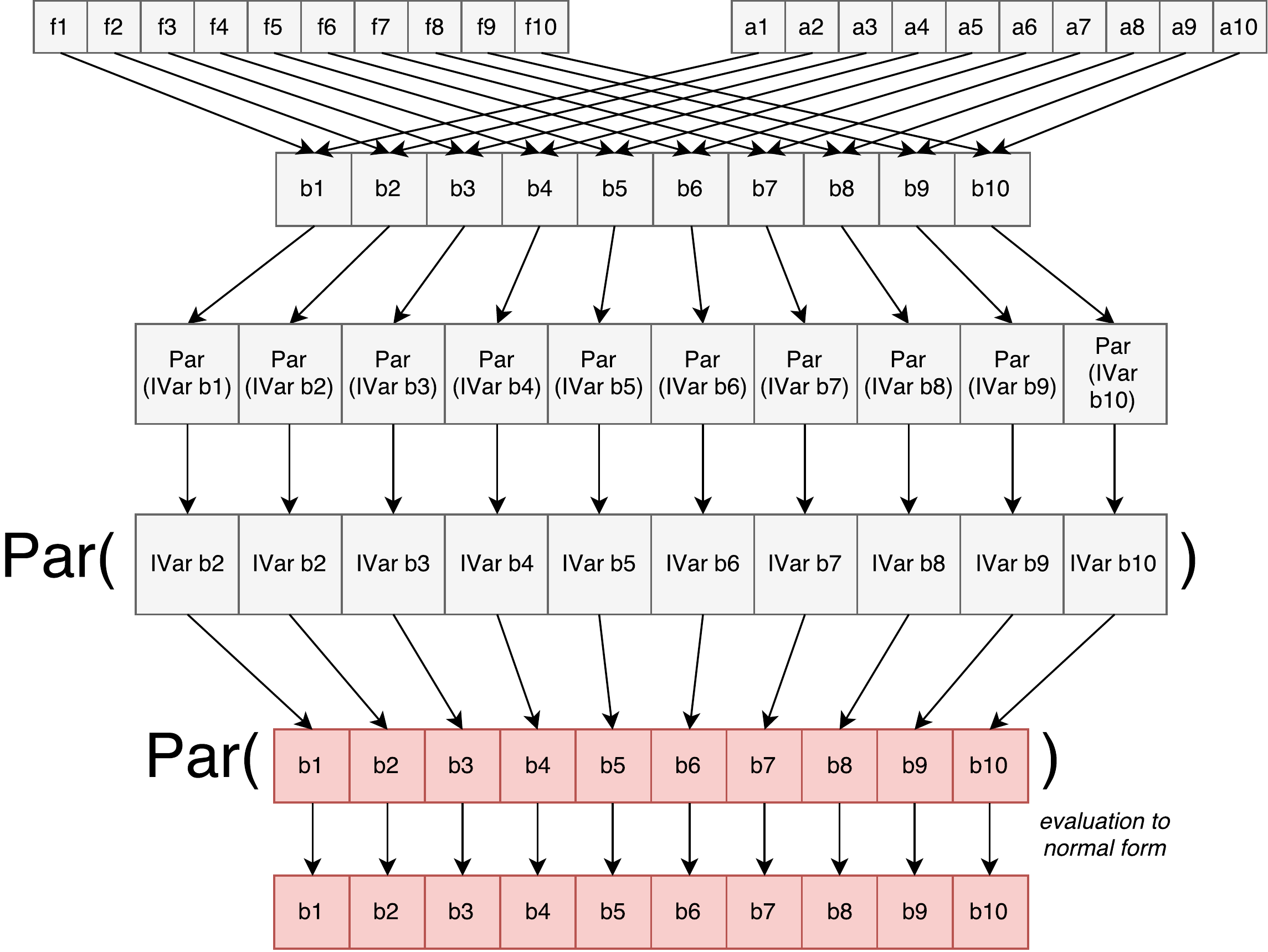}
	\caption{\ensuremath{\Varid{parEvalN}} (\ensuremath{\Conid{Par}} Monad).}
	\label{fig:parEvalNParMonadImg}
\end{figure}

The \ensuremath{\Conid{Par}} Monad version of our parallel evaluation function \ensuremath{\Varid{parEvalN}} is defined by zipping the list of \ensuremath{[\mskip1.5mu \Varid{a}\to \Varid{b}\mskip1.5mu]} with the list of inputs \ensuremath{[\mskip1.5mu \Varid{a}\mskip1.5mu]} with the application operator \ensuremath{\mathbin{\$}} just like with GpH. 
Then, we map over this not yet evaluated lazy list of results \ensuremath{[\mskip1.5mu \Varid{b}\mskip1.5mu]} with \ensuremath{\Varid{spawn}\mathbin{::}\Conid{NFData}\;\Varid{a}\Rightarrow \Conid{Par}\;\Varid{a}\to \Conid{Par}\;(\Conid{IVar}\;\Varid{a})} to transform them to a list of not yet evaluated forked away computations \ensuremath{[\mskip1.5mu \Conid{Par}\;(\Conid{IVar}\;\Varid{b})\mskip1.5mu]}, which we convert to \ensuremath{\Conid{Par}\;[\mskip1.5mu \Conid{IVar}\;\Varid{b}\mskip1.5mu]} with \ensuremath{\Varid{sequenceA}}. We wait for the computations to finish by mapping over the \ensuremath{\Conid{IVar}\;\Varid{b}} values inside the \ensuremath{\Conid{Par}} Monad with \ensuremath{\Varid{get}}. This results in \ensuremath{\Conid{Par}\;[\mskip1.5mu \Varid{b}\mskip1.5mu]}. We execute this process with \ensuremath{\Varid{runPar}} to finally get \ensuremath{[\mskip1.5mu \Varid{b}\mskip1.5mu]}. While we used \ensuremath{\Varid{spawn}} in the definition above, a head-strict variant can easily be defined by replacing \ensuremath{\Varid{spawn}} with \ensuremath{\Varid{spawn\char95 }\mathbin{::}\Conid{Par}\;\Varid{a}\to \Conid{Par}\;(\Conid{IVar}\;\Varid{a})}.
Fig.~\ref{fig:parEvalNParMonadImg} shows a graphical representation.

\subsubsection{Eden}
Eden \cite{eden,Loogen2012} is a parallel Haskell for distributed memory and comes with MPI and PVM as distributed backends.\footnote{The projects homepage can be found at \url{http://www.mathematik.uni-marburg.de/~eden/}. The Hackage page is at \url{https://hackage.haskell.org/package/edenmodules-1.2.0.0/}.} It is targeted towards clusters, but also functions well in a shared-memory setting with a further simple backend. However, in contrast to many other parallel Haskells, in Eden each process has its own heap. This seems to be a waste of memory, but with distributed programming paradigm and individual GC per process, Eden yields good performance results on multicores, as well \cite{arcs-dc,aswad2009low}.

While Eden comes with a Monad \ensuremath{\Conid{PA}} for parallel evaluation, it also ships with a completely functional interface that includes
a \ensuremath{\Varid{spawnF}\mathbin{::}(\Conid{Trans}\;\Varid{a},\Conid{Trans}\;\Varid{b})\Rightarrow [\mskip1.5mu \Varid{a}\to \Varid{b}\mskip1.5mu]\to [\mskip1.5mu \Varid{a}\mskip1.5mu]\to [\mskip1.5mu \Varid{b}\mskip1.5mu]}
function that
allows us to define \ensuremath{\Varid{parEvalN}} directly:

\begin{hscode}\SaveRestoreHook
\column{B}{@{}>{\hspre}l<{\hspost}@{}}%
\column{E}{@{}>{\hspre}l<{\hspost}@{}}%
\>[B]{}\Varid{parEvalN}\mathbin{::}(\Conid{Trans}\;\Varid{a},\Conid{Trans}\;\Varid{b})\Rightarrow [\mskip1.5mu \Varid{a}\to \Varid{b}\mskip1.5mu]\to [\mskip1.5mu \Varid{a}\mskip1.5mu]\to [\mskip1.5mu \Varid{b}\mskip1.5mu]{}\<[E]%
\\
\>[B]{}\Varid{parEvalN}\mathrel{=}\Varid{spawnF}{}\<[E]%
\ColumnHook
\end{hscode}\resethooks

\paragraph{Eden TraceViewer.}
\label{sec:edentv}
To comprehend the efficiency and the lack thereof in a parallel program, an inspection of its execution is extremely helpful. While some large-scale solutions exist \cite{Geimer2010}, the parallel Haskell community mainly utilises the tools Threadscope \cite{Wheeler2009} and Eden TraceViewer\footnote{See \url{http://hackage.haskell.org/package/edentv} on Hackage for the last available version of Eden TraceViewer.} \cite{Berthold2007a}. In the next sections we will present some \emph{trace visualisations}, the post-mortem process diagrams of Eden processes and their activity.

The trace visualisations are colour-coded. In such a visualisation (Fig.~\ref{fig:withoutFutures}), the $x$ axis shows the time, the $y$ axis enumerates the machines and processes. The visualisation shows a running process in green, a blocked process is red. If the process is \enquote{runnable}, \ie it may run, but does not, it is yellow. The typical reason for this is GC. An inactive machine, where no processes are started yet, or all are already terminated, shows as a blue bar. A~communication from one process to another is represented with a black arrow. A~stream of communications, \eg a transmitted list is shows as a dark shading between sender and receiver processes.

	\section{Parallel Arrows}
\label{sec:parallel-arrows}
While Arrows are a general interface to computation, we introduce here specialised Arrows as a general interface to \textit{parallel computations}. We present the \ensuremath{\Conid{ArrowParallel}} type class and explain the reasoning behind it before discussing some parallel Haskell implementations and basic extensions.
\subsection{The \ensuremath{\Conid{ArrowParallel}} type class}
A~parallel computation (on functions) can be seen as execution of some functions \ensuremath{\Varid{a}\to \Varid{b}} in parallel, as our \ensuremath{\Varid{parEvalN}} prototype shows (Section~\ref{sec:parEvalNIntro}).
Translating this into Arrow terms gives us a new operator \ensuremath{\Varid{parEvalN}} that lifts a list of Arrows \ensuremath{[\mskip1.5mu \Varid{arr}\;\Varid{a}\;\Varid{b}\mskip1.5mu]} to a parallel Arrow \ensuremath{\Varid{arr}\;[\mskip1.5mu \Varid{a}\mskip1.5mu]\;[\mskip1.5mu \Varid{b}\mskip1.5mu]}. 
This combinator is similar to \ensuremath{\Varid{evalN}} from Appendix~\ref{utilfns}, but does parallel instead of serial evaluation.
\begin{hscode}\SaveRestoreHook
\column{B}{@{}>{\hspre}l<{\hspost}@{}}%
\column{E}{@{}>{\hspre}l<{\hspost}@{}}%
\>[B]{}\Varid{parEvalN}\mathbin{::}(\Conid{Arrow}\;\Varid{arr})\Rightarrow [\mskip1.5mu \Varid{arr}\;\Varid{a}\;\Varid{b}\mskip1.5mu]\to \Varid{arr}\;[\mskip1.5mu \Varid{a}\mskip1.5mu]\;[\mskip1.5mu \Varid{b}\mskip1.5mu]{}\<[E]%
\ColumnHook
\end{hscode}\resethooks
With this definition of \ensuremath{\Varid{parEvalN}}, parallel execution is yet another Arrow combinator. But as the implementation may differ depending on the actual type of the Arrow \ensuremath{\Varid{arr}} - or even the input \ensuremath{\Varid{a}} and output \ensuremath{\Varid{b}} - and we want this to be an interface for different backends, we introduce a new type class \ensuremath{\Conid{ArrowParallel}\;\Varid{arr}\;\Varid{a}\;\Varid{b}}: 
\begin{hscode}\SaveRestoreHook
\column{B}{@{}>{\hspre}l<{\hspost}@{}}%
\column{9}{@{}>{\hspre}l<{\hspost}@{}}%
\column{E}{@{}>{\hspre}l<{\hspost}@{}}%
\>[B]{}\mathbf{class}\;\Conid{Arrow}\;\Varid{arr}\Rightarrow \Conid{ArrowParallel}\;\Varid{arr}\;\Varid{a}\;\Varid{b}\;\mathbf{where}{}\<[E]%
\\
\>[B]{}\hsindent{9}{}\<[9]%
\>[9]{}\Varid{parEvalN}\mathbin{::}[\mskip1.5mu \Varid{arr}\;\Varid{a}\;\Varid{b}\mskip1.5mu]\to \Varid{arr}\;[\mskip1.5mu \Varid{a}\mskip1.5mu]\;[\mskip1.5mu \Varid{b}\mskip1.5mu]{}\<[E]%
\ColumnHook
\end{hscode}\resethooks
Sometimes parallel Haskells require or allow for additional configuration parameters, \eg an information about the execution environment or the level of evaluation (weak head normal form vs. normal form). For this reason we introduce an additional \ensuremath{\Varid{conf}} parameter as we do not want \ensuremath{\Varid{conf}} to be a fixed type, as the configuration parameters can differ for different instances of \ensuremath{\Conid{ArrowParallel}}. 
\begin{hscode}\SaveRestoreHook
\column{B}{@{}>{\hspre}l<{\hspost}@{}}%
\column{9}{@{}>{\hspre}l<{\hspost}@{}}%
\column{E}{@{}>{\hspre}l<{\hspost}@{}}%
\>[B]{}\mathbf{class}\;\Conid{Arrow}\;\Varid{arr}\Rightarrow \Conid{ArrowParallel}\;\Varid{arr}\;\Varid{a}\;\Varid{b}\;\Varid{conf}\;\mathbf{where}{}\<[E]%
\\
\>[B]{}\hsindent{9}{}\<[9]%
\>[9]{}\Varid{parEvalN}\mathbin{::}\Varid{conf}\to [\mskip1.5mu \Varid{arr}\;\Varid{a}\;\Varid{b}\mskip1.5mu]\to \Varid{arr}\;[\mskip1.5mu \Varid{a}\mskip1.5mu]\;[\mskip1.5mu \Varid{b}\mskip1.5mu]{}\<[E]%
\ColumnHook
\end{hscode}\resethooks
By restricting the implementations of our backends to a specific \ensuremath{\Varid{conf}} type, we also get interoperability between backends for free. We can parallelize one part of a program using one backend, and parallelize the next with another one.

\subsection{\ensuremath{\Conid{ArrowParallel}} instances}
With the type class defined, we will now give implementations of it with GpH, the \ensuremath{\Conid{Par}} Monad and Eden.
\subsubsection{Glasgow parallel Haskell} \label{sec:parrows:multicore}
The GpH implementation of \ensuremath{\Conid{ArrowParallel}} is implemented in a straightforward manner in Fig.~\ref{fig:ArrowParallelMulticore}, but a bit different compared to the variant from Section \ref{sec:GpHIntro}. We use \ensuremath{\Varid{evalN}\mathbin{::}[\mskip1.5mu \Varid{arr}\;\Varid{a}\;\Varid{b}\mskip1.5mu]\to \Varid{arr}\;[\mskip1.5mu \Varid{a}\mskip1.5mu]\;[\mskip1.5mu \Varid{b}\mskip1.5mu]} (definition in Appendix~\ref{utilfns}, think \ensuremath{\Varid{zipWith}\;(\mathbin{\$})} on Arrows) combined with \ensuremath{\Varid{withStrategy}\mathbin{::}\Conid{Strategy}\;\Varid{a}\to \Varid{a}\to \Varid{a}} from GpH, where \ensuremath{\Varid{withStrategy}} is the same as \ensuremath{\Varid{using}\mathbin{::}\Varid{a}\to \Conid{Strategy}\;\Varid{a}\to \Varid{a}}, but with flipped parameters. Our \ensuremath{\Conid{Conf}\;\Varid{a}} datatype simply wraps a \ensuremath{\Conid{Strategy}\;\Varid{a}}, but could be extended in future versions of our DSL.
\begin{figure}[t]
\begin{hscode}\SaveRestoreHook
\column{B}{@{}>{\hspre}l<{\hspost}@{}}%
\column{3}{@{}>{\hspre}l<{\hspost}@{}}%
\column{5}{@{}>{\hspre}l<{\hspost}@{}}%
\column{9}{@{}>{\hspre}l<{\hspost}@{}}%
\column{E}{@{}>{\hspre}l<{\hspost}@{}}%
\>[B]{}\mathbf{data}\;\Conid{Conf}\;\Varid{a}\mathrel{=}\Conid{Conf}\;(\Conid{Strategy}\;\Varid{a}){}\<[E]%
\\[\blanklineskip]%
\>[B]{}\mathbf{instance}\;(\Conid{ArrowChoice}\;\Varid{arr})\Rightarrow {}\<[E]%
\\
\>[B]{}\hsindent{3}{}\<[3]%
\>[3]{}\Conid{ArrowParallel}\;\Varid{arr}\;\Varid{a}\;\Varid{b}\;(\Conid{Conf}\;\Varid{b})\;\mathbf{where}{}\<[E]%
\\
\>[3]{}\hsindent{2}{}\<[5]%
\>[5]{}\Varid{parEvalN}\;(\Conid{Conf}\;\Varid{strat})\;\Varid{fs}\mathrel{=}{}\<[E]%
\\
\>[5]{}\hsindent{4}{}\<[9]%
\>[9]{}\Varid{evalN}\;\Varid{fs}\mathbin{>\!\!>\!\!>}{}\<[E]%
\\
\>[5]{}\hsindent{4}{}\<[9]%
\>[9]{}\Varid{arr}\;(\Varid{withStrategy}\;(\Varid{parList}\;\Varid{strat})){}\<[E]%
\ColumnHook
\end{hscode}\resethooks
\caption{GpH \ensuremath{\Conid{ArrowParallel}} instance.}
\label{fig:ArrowParallelMulticore}
\end{figure}

\subsubsection{\ensuremath{\Conid{Par}} Monad}
\olcomment{introduce a newcommand for par-monad, "Arrows", "parrows" and replace all mentions to them to ensure uniform typesetting \done, we write Arrows. also "Monad"? \done}
As for GpH we can easily lift the definition of \ensuremath{\Varid{parEvalN}} for the \ensuremath{\Conid{Par}} Monad to Arrows in Fig.~\ref{fig:ArrowParallelParMonad}. To start off, we define the \ensuremath{\Conid{Strategy}\;\Varid{a}} and \ensuremath{\Conid{Conf}\;\Varid{a}} type so we can have a configurable instance of ArrowParallel:
\begin{hscode}\SaveRestoreHook
\column{B}{@{}>{\hspre}l<{\hspost}@{}}%
\column{E}{@{}>{\hspre}l<{\hspost}@{}}%
\>[B]{}\mathbf{type}\;\Conid{Strategy}\;\Varid{a}\mathrel{=}\Varid{a}\to \Conid{Par}\;(\Conid{IVar}\;\Varid{a}){}\<[E]%
\\
\>[B]{}\mathbf{data}\;\Conid{Conf}\;\Varid{a}\mathrel{=}\Conid{Conf}\;(\Conid{Strategy}\;\Varid{a}){}\<[E]%
\ColumnHook
\end{hscode}\resethooks
Now we can once again define our \ensuremath{\Conid{ArrowParallel}} instance as follows: First, we convert our Arrows \ensuremath{[\mskip1.5mu \Varid{arr}\;\Varid{a}\;\Varid{b}\mskip1.5mu]} with \ensuremath{\Varid{evalN}\;(\Varid{map}\;(\mathbin{>\!\!>\!\!>}\Varid{arr}\;\Varid{strat})\;\Varid{fs})} into an Arrow \ensuremath{\Varid{arr}\;[\mskip1.5mu \Varid{a}\mskip1.5mu]\;[\mskip1.5mu (\Conid{Par}\;(\Conid{IVar}\;\Varid{b}))\mskip1.5mu]} that yields composable computations in the \ensuremath{\Conid{Par}} monad. By combining the result of this Arrow with \ensuremath{\Varid{arr}\;\Varid{sequenceA}}, we get an Arrow \ensuremath{\Varid{arr}\;[\mskip1.5mu \Varid{a}\mskip1.5mu]\;(\Conid{Par}\;[\mskip1.5mu \Conid{IVar}\;\Varid{b}\mskip1.5mu])}. Then, in order to fetch the results of the different threads, we map over the \ensuremath{\Conid{IVar}}s inside the \ensuremath{\Conid{Par}} Monad with \ensuremath{\Varid{arr}\;(\bind \Varid{mapM}\;\Varid{get})} -- our intermediary Arrow is of type \ensuremath{\Varid{arr}\;[\mskip1.5mu \Varid{a}\mskip1.5mu]\;(\Conid{Par}\;[\mskip1.5mu \Varid{b}\mskip1.5mu])}. Finally, we execute the computation \ensuremath{\Conid{Par}\;[\mskip1.5mu \Varid{b}\mskip1.5mu]} by composing with \ensuremath{\Varid{arr}\;\Varid{runPar}} and get the final Arrow \ensuremath{\Varid{arr}\;[\mskip1.5mu \Varid{a}\mskip1.5mu]\;[\mskip1.5mu \Varid{b}\mskip1.5mu]}.
\begin{figure}[h]
\begin{hscode}\SaveRestoreHook
\column{B}{@{}>{\hspre}l<{\hspost}@{}}%
\column{5}{@{}>{\hspre}l<{\hspost}@{}}%
\column{9}{@{}>{\hspre}l<{\hspost}@{}}%
\column{E}{@{}>{\hspre}l<{\hspost}@{}}%
\>[B]{}\mathbf{instance}\;(\Conid{ArrowChoice}\;\Varid{arr})\Rightarrow \Conid{ArrowParallel}\;\Varid{arr}\;\Varid{a}\;\Varid{b}\;(\Conid{Conf}\;\Varid{b})\;\mathbf{where}{}\<[E]%
\\
\>[B]{}\hsindent{5}{}\<[5]%
\>[5]{}\Varid{parEvalN}\;(\Conid{Conf}\;\Varid{strat})\;\Varid{fs}\mathrel{=}{}\<[E]%
\\
\>[5]{}\hsindent{4}{}\<[9]%
\>[9]{}\Varid{evalN}\;(\Varid{map}\;(\mathbin{>\!\!>\!\!>}\Varid{arr}\;\Varid{strat})\;\Varid{fs})\mathbin{>\!\!>\!\!>}{}\<[E]%
\\
\>[5]{}\hsindent{4}{}\<[9]%
\>[9]{}\Varid{arr}\;\Varid{sequenceA}\mathbin{>\!\!>\!\!>}{}\<[E]%
\\
\>[5]{}\hsindent{4}{}\<[9]%
\>[9]{}\Varid{arr}\;(\bind \Varid{mapM}\;\Varid{\Conid{Control}.\Conid{Monad}.\Conid{Par}.get})\mathbin{>\!\!>\!\!>}{}\<[E]%
\\
\>[5]{}\hsindent{4}{}\<[9]%
\>[9]{}\Varid{arr}\;\Varid{runPar}{}\<[E]%
\ColumnHook
\end{hscode}\resethooks
\caption{\ensuremath{\Conid{Par}} Monad \ensuremath{\Conid{ArrowParallel}} instance.}
\label{fig:ArrowParallelParMonad}
\end{figure}

\subsubsection{Eden}
For both the GpH Haskell and \ensuremath{\Conid{Par}} Monad implementations we could use general instances of \ensuremath{\Conid{ArrowParallel}} that just require the \ensuremath{\Conid{ArrowChoice}} type class. With Eden this is not the case as we can only spawn a list of functions, which we cannot extract from general Arrows. While we could still manage to have only one instance in the module by introducing a type class 
\begin{hscode}\SaveRestoreHook
\column{B}{@{}>{\hspre}l<{\hspost}@{}}%
\column{9}{@{}>{\hspre}l<{\hspost}@{}}%
\column{E}{@{}>{\hspre}l<{\hspost}@{}}%
\>[B]{}\mathbf{class}\;(\Conid{Arrow}\;\Varid{arr})\Rightarrow \Conid{ArrowUnwrap}\;\Varid{arr}\;\mathbf{where}{}\<[E]%
\\
\>[B]{}\hsindent{9}{}\<[9]%
\>[9]{}\Varid{arr}\;\Varid{a}\;\Varid{b}\to (\Varid{a}\to \Varid{b}){}\<[E]%
\ColumnHook
\end{hscode}\resethooks
we avoid doing so for aesthetic reasons. For now, we just implement \ensuremath{\Conid{ArrowParallel}} for normal functions and the Kleisli type in Fig.~\ref{fig:ArrowParallelEden}, where \ensuremath{\Conid{Conf}} is simply defined as \ensuremath{\mathbf{data}\;\Conid{Conf}\mathrel{=}\Conid{Nil}} since Eden does not have a configurable \ensuremath{\Varid{spawnF}} variant. 
\begin{figure}[t]
\begin{hscode}\SaveRestoreHook
\column{B}{@{}>{\hspre}l<{\hspost}@{}}%
\column{3}{@{}>{\hspre}l<{\hspost}@{}}%
\column{5}{@{}>{\hspre}l<{\hspost}@{}}%
\column{7}{@{}>{\hspre}l<{\hspost}@{}}%
\column{E}{@{}>{\hspre}l<{\hspost}@{}}%
\>[B]{}\mathbf{instance}\;(\Conid{Trans}\;\Varid{a},\Conid{Trans}\;\Varid{b})\Rightarrow \Conid{ArrowParallel}\;(\to )\;\Varid{a}\;\Varid{b}\;\Conid{Conf}\;\mathbf{where}{}\<[E]%
\\
\>[B]{}\hsindent{5}{}\<[5]%
\>[5]{}\Varid{parEvalN}\;\anonymous \mathrel{=}\Varid{spawnF}{}\<[E]%
\\[\blanklineskip]%
\>[B]{}\mathbf{instance}\;(\Conid{ArrowParallel}\;(\to )\;\Varid{a}\;(\Varid{m}\;\Varid{b})\;\Conid{Conf},{}\<[E]%
\\
\>[B]{}\hsindent{3}{}\<[3]%
\>[3]{}\Conid{Monad}\;\Varid{m},\Conid{Trans}\;\Varid{a},\Conid{Trans}\;\Varid{b},\Conid{Trans}\;(\Varid{m}\;\Varid{b}))\Rightarrow {}\<[E]%
\\
\>[B]{}\hsindent{3}{}\<[3]%
\>[3]{}\Conid{ArrowParallel}\;(\Conid{Kleisli}\;\Varid{m})\;\Varid{a}\;\Varid{b}\;\Varid{conf}\;\mathbf{where}{}\<[E]%
\\
\>[3]{}\hsindent{2}{}\<[5]%
\>[5]{}\Varid{parEvalN}\;\Varid{conf}\;\Varid{fs}\mathrel{=}{}\<[E]%
\\
\>[5]{}\hsindent{2}{}\<[7]%
\>[7]{}\Varid{arr}\;(\Varid{parEvalN}\;\Varid{conf}\;(\Varid{map}\;(\lambda (\Conid{Kleisli}\;\Varid{f})\to \Varid{f})\;\Varid{fs}))\mathbin{>\!\!>\!\!>}{}\<[E]%
\\
\>[5]{}\hsindent{2}{}\<[7]%
\>[7]{}\Conid{Kleisli}\;\Varid{sequence}{}\<[E]%
\ColumnHook
\end{hscode}\resethooks
\caption{Eden \ensuremath{\Conid{ArrowParallel}} instance.}
\label{fig:ArrowParallelEden}
\end{figure}

\subsubsection{Default configuration instances}
While the configurability in the instances of the \ensuremath{\Conid{ArrowParallel}} instances above is nice, users probably would like to have proper default configurations for many parallel programs as well. These can also easily be defined as we can see by the example of the default implementation of \ensuremath{\Conid{ArrowParallel}} for the GpH:

\begin{hscode}\SaveRestoreHook
\column{B}{@{}>{\hspre}l<{\hspost}@{}}%
\column{3}{@{}>{\hspre}l<{\hspost}@{}}%
\column{5}{@{}>{\hspre}l<{\hspost}@{}}%
\column{E}{@{}>{\hspre}l<{\hspost}@{}}%
\>[B]{}\mathbf{instance}\;(\Conid{NFData}\;\Varid{b},\Conid{ArrowChoice}\;\Varid{arr},\Conid{ArrowParallel}\;\Varid{arr}\;\Varid{a}\;\Varid{b}\;(\Conid{Conf}\;\Varid{b}))\Rightarrow {}\<[E]%
\\
\>[B]{}\hsindent{3}{}\<[3]%
\>[3]{}\Conid{ArrowParallel}\;\Varid{arr}\;\Varid{a}\;\Varid{b}\;()\;\mathbf{where}{}\<[E]%
\\
\>[3]{}\hsindent{2}{}\<[5]%
\>[5]{}\Varid{parEvalN}\;\anonymous \;\Varid{fs}\mathrel{=}\Varid{parEvalN}\;(\Varid{defaultConf}\;\Varid{fs})\;\Varid{fs}{}\<[E]%
\\[\blanklineskip]%
\>[B]{}\Varid{defaultConf}\mathbin{::}(\Conid{NFData}\;\Varid{b})\Rightarrow [\mskip1.5mu \Varid{arr}\;\Varid{a}\;\Varid{b}\mskip1.5mu]\to \Conid{Conf}\;\Varid{b}{}\<[E]%
\\
\>[B]{}\Varid{defaultConf}\;\Varid{fs}\mathrel{=}\Varid{stratToConf}\;\Varid{fs}\;\Varid{rdeepseq}{}\<[E]%
\\[\blanklineskip]%
\>[B]{}\Varid{stratToConf}\mathbin{::}[\mskip1.5mu \Varid{arr}\;\Varid{a}\;\Varid{b}\mskip1.5mu]\to \Conid{Strategy}\;\Varid{b}\to \Conid{Conf}\;\Varid{b}{}\<[E]%
\\
\>[B]{}\Varid{stratToConf}\;\anonymous \;\Varid{strat}\mathrel{=}\Conid{Conf}\;\Varid{strat}{}\<[E]%
\ColumnHook
\end{hscode}\resethooks

The other backends have similarly structured implementations which we do not discuss here for the sake of brevity. We can, however, only have one instance of \ensuremath{\Conid{ArrowParallel}\;\Varid{arr}\;\Varid{a}\;\Varid{b}\;()} present at a time, which should not be a problem, though.

Up until now we discussed Arrow operations more in detail, but in the following sections we focus more on the data-flow between the Arrows, now that we have seen that Arrows are capable of expressing parallelism. We do explain new concepts with more details if required for better understanding, though.
	\subsection{Extending the interface}
\label{sec:extending-interface}
With the \ensuremath{\Conid{ArrowParallel}} type class in place and implemented, we can now define other parallel interface functions. These are basic algorithmic skeletons that are used to define more sophisticated  skeletons.

\subsubsection{Lazy \ensuremath{\Varid{parEvalN}}}
\begin{figure}[tb]
	\includegraphics[scale=0.7]{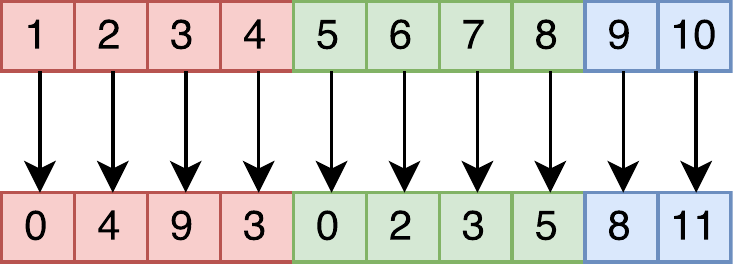}
	\caption{\ensuremath{\Varid{parEvalNLazy}} depiction.}
	\label{fig:parEvalNLazyImg}
\end{figure}
The function \ensuremath{\Varid{parEvalN}} fully traverses the list of passed Arrows as well as their inputs. Sometimes this might not be feasible, as it will not work on infinite lists of functions like \eg \ensuremath{\Varid{map}\;(\Varid{arr}\mathbin{\circ}(\mathbin{+}))\;[\mskip1.5mu \mathrm{1}\mathinner{\ldotp\ldotp}\mskip1.5mu]} or just because we need the Arrows evaluated in chunks. \ensuremath{\Varid{parEvalNLazy}} (Figs.~\ref{fig:parEvalNLazyImg},~\ref{fig:parEvalNLazy}) fixes this. It works by first chunking the input from \ensuremath{[\mskip1.5mu \Varid{a}\mskip1.5mu]} to \ensuremath{[\mskip1.5mu [\mskip1.5mu \Varid{a}\mskip1.5mu]\mskip1.5mu]} with the given \ensuremath{\Varid{chunkSize}} in \ensuremath{\Varid{arr}\;(\Varid{chunksOf}\;\Varid{chunkSize})}. These chunks are then fed into a list \ensuremath{[\mskip1.5mu \Varid{arr}\;[\mskip1.5mu \Varid{a}\mskip1.5mu]\;[\mskip1.5mu \Varid{b}\mskip1.5mu]\mskip1.5mu]} of chunk-wise parallel Arrows with the help of our lazy and sequential \ensuremath{\Varid{evalN}}. The resulting \ensuremath{[\mskip1.5mu [\mskip1.5mu \Varid{b}\mskip1.5mu]\mskip1.5mu]} is lastly converted into \ensuremath{[\mskip1.5mu \Varid{b}\mskip1.5mu]} with \ensuremath{\Varid{arr}\;\Varid{concat}}.
\begin{figure}[t]
\begin{hscode}\SaveRestoreHook
\column{B}{@{}>{\hspre}l<{\hspost}@{}}%
\column{5}{@{}>{\hspre}l<{\hspost}@{}}%
\column{7}{@{}>{\hspre}l<{\hspost}@{}}%
\column{9}{@{}>{\hspre}l<{\hspost}@{}}%
\column{E}{@{}>{\hspre}l<{\hspost}@{}}%
\>[B]{}\Varid{parEvalNLazy}\mathbin{::}(\Conid{ArrowParallel}\;\Varid{arr}\;\Varid{a}\;\Varid{b}\;\Varid{conf},\Conid{ArrowChoice}\;\Varid{arr},\Conid{ArrowApply}\;\Varid{arr})\Rightarrow {}\<[E]%
\\
\>[B]{}\hsindent{9}{}\<[9]%
\>[9]{}\Varid{conf}\to \Conid{ChunkSize}\to [\mskip1.5mu \Varid{arr}\;\Varid{a}\;\Varid{b}\mskip1.5mu]\to (\Varid{arr}\;[\mskip1.5mu \Varid{a}\mskip1.5mu]\;[\mskip1.5mu \Varid{b}\mskip1.5mu]){}\<[E]%
\\
\>[B]{}\Varid{parEvalNLazy}\;\Varid{conf}\;\Varid{chunkSize}\;\Varid{fs}\mathrel{=}{}\<[E]%
\\
\>[B]{}\hsindent{9}{}\<[9]%
\>[9]{}\Varid{arr}\;(\Varid{chunksOf}\;\Varid{chunkSize})\mathbin{>\!\!>\!\!>}{}\<[E]%
\\
\>[B]{}\hsindent{5}{}\<[5]%
\>[5]{}\Varid{evalN}\;\Varid{fchunks}\mathbin{>\!\!>\!\!>}{}\<[E]%
\\
\>[B]{}\hsindent{5}{}\<[5]%
\>[5]{}\Varid{arr}\;\Varid{concat}{}\<[E]%
\\
\>[B]{}\hsindent{5}{}\<[5]%
\>[5]{}\mathbf{where}{}\<[E]%
\\
\>[5]{}\hsindent{2}{}\<[7]%
\>[7]{}\Varid{fchunks}\mathrel{=}\Varid{map}\;(\Varid{parEvalN}\;\Varid{conf})\;(\Varid{chunksOf}\;\Varid{chunkSize}\;\Varid{fs}){}\<[E]%
\ColumnHook
\end{hscode}\resethooks
\caption{Definition of \ensuremath{\Varid{parEvalNLazy}}.}
\label{fig:parEvalNLazy}
\end{figure}

\subsubsection{Heterogeneous tasks}
\begin{figure}[tb]
	\includegraphics[scale=0.7]{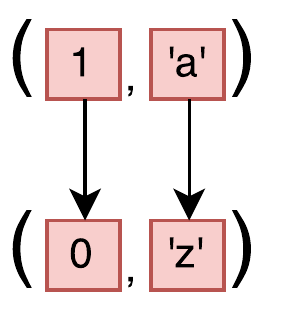}
	\caption{\ensuremath{\Varid{parEval2}} depiction.}
	\label{fig:parEval2Img}
\end{figure}
We have only talked about the parallelization of Arrows of the same set of input and output types until now. But sometimes we want to parallelize heterogeneous types as well. We can implement such a \ensuremath{\Varid{parEval2}} combinator (Figs.~\ref{fig:parEval2Img},~\ref{fig:parEval2}) which combines two Arrows \ensuremath{\Varid{arr}\;\Varid{a}\;\Varid{b}} and \ensuremath{\Varid{arr}\;\Varid{c}\;\Varid{d}} into a new parallel Arrow \ensuremath{\Varid{arr}\;(\Varid{a},\Varid{c})\;(\Varid{b},\Varid{d})} quite easily with the help of the \ensuremath{\Conid{ArrowChoice}} type class. Here, the general idea is to use the \ensuremath{\mathbin{+\!\!+\!\!+}} combinator which combines two Arrows \ensuremath{\Varid{arr}\;\Varid{a}\;\Varid{b}} and \ensuremath{\Varid{arr}\;\Varid{c}\;\Varid{d}} and transforms them into \ensuremath{\Varid{arr}\;(\Conid{Either}\;\Varid{a}\;\Varid{c})\;(\Conid{Either}\;\Varid{b}\;\Varid{d})} to get a common Arrow type that we can then feed into \ensuremath{\Varid{parEvalN}}.

	\section{Futures} \label{sec:futures}\label{futures}
\begin{figure}[t]
\begin{hscode}\SaveRestoreHook
\column{B}{@{}>{\hspre}l<{\hspost}@{}}%
\column{9}{@{}>{\hspre}l<{\hspost}@{}}%
\column{E}{@{}>{\hspre}l<{\hspost}@{}}%
\>[B]{}\Varid{someCombinator}\mathbin{::}(\Conid{ArrowChoice}\;\Varid{arr},{}\<[E]%
\\
\>[B]{}\hsindent{9}{}\<[9]%
\>[9]{}\Conid{ArrowParallel}\;\Varid{arr}\;\Varid{a}\;\Varid{b}\;(),{}\<[E]%
\\
\>[B]{}\hsindent{9}{}\<[9]%
\>[9]{}\Conid{ArrowParallel}\;\Varid{arr}\;\Varid{b}\;\Varid{c}\;())\Rightarrow {}\<[E]%
\\
\>[B]{}\hsindent{9}{}\<[9]%
\>[9]{}[\mskip1.5mu \Varid{arr}\;\Varid{a}\;\Varid{b}\mskip1.5mu]\to [\mskip1.5mu \Varid{arr}\;\Varid{b}\;\Varid{c}\mskip1.5mu]\to \Varid{arr}\;[\mskip1.5mu \Varid{a}\mskip1.5mu]\;[\mskip1.5mu \Varid{c}\mskip1.5mu]{}\<[E]%
\\
\>[B]{}\Varid{someCombinator}\;\Varid{fs1}\;\Varid{fs2}\mathrel{=}{}\<[E]%
\\
\>[B]{}\hsindent{9}{}\<[9]%
\>[9]{}\Varid{parEvalN}\;()\;\Varid{fs1}\mathbin{>\!\!>\!\!>}{}\<[E]%
\\
\>[B]{}\hsindent{9}{}\<[9]%
\>[9]{}\Varid{rightRotate}\mathbin{>\!\!>\!\!>}{}\<[E]%
\\
\>[B]{}\hsindent{9}{}\<[9]%
\>[9]{}\Varid{parEvalN}\;()\;\Varid{fs2}{}\<[E]%
\ColumnHook
\end{hscode}\resethooks
\caption{The outline combinator.}
\label{fig:someCombinator}
\end{figure}

Consider the outline parallel Arrow combinator in Fig.~\ref{fig:someCombinator}. In a distributed environment this first evaluates all \ensuremath{[\mskip1.5mu \Varid{arr}\;\Varid{a}\;\Varid{b}\mskip1.5mu]} in parallel, sends the results back to the master node, rotates the input once (in the example we require \ensuremath{\Conid{ArrowChoice}} for this) and then evaluates the \ensuremath{[\mskip1.5mu \Varid{arr}\;\Varid{b}\;\Varid{c}\mskip1.5mu]} in parallel to then gather the input once again on the master node.
Such situations arise, \eg in scientific computations when the data distributed across the nodes needs to be transposed. A concrete example is 2D FFT computation \cite{Gorlatch,Berthold2009-fft}.

\begin{figure}[ht]
	\centering
	\includegraphics[width=0.9\textwidth]{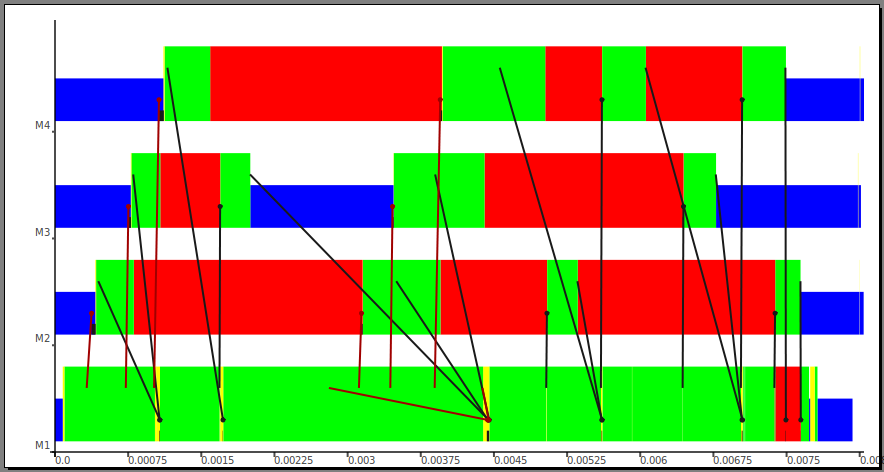}
	\caption[without Futures]{Communication between 4 Eden processes without Futures. All communication goes through the master node. Each bar represents one process. Black lines represent communication. Colours: blue $\hat{=}$ idle, green $\hat{=}$ running, red  $\hat{=}$ blocked, yellow $\hat{=}$ suspended.}
	\label{fig:withoutFutures}
\end{figure}

While the example could be rewritten into a single \ensuremath{\Varid{parEvalN}} call by directly wiring the Arrows together before spawning, it illustrates an important problem. When using a \ensuremath{\Conid{ArrowParallel}} backend that resides on multiple computers, all communication between the nodes is done via the master node, as shown in the Eden trace in Figure~\ref{fig:withoutFutures}. This can become a serious bottleneck
for a larger amount of data and number of processes \citep[as e.g.][showcases]{Berthold2009-fft}.

This is only a problem in distributed memory (in the scope of this paper) and we should allow nodes to communicate directly with each other. Eden already provides \enquote{remote data} that enable this \cite{AlGo03a,Dieterle2010}.
But as we want code with our DSL to be implementation agnostic, we have to wrap this concept. We do this with the \ensuremath{\Conid{Future}} type class (Fig.~\ref{fig:future}). We require a \ensuremath{\Varid{conf}} parameter here as well, but only so that Haskells type system allows us to have multiple Future implementations imported at once without breaking any dependencies similar to what we did with the \ensuremath{\Conid{ArrowParallel}} type class earlier.
\begin{figure}[h]
\begin{hscode}\SaveRestoreHook
\column{B}{@{}>{\hspre}l<{\hspost}@{}}%
\column{5}{@{}>{\hspre}l<{\hspost}@{}}%
\column{E}{@{}>{\hspre}l<{\hspost}@{}}%
\>[B]{}\mathbf{class}\;\Conid{Future}\;\Varid{fut}\;\Varid{a}\;\Varid{conf}\mid \Varid{a}\;\Varid{conf}\to \Varid{fut}\;\mathbf{where}{}\<[E]%
\\
\>[B]{}\hsindent{5}{}\<[5]%
\>[5]{}\Varid{put}\mathbin{::}(\Conid{Arrow}\;\Varid{arr})\Rightarrow \Varid{conf}\to \Varid{arr}\;\Varid{a}\;(\Varid{fut}\;\Varid{a}){}\<[E]%
\\
\>[B]{}\hsindent{5}{}\<[5]%
\>[5]{}\Varid{get}\mathbin{::}(\Conid{Arrow}\;\Varid{arr})\Rightarrow \Varid{conf}\to \Varid{arr}\;(\Varid{fut}\;\Varid{a})\;\Varid{a}{}\<[E]%
\ColumnHook
\end{hscode}\resethooks
\caption{The \ensuremath{\Conid{Future}} type class.}
\label{fig:future}
\end{figure}
Since \ensuremath{\Conid{RD}} is only a type synonym for a communication type that Eden uses internally, we have to use some wrapper classes to fit that definition, though, as Fig.~\ref{fig:RDFuture} shows. 
Technical details are in Appendix, in Section~\ref{app:omitted}.

For GpH and \ensuremath{\Conid{Par}} Monad, we can simply use \ensuremath{\Conid{BasicFuture}}s (Fig.~\ref{fig:BasicFuture}), which are just simple wrappers around the actual data with boiler-plate logic so that the type class is satisfied. This is because the concept of a \ensuremath{\Conid{Future}} does not change anything for shared-memory execution as there are no communication problems to fix. Nevertheless, we require a common interface so the parallel Arrows are portable across backends. The implementation can also be found in Section ~\ref{app:omitted}.

\begin{figure}[ht]
	\centering
	\includegraphics[width=0.9\textwidth]{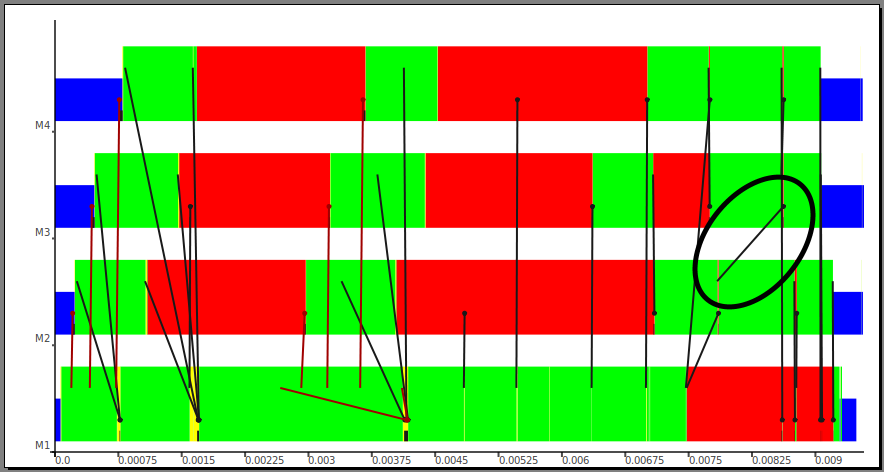}
	\caption[with Futures]{Communication between 4 Eden processes with Futures. Other than in Fig.~\ref{fig:withoutFutures}, processes communicate directly (one example message is highlighted) instead of always going through the master node (bottom bar).}
	\label{fig:withFutures}
\end{figure}

In our communication example we can use this \ensuremath{\Conid{Future}} concept for direct communication between nodes as shown in Fig.~\ref{fig:someCombinatorParallel}.
\begin{figure}[t]
\begin{hscode}\SaveRestoreHook
\column{B}{@{}>{\hspre}l<{\hspost}@{}}%
\column{9}{@{}>{\hspre}l<{\hspost}@{}}%
\column{E}{@{}>{\hspre}l<{\hspost}@{}}%
\>[B]{}\Varid{someCombinator}\mathbin{::}(\Conid{ArrowChoice}\;\Varid{arr},{}\<[E]%
\\
\>[B]{}\hsindent{9}{}\<[9]%
\>[9]{}\Conid{ArrowParallel}\;\Varid{arr}\;\Varid{a}\;(\Varid{fut}\;\Varid{b})\;(),{}\<[E]%
\\
\>[B]{}\hsindent{9}{}\<[9]%
\>[9]{}\Conid{ArrowParallel}\;\Varid{arr}\;(\Varid{fut}\;\Varid{b})\;\Varid{c}\;(),{}\<[E]%
\\
\>[B]{}\hsindent{9}{}\<[9]%
\>[9]{}\Conid{Future}\;\Varid{fut}\;\Varid{b}\;())\Rightarrow {}\<[E]%
\\
\>[B]{}\hsindent{9}{}\<[9]%
\>[9]{}[\mskip1.5mu \Varid{arr}\;\Varid{a}\;\Varid{b}\mskip1.5mu]\to [\mskip1.5mu \Varid{arr}\;\Varid{b}\;\Varid{c}\mskip1.5mu]\to \Varid{arr}\;[\mskip1.5mu \Varid{a}\mskip1.5mu]\;[\mskip1.5mu \Varid{c}\mskip1.5mu]{}\<[E]%
\\
\>[B]{}\Varid{someCombinator}\;\Varid{fs1}\;\Varid{fs2}\mathrel{=}{}\<[E]%
\\
\>[B]{}\hsindent{9}{}\<[9]%
\>[9]{}\Varid{parEvalN}\;()\;(\Varid{map}\;(\mathbin{>\!\!>\!\!>}\Varid{put}\;())\;\Varid{fs1})\mathbin{>\!\!>\!\!>}{}\<[E]%
\\
\>[B]{}\hsindent{9}{}\<[9]%
\>[9]{}\Varid{rightRotate}\mathbin{>\!\!>\!\!>}{}\<[E]%
\\
\>[B]{}\hsindent{9}{}\<[9]%
\>[9]{}\Varid{parEvalN}\;()\;(\Varid{map}\;(\Varid{get}\;()\mathbin{>\!\!>\!\!>})\;\Varid{fs2}){}\<[E]%
\ColumnHook
\end{hscode}\resethooks
\caption{The outline combinator in parallel.}
\label{fig:someCombinatorParallel}
\end{figure}
In a distributed environment, this gives us a communication scheme with messages going through the master node only if it is needed -- similar to what is shown in the trace visualisation in Fig.~\ref{fig:withFutures}. One especially elegant aspect of the definition in Fig.~\ref{fig:future} is that we can specify the type of \ensuremath{\Conid{Future}} to be used per backend with full interoperability between code using different backends, without even requiring to know about the actual type used for communication. We only specify that there has to be a compatible Future and do not care about any specifics as can be seen in Fig.~\ref{fig:someCombinatorParallel}. With the PArrows DSL we can also define default instances \ensuremath{\Conid{Future}\;\Varid{fut}\;\Varid{a}\;()} for each backend similar to how \ensuremath{\Conid{ArrowParallel}\;\Varid{arr}\;\Varid{a}\;\Varid{b}\;()} was defined in Section~\ref{sec:parallel-arrows}. Details can be found in Section~\ref{app:omitted}. \olcomment{Fig. is not really clear. Do Figs with a lot of load? --- fft?}
	\section{Skeletons}
\label{sec:skeletons}
Now we have developed Parallel Arrows far enough to define some useful algorithmic skeletons that abstract typical parallel computations. While there are many possible skeletons to implement, we demonstrate the expressive power of PArrows here using four \ensuremath{\Varid{map}}-based and three toplogical skeletons.
\subsection{\ensuremath{\Varid{map}}-based skeletons}
\label{sec:map-skeletons}
The essential differences between the mapping skeletons presented here are in terms of order of evaluation and work distribution but still provide the same semantics as a sequential \ensuremath{\Varid{map}}.

\paragraph{Parallel \ensuremath{\Varid{map}} and laziness.}
The \ensuremath{\Varid{parMap}} skeleton (Figs.~\ref{fig:parMapImg},~\ref{fig:parMap}) is probably the most common skeleton for parallel programs. We can implement it with \ensuremath{\Conid{ArrowParallel}} by repeating an Arrow \ensuremath{\Varid{arr}\;\Varid{a}\;\Varid{b}} and then passing it into \ensuremath{\Varid{parEvalN}} to obtain an Arrow \ensuremath{\Varid{arr}\;[\mskip1.5mu \Varid{a}\mskip1.5mu]\;[\mskip1.5mu \Varid{b}\mskip1.5mu]}.
Just like \ensuremath{\Varid{parEvalN}}, \ensuremath{\Varid{parMap}} traverses all input Arrows as well as the inputs.
Because of this, it has the same restrictions as \ensuremath{\Varid{parEvalN}} as compared to \ensuremath{\Varid{parEvalNLazy}}. So it makes sense to also have a \ensuremath{\Varid{parMapStream}} (Figs.~\ref{fig:parMapStreamImg},~\ref{fig:parMapStream}) which behaves like \ensuremath{\Varid{parMap}}, but uses \ensuremath{\Varid{parEvalNLazy}} instead of \ensuremath{\Varid{parEvalN}}. Implementing these skeletons is straightforward as in Appendix \ref{app:omitted} in Figs.\ref{fig:parMap} and \ref{fig:parMapStream}.

\begin{figure}[thb]
\includegraphics[scale=0.7]{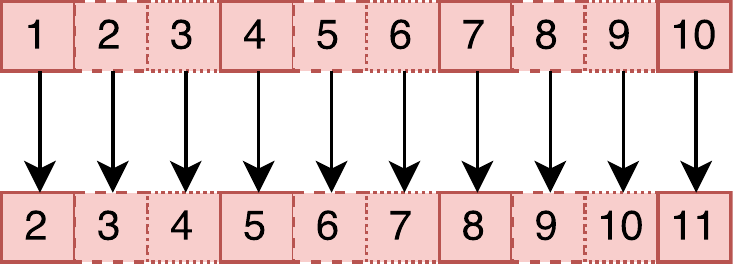}
\caption{\ensuremath{\Varid{farm}} depiction.}
\label{fig:farmImg}

\begin{hscode}\SaveRestoreHook
\column{B}{@{}>{\hspre}l<{\hspost}@{}}%
\column{9}{@{}>{\hspre}l<{\hspost}@{}}%
\column{E}{@{}>{\hspre}l<{\hspost}@{}}%
\>[B]{}\Varid{farm}\mathbin{::}(\Conid{ArrowParallel}\;\Varid{arr}\;\Varid{a}\;\Varid{b}\;\Varid{conf},{}\<[E]%
\\
\>[B]{}\hsindent{9}{}\<[9]%
\>[9]{}\Conid{ArrowParallel}\;\Varid{arr}\;[\mskip1.5mu \Varid{a}\mskip1.5mu]\;[\mskip1.5mu \Varid{b}\mskip1.5mu]\;\Varid{conf},\Conid{ArrowChoice}\;\Varid{arr})\Rightarrow {}\<[E]%
\\
\>[B]{}\hsindent{9}{}\<[9]%
\>[9]{}\Varid{conf}\to \Conid{NumCores}\to \Varid{arr}\;\Varid{a}\;\Varid{b}\to \Varid{arr}\;[\mskip1.5mu \Varid{a}\mskip1.5mu]\;[\mskip1.5mu \Varid{b}\mskip1.5mu]{}\<[E]%
\\
\>[B]{}\Varid{farm}\;\Varid{conf}\;\Varid{numCores}\;\Varid{f}\mathrel{=}{}\<[E]%
\\
\>[B]{}\hsindent{9}{}\<[9]%
\>[9]{}\Varid{unshuffle}\;\Varid{numCores}\mathbin{>\!\!>\!\!>}{}\<[E]%
\\
\>[B]{}\hsindent{9}{}\<[9]%
\>[9]{}\Varid{parEvalN}\;\Varid{conf}\;(\Varid{repeat}\;(\Varid{mapArr}\;\Varid{f}))\mathbin{>\!\!>\!\!>}{}\<[E]%
\\
\>[B]{}\hsindent{9}{}\<[9]%
\>[9]{}\Varid{shuffle}{}\<[E]%
\ColumnHook
\end{hscode}\resethooks
\caption{\ensuremath{\Varid{farm}} definition.}
\label{fig:farm}

\includegraphics[scale=0.7]{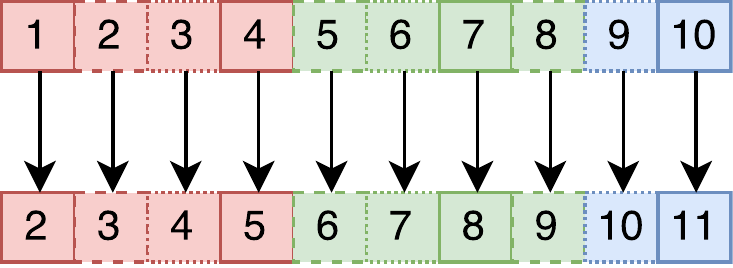}
\caption{\ensuremath{\Varid{farmChunk}} depiction.}
\label{fig:farmChunkImg}
\end{figure}

\paragraph{Statically load-balancing parallel \ensuremath{\Varid{map}}.}
Our \ensuremath{\Varid{parMap}} spawns every single computation in a new thread (at least for the instances of \ensuremath{\Conid{ArrowParallel}} we presented in this paper). This can be quite wasteful and a statically load-balancing \ensuremath{\Varid{farm}} (Figs.~\ref{fig:farmImg},~\ref{fig:farm}) that equally distributes the workload over \ensuremath{\Varid{numCores}} workers seems useful.
The definitions of the helper functions \ensuremath{\Varid{unshuffle}}, \ensuremath{\Varid{takeEach}}, \ensuremath{\Varid{shuffle}} (Fig.~\ref{fig:edenshuffleetc}) originate from an Eden skeleton\footnote{Available on Hackage under \url{https://hackage.haskell.org/package/edenskel-2.1.0.0/docs/src/Control-Parallel-Eden-Map.html}.}.

Since a \ensuremath{\Varid{farm}}  is basically just \ensuremath{\Varid{parMap}} with a different work distribution, it has the same restrictions as \ensuremath{\Varid{parEvalN}} and \ensuremath{\Varid{parMap}}. We can, however, define \ensuremath{\Varid{farmChunk}} (Figs.~\ref{fig:farmChunkImg},~\ref{fig:farmChunk}) which uses \ensuremath{\Varid{parEvalNLazy}} instead of \ensuremath{\Varid{parEvalN}}. It is basically the same definition as for \ensuremath{\Varid{farm}}, but with \ensuremath{\Varid{parEvalNLazy}} instead of \ensuremath{\Varid{parEvalN}}.

\subsection{Topological skeletons}
\label{sec:topology-skeletons}
Even though many algorithms can be expressed by parallel maps, some problems require more sophisticated skeletons. The Eden library leverages this problem and already comes with more predefined skeletons\footnote{Available on Hackage: \url{https://hackage.haskell.org/package/edenskel-2.1.0.0/docs/Control-Parallel-Eden-Topology.html}.}, among them a \ensuremath{\Varid{pipe}}, a \ensuremath{\Varid{ring}}, and a \ensuremath{\Varid{torus}} implementations \citep{Eden:SkeletonBookChapter02}. These seem like reasonable candidates to be ported to our Arrow-based parallel Haskell. We aim to showcase that we can express more sophisticated skeletons with parallel Arrows as well.

If we used the original definition of \ensuremath{\Varid{parEvalN}}, however, these skeletons would produce an infinite loop with the GpH and \ensuremath{\Conid{Par}} Monad which during runtime would result in the program crash. This materialises with the usage of \ensuremath{\Varid{loop}} of the \ensuremath{\Conid{ArrowLoop}} type class and we think that this is due to difference of how evaluation is done in these backends when compared to Eden. An investigation of why this difference exists is beyond the scope of this work, we only provide a workaround for these types of skeletons as such they probably are not of much importance outside of a distributed memory environment. However our workaround enables users of the DSL to test their code within a shared memory setting.
The idea of the fix is to provide a \ensuremath{\Conid{ArrowLoopParallel}} type class that has two functions -- \ensuremath{\Varid{loopParEvalN}} and \ensuremath{\Varid{postLoopParEvalN}}, where the first is to be used inside an \ensuremath{\Varid{loop}} construct while the latter will be used right outside of the \ensuremath{\Varid{loop}}. This way we can delegate to the actual \ensuremath{\Varid{parEvalN}} in the spot where the backend supports it.
\begin{hscode}\SaveRestoreHook
\column{B}{@{}>{\hspre}l<{\hspost}@{}}%
\column{5}{@{}>{\hspre}l<{\hspost}@{}}%
\column{9}{@{}>{\hspre}l<{\hspost}@{}}%
\column{E}{@{}>{\hspre}l<{\hspost}@{}}%
\>[B]{}\mathbf{class}\;\Conid{ArrowParallel}\;\Varid{arr}\;\Varid{a}\;\Varid{b}\;\Varid{conf}\Rightarrow {}\<[E]%
\\
\>[B]{}\hsindent{9}{}\<[9]%
\>[9]{}\Conid{ArrowLoopParallel}\;\Varid{arr}\;\Varid{a}\;\Varid{b}\;\Varid{conf}\;\mathbf{where}{}\<[E]%
\\
\>[B]{}\hsindent{5}{}\<[5]%
\>[5]{}\Varid{loopParEvalN}\mathbin{::}\Varid{conf}\to [\mskip1.5mu \Varid{arr}\;\Varid{a}\;\Varid{b}\mskip1.5mu]\to \Varid{arr}\;[\mskip1.5mu \Varid{a}\mskip1.5mu]\;[\mskip1.5mu \Varid{b}\mskip1.5mu]{}\<[E]%
\\
\>[B]{}\hsindent{5}{}\<[5]%
\>[5]{}\Varid{postLoopParEvalN}\mathbin{::}\Varid{conf}\to [\mskip1.5mu \Varid{arr}\;\Varid{a}\;\Varid{b}\mskip1.5mu]\to \Varid{arr}\;[\mskip1.5mu \Varid{a}\mskip1.5mu]\;[\mskip1.5mu \Varid{b}\mskip1.5mu]{}\<[E]%
\ColumnHook
\end{hscode}\resethooks
As Eden has no problems with the looping skeletons, we use this instance:
\begin{hscode}\SaveRestoreHook
\column{B}{@{}>{\hspre}l<{\hspost}@{}}%
\column{5}{@{}>{\hspre}l<{\hspost}@{}}%
\column{9}{@{}>{\hspre}l<{\hspost}@{}}%
\column{E}{@{}>{\hspre}l<{\hspost}@{}}%
\>[B]{}\mathbf{instance}\;(\Conid{ArrowChoice}\;\Varid{arr},\Conid{ArrowParallel}\;\Varid{arr}\;\Varid{a}\;\Varid{b}\;\Conid{Conf})\Rightarrow {}\<[E]%
\\
\>[B]{}\hsindent{9}{}\<[9]%
\>[9]{}\Conid{ArrowLoopParallel}\;\Varid{arr}\;\Varid{a}\;\Varid{b}\;\Conid{Conf}\;\mathbf{where}{}\<[E]%
\\
\>[B]{}\hsindent{5}{}\<[5]%
\>[5]{}\Varid{loopParEvalN}\mathrel{=}\Varid{parEvalN}{}\<[E]%
\\
\>[B]{}\hsindent{5}{}\<[5]%
\>[5]{}\Varid{postLoopParEvalN}\;\anonymous \mathrel{=}\Varid{evalN}{}\<[E]%
\ColumnHook
\end{hscode}\resethooks
As \ensuremath{\Conid{Par}} Monad and GpH have problems with \ensuremath{\Varid{parEvalN}} inside of \ensuremath{\Varid{loop}} their respective instances for \ensuremath{\Conid{ArrowLoopParallel}} look like this:
\begin{hscode}\SaveRestoreHook
\column{B}{@{}>{\hspre}l<{\hspost}@{}}%
\column{5}{@{}>{\hspre}l<{\hspost}@{}}%
\column{9}{@{}>{\hspre}l<{\hspost}@{}}%
\column{E}{@{}>{\hspre}l<{\hspost}@{}}%
\>[B]{}\mathbf{instance}\;(\Conid{ArrowChoice}\;\Varid{arr},\Conid{ArrowParallel}\;\Varid{arr}\;\Varid{a}\;\Varid{b}\;(\Conid{Conf}\;\Varid{b}))\Rightarrow {}\<[E]%
\\
\>[B]{}\hsindent{9}{}\<[9]%
\>[9]{}\Conid{ArrowLoopParallel}\;\Varid{arr}\;\Varid{a}\;\Varid{b}\;(\Conid{Conf}\;\Varid{b})\;\mathbf{where}{}\<[E]%
\\
\>[B]{}\hsindent{5}{}\<[5]%
\>[5]{}\Varid{loopParEvalN}\;\anonymous \mathrel{=}\Varid{evalN}{}\<[E]%
\\
\>[B]{}\hsindent{5}{}\<[5]%
\>[5]{}\Varid{postLoopParEvalN}\mathrel{=}\Varid{parEvalN}{}\<[E]%
\ColumnHook
\end{hscode}\resethooks

\subsubsection{Parallel pipe}\label{sec:pipe}

The parallel \ensuremath{\Varid{pipe}} skeleton is semantically equivalent to folding over a list \ensuremath{[\mskip1.5mu \Varid{arr}\;\Varid{a}\;\Varid{a}\mskip1.5mu]} of Arrows with \ensuremath{\mathbin{>\!\!>\!\!>}}, but does this in parallel, meaning that the Arrows do not have to reside on the same thread/machine. We implement this skeleton using the \ensuremath{\Conid{ArrowLoop}} type class which gives us the \ensuremath{\Varid{loop}\mathbin{::}\Varid{arr}\;(\Varid{a},\Varid{b})\;(\Varid{c},\Varid{b})\to \Varid{arr}\;\Varid{a}\;\Varid{c}} combinator which allows us to express recursive fix-point computations in which output values are fed back as input. For example 
\begin{hscode}\SaveRestoreHook
\column{B}{@{}>{\hspre}l<{\hspost}@{}}%
\column{E}{@{}>{\hspre}l<{\hspost}@{}}%
\>[B]{}\Varid{loop}\;(\Varid{arr}\;(\lambda (\Varid{a},\Varid{b})\to (\Varid{b},\Varid{a}\mathbin{:}\Varid{b}))){}\<[E]%
\ColumnHook
\end{hscode}\resethooks
which is the same as
\begin{hscode}\SaveRestoreHook
\column{B}{@{}>{\hspre}l<{\hspost}@{}}%
\column{E}{@{}>{\hspre}l<{\hspost}@{}}%
\>[B]{}\Varid{loop}\;(\Varid{arr}\;\Varid{snd}\mathbin{\&\!\&\!\&}\Varid{arr}\;(\Varid{uncurry}\;(\mathbin{:}))){}\<[E]%
\ColumnHook
\end{hscode}\resethooks
defines an Arrow that takes its input \ensuremath{\Varid{a}} and converts it into an infinite stream \ensuremath{[\mskip1.5mu \Varid{a}\mskip1.5mu]} of it. Using \ensuremath{\Varid{loop}} to our advantage gives us a first draft of a pipe implementation (Fig.~\ref{fig:pipeSimple}) by plugging in the parallel evaluation call \ensuremath{\Varid{evalN}\;\Varid{conf}\;\Varid{fs}} inside the second argument of \ensuremath{\mathbin{\&\!\&\!\&}} and then only picking the first element of the resulting list with \ensuremath{\Varid{arr}\;\Varid{last}}.

\begin{figure}[t]
\begin{hscode}\SaveRestoreHook
\column{B}{@{}>{\hspre}l<{\hspost}@{}}%
\column{5}{@{}>{\hspre}l<{\hspost}@{}}%
\column{9}{@{}>{\hspre}l<{\hspost}@{}}%
\column{E}{@{}>{\hspre}l<{\hspost}@{}}%
\>[B]{}\Varid{pipeSimple}\mathbin{::}(\Conid{ArrowLoop}\;\Varid{arr},\Conid{ArrowLoopParallel}\;\Varid{arr}\;\Varid{a}\;\Varid{a}\;\Varid{conf})\Rightarrow {}\<[E]%
\\
\>[B]{}\hsindent{9}{}\<[9]%
\>[9]{}\Varid{conf}\to [\mskip1.5mu \Varid{arr}\;\Varid{a}\;\Varid{a}\mskip1.5mu]\to \Varid{arr}\;\Varid{a}\;\Varid{a}{}\<[E]%
\\
\>[B]{}\Varid{pipeSimple}\;\Varid{conf}\;\Varid{fs}\mathrel{=}{}\<[E]%
\\
\>[B]{}\hsindent{5}{}\<[5]%
\>[5]{}\Varid{loop}\;(\Varid{arr}\;\Varid{snd}\mathbin{\&\!\&\!\&}{}\<[E]%
\\
\>[5]{}\hsindent{4}{}\<[9]%
\>[9]{}(\Varid{arr}\;(\Varid{uncurry}\;(\mathbin{:})\mathbin{>\!\!>\!\!>}\Varid{lazy})\mathbin{>\!\!>\!\!>}\Varid{loopParEvalN}\;\Varid{conf}\;\Varid{fs}))\mathbin{>\!\!>\!\!>}{}\<[E]%
\\
\>[B]{}\hsindent{5}{}\<[5]%
\>[5]{}\Varid{arr}\;\Varid{last}{}\<[E]%
\ColumnHook
\end{hscode}\resethooks
\caption{Simple \ensuremath{\Varid{pipe}} skeleton. The use of \ensuremath{\Varid{lazy}} (Fig.~\ref{fig:edenlazyrightrotate}) is essential as without it programs using this definition would never halt. We need to enforce that the evaluation of the input \ensuremath{[\mskip1.5mu \Varid{a}\mskip1.5mu]} terminates before passing it into \ensuremath{\Varid{evalN}}.}
\label{fig:pipeSimple}
\end{figure}

However, using this definition directly will make the master node a potential bottleneck in distributed environments as described in Section~\ref{futures}. Therefore, we introduce a more sophisticated version that internally uses Futures and obtain the final definition of \ensuremath{\Varid{pipe}} in Fig.~\ref{fig:pipe}.

\begin{figure}[t]
\begin{hscode}\SaveRestoreHook
\column{B}{@{}>{\hspre}l<{\hspost}@{}}%
\column{9}{@{}>{\hspre}l<{\hspost}@{}}%
\column{E}{@{}>{\hspre}l<{\hspost}@{}}%
\>[B]{}\Varid{pipe}\mathbin{::}(\Conid{ArrowLoop}\;\Varid{arr},\Conid{ArrowLoopParallel}\;\Varid{arr}\;(\Varid{fut}\;\Varid{a})\;(\Varid{fut}\;\Varid{a})\;\Varid{conf},{}\<[E]%
\\
\>[B]{}\hsindent{9}{}\<[9]%
\>[9]{}\Conid{Future}\;\Varid{fut}\;\Varid{a}\;\Varid{conf})\Rightarrow {}\<[E]%
\\
\>[B]{}\hsindent{9}{}\<[9]%
\>[9]{}\Varid{conf}\to [\mskip1.5mu \Varid{arr}\;\Varid{a}\;\Varid{a}\mskip1.5mu]\to \Varid{arr}\;\Varid{a}\;\Varid{a}{}\<[E]%
\\
\>[B]{}\Varid{pipe}\;\Varid{conf}\;\Varid{fs}\mathrel{=}\Varid{unliftFut}\;\Varid{conf}\;(\Varid{pipeSimple}\;\Varid{conf}\;(\Varid{map}\;(\Varid{liftFut}\;\Varid{conf})\;\Varid{fs})){}\<[E]%
\\[\blanklineskip]%
\>[B]{}\Varid{liftFut}\mathbin{::}(\Conid{Arrow}\;\Varid{arr},\Conid{Future}\;\Varid{fut}\;\Varid{a}\;\Varid{conf},\Conid{Future}\;\Varid{fut}\;\Varid{b}\;\Varid{conf})\Rightarrow {}\<[E]%
\\
\>[B]{}\hsindent{9}{}\<[9]%
\>[9]{}\Varid{conf}\to \Varid{arr}\;\Varid{a}\;\Varid{b}\to \Varid{arr}\;(\Varid{fut}\;\Varid{a})\;(\Varid{fut}\;\Varid{b}){}\<[E]%
\\
\>[B]{}\Varid{liftFut}\;\Varid{conf}\;\Varid{f}\mathrel{=}\Varid{get}\;\Varid{conf}\mathbin{>\!\!>\!\!>}\Varid{f}\mathbin{>\!\!>\!\!>}\Varid{put}\;\Varid{conf}{}\<[E]%
\\[\blanklineskip]%
\>[B]{}\Varid{unliftFut}\mathbin{::}(\Conid{Arrow}\;\Varid{arr},\Conid{Future}\;\Varid{fut}\;\Varid{a}\;\Varid{conf},\Conid{Future}\;\Varid{fut}\;\Varid{b}\;\Varid{conf})\Rightarrow {}\<[E]%
\\
\>[B]{}\hsindent{9}{}\<[9]%
\>[9]{}\Varid{conf}\to \Varid{arr}\;(\Varid{fut}\;\Varid{a})\;(\Varid{fut}\;\Varid{b})\to \Varid{arr}\;\Varid{a}\;\Varid{b}{}\<[E]%
\\
\>[B]{}\Varid{unliftFut}\;\Varid{conf}\;\Varid{f}\mathrel{=}\Varid{put}\;\Varid{conf}\mathbin{>\!\!>\!\!>}\Varid{f}\mathbin{>\!\!>\!\!>}\Varid{get}\;\Varid{conf}{}\<[E]%
\ColumnHook
\end{hscode}\resethooks
\caption{\ensuremath{\Varid{pipe}} skeleton definition with Futures.}
\label{fig:pipe}
\end{figure}

Sometimes, this \ensuremath{\Varid{pipe}} definition can be a bit inconvenient, especially if we want to pipe Arrows of mixed types together, \ie  \ensuremath{\Varid{arr}\;\Varid{a}\;\Varid{b}} and \ensuremath{\Varid{arr}\;\Varid{b}\;\Varid{c}}. By wrapping these two Arrows inside a bigger Arrow \ensuremath{\Varid{arr}\;(([\mskip1.5mu \Varid{a}\mskip1.5mu],[\mskip1.5mu \Varid{b}\mskip1.5mu]),[\mskip1.5mu \Varid{c}\mskip1.5mu])\;(([\mskip1.5mu \Varid{a}\mskip1.5mu],[\mskip1.5mu \Varid{b}\mskip1.5mu]),[\mskip1.5mu \Varid{c}\mskip1.5mu])} suitable for \ensuremath{\Varid{pipe}}, we can define \ensuremath{\Varid{pipe2}} as in Fig.~\ref{fig:pipe2}.
\begin{figure}[tb]
\begin{hscode}\SaveRestoreHook
\column{B}{@{}>{\hspre}l<{\hspost}@{}}%
\column{5}{@{}>{\hspre}l<{\hspost}@{}}%
\column{9}{@{}>{\hspre}l<{\hspost}@{}}%
\column{25}{@{}>{\hspre}l<{\hspost}@{}}%
\column{E}{@{}>{\hspre}l<{\hspost}@{}}%
\>[B]{}\Varid{pipe2}\mathbin{::}(\Conid{ArrowLoop}\;\Varid{arr},\Conid{ArrowChoice}\;\Varid{arr},{}\<[E]%
\\
\>[B]{}\hsindent{5}{}\<[5]%
\>[5]{}\Conid{ArrowLoopParallel}\;\Varid{arr}\;(\Varid{fut}\;(([\mskip1.5mu \Varid{a}\mskip1.5mu],[\mskip1.5mu \Varid{b}\mskip1.5mu]),[\mskip1.5mu \Varid{c}\mskip1.5mu]))\;(\Varid{fut}\;(([\mskip1.5mu \Varid{a}\mskip1.5mu],[\mskip1.5mu \Varid{b}\mskip1.5mu]),[\mskip1.5mu \Varid{c}\mskip1.5mu]))\;\Varid{conf},{}\<[E]%
\\
\>[B]{}\hsindent{5}{}\<[5]%
\>[5]{}\Conid{Future}\;\Varid{fut}\;(([\mskip1.5mu \Varid{a}\mskip1.5mu],[\mskip1.5mu \Varid{b}\mskip1.5mu]),[\mskip1.5mu \Varid{c}\mskip1.5mu])\;\Varid{conf})\Rightarrow {}\<[E]%
\\
\>[5]{}\hsindent{4}{}\<[9]%
\>[9]{}\Varid{conf}\to \Varid{arr}\;\Varid{a}\;\Varid{b}\to \Varid{arr}\;\Varid{b}\;\Varid{c}\to \Varid{arr}\;\Varid{a}\;\Varid{c}{}\<[E]%
\\
\>[B]{}\Varid{pipe2}\;\Varid{conf}\;\Varid{f}\;\Varid{g}\mathrel{=}{}\<[E]%
\\
\>[B]{}\hsindent{5}{}\<[5]%
\>[5]{}(\Varid{arr}\;\Varid{return}\mathbin{\&\!\&\!\&}\Varid{arr}\;(\Varid{const}\;[\mskip1.5mu \mskip1.5mu]))\mathbin{\&\!\&\!\&}\Varid{arr}\;(\Varid{const}\;[\mskip1.5mu \mskip1.5mu])\mathbin{>\!\!>\!\!>}{}\<[E]%
\\
\>[B]{}\hsindent{5}{}\<[5]%
\>[5]{}\Varid{pipe}\;\Varid{conf}\;(\Varid{replicate}\;\mathrm{2}\;(\Varid{unify}\;\Varid{f}\;\Varid{g}))\mathbin{>\!\!>\!\!>}{}\<[E]%
\\
\>[B]{}\hsindent{5}{}\<[5]%
\>[5]{}\Varid{arr}\;\Varid{snd}\mathbin{>\!\!>\!\!>}{}\<[E]%
\\
\>[B]{}\hsindent{5}{}\<[5]%
\>[5]{}\Varid{arr}\;\Varid{head}{}\<[E]%
\\
\>[B]{}\hsindent{5}{}\<[5]%
\>[5]{}\mathbf{where}{}\<[E]%
\\
\>[5]{}\hsindent{4}{}\<[9]%
\>[9]{}\Varid{unify}\mathbin{::}(\Conid{ArrowChoice}\;\Varid{arr})\Rightarrow {}\<[E]%
\\
\>[9]{}\hsindent{16}{}\<[25]%
\>[25]{}\Varid{arr}\;\Varid{a}\;\Varid{b}\to \Varid{arr}\;\Varid{b}\;\Varid{c}\to \Varid{arr}\;(([\mskip1.5mu \Varid{a}\mskip1.5mu],[\mskip1.5mu \Varid{b}\mskip1.5mu]),[\mskip1.5mu \Varid{c}\mskip1.5mu])\;(([\mskip1.5mu \Varid{a}\mskip1.5mu],[\mskip1.5mu \Varid{b}\mskip1.5mu]),[\mskip1.5mu \Varid{c}\mskip1.5mu]){}\<[E]%
\\
\>[5]{}\hsindent{4}{}\<[9]%
\>[9]{}\Varid{unify}\;\Varid{f'}\;\Varid{g'}\mathrel{=}{}\<[E]%
\\
\>[9]{}\hsindent{16}{}\<[25]%
\>[25]{}(\Varid{mapArr}\;\Varid{f'}\mathbin{*\!*\!*}\Varid{mapArr}\;\Varid{g'})\mathbin{*\!*\!*}\Varid{arr}\;(\Varid{const}\;[\mskip1.5mu \mskip1.5mu])\mathbin{>\!\!>\!\!>}{}\<[E]%
\\
\>[9]{}\hsindent{16}{}\<[25]%
\>[25]{}\Varid{arr}\;(\lambda ((\Varid{b},\Varid{c}),\Varid{a})\to ((\Varid{a},\Varid{b}),\Varid{c})){}\<[E]%
\\[\blanklineskip]%
\>[B]{}(\mathbin{|\!>\!\!>\!\!>\!|})\mathbin{::}(\Conid{ArrowLoop}\;\Varid{arr},\Conid{ArrowChoice}\;\Varid{arr},{}\<[E]%
\\
\>[B]{}\hsindent{5}{}\<[5]%
\>[5]{}\Conid{ArrowLoopParallel}\;\Varid{arr}\;(\Varid{fut}\;(([\mskip1.5mu \Varid{a}\mskip1.5mu],[\mskip1.5mu \Varid{b}\mskip1.5mu]),[\mskip1.5mu \Varid{c}\mskip1.5mu]))\;(\Varid{fut}\;(([\mskip1.5mu \Varid{a}\mskip1.5mu],[\mskip1.5mu \Varid{b}\mskip1.5mu]),[\mskip1.5mu \Varid{c}\mskip1.5mu]))\;(),{}\<[E]%
\\
\>[B]{}\hsindent{5}{}\<[5]%
\>[5]{}\Conid{Future}\;\Varid{fut}\;(([\mskip1.5mu \Varid{a}\mskip1.5mu],[\mskip1.5mu \Varid{b}\mskip1.5mu]),[\mskip1.5mu \Varid{c}\mskip1.5mu])\;())\Rightarrow {}\<[E]%
\\
\>[B]{}\hsindent{5}{}\<[5]%
\>[5]{}\Varid{arr}\;\Varid{a}\;\Varid{b}\to \Varid{arr}\;\Varid{b}\;\Varid{c}\to \Varid{arr}\;\Varid{a}\;\Varid{c}{}\<[E]%
\\
\>[B]{}(\mathbin{|\!>\!\!>\!\!>\!|})\mathrel{=}\Varid{pipe2}\;(){}\<[E]%
\ColumnHook
\end{hscode}\resethooks
\caption{Definition of \ensuremath{\Varid{pipe2}} and \ensuremath{(\mathbin{|\!>\!\!>\!\!>\!|})}, a parallel \ensuremath{\mathbin{>\!\!>\!\!>}}.}
\label{fig:pipe2}
\end{figure}

Extensive use of \ensuremath{\Varid{pipe2}} over \ensuremath{\Varid{pipe}} with a hand-written combination data type will probably result in worse performance because of more communication overhead from the many calls to \ensuremath{\Varid{parEvalN}} inside of \ensuremath{\Varid{evalN}}. Nonetheless, we can define a parallel piping operator \ensuremath{\mathbin{|\!>\!\!>\!\!>\!|}}, which is semantically equivalent to \ensuremath{\mathbin{>\!\!>\!\!>}} similarly to other parallel syntactic sugar from Appendix~\ref{syntacticSugar}.


\subsubsection{Ring skeleton} \label{sec:ring}
\begin{figure}[tb]
	\includegraphics[scale=0.75]{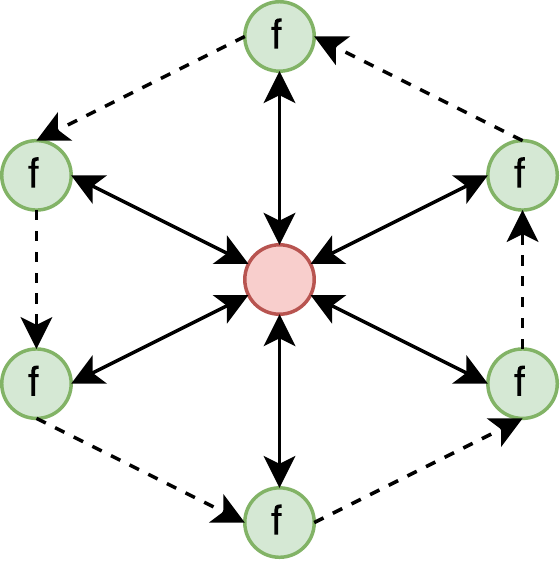}
	\caption{\ensuremath{\Varid{ring}} skeleton depiction.}
	\label{fig:ringImg}
\end{figure}
Eden comes with a ring skeleton\footnote{Available on Hackage: \url{https://hackage.haskell.org/package/edenskel-2.1.0.0/docs/Control-Parallel-Eden-Topology.html}} (Fig.~\ref{fig:ringImg}) implementation that allows the computation of a function \ensuremath{[\mskip1.5mu \Varid{i}\mskip1.5mu]\to [\mskip1.5mu \Varid{o}\mskip1.5mu]} with a ring of nodes that communicate with each other. Its input is a node function \ensuremath{\Varid{i}\to \Varid{r}\to (\Varid{o},\Varid{r})} in which \ensuremath{\Varid{r}} serves as the intermediary output that gets send to the neighbour of each node. This data is sent over direct communication channels, the so called \enquote{remote data}. We depict it in Appendix, Fig.~\ref{fig:ringEden}.

We can rewrite this functionality easily with the use of \ensuremath{\Varid{loop}} as the definition of the node function, \ensuremath{\Varid{arr}\;(\Varid{i},\Varid{r})\;(\Varid{o},\Varid{r})}, after being transformed into an Arrow, already fits quite neatly into \ensuremath{\Varid{loop}}'s signature: \ensuremath{\Varid{arr}\;(\Varid{a},\Varid{b})\;(\Varid{c},\Varid{b})\to \Varid{arr}\;\Varid{a}\;\Varid{c}}. In each iteration we start by rotating the intermediary input from the nodes \ensuremath{[\mskip1.5mu \Varid{fut}\;\Varid{r}\mskip1.5mu]} with \ensuremath{\Varid{second}\;(\Varid{rightRotate}\mathbin{>\!\!>\!\!>}\Varid{lazy})} (Fig.~\ref{fig:edenlazyrightrotate}). Similarly to the \ensuremath{\Varid{pipe}} from Section~\ref{sec:pipe} (Fig.~\ref{fig:pipeSimple}), we have to feed the intermediary input into our \ensuremath{\Varid{lazy}} (Fig.~\ref{fig:edenlazyrightrotate}) Arrow here, or the evaluation would fail to terminate. The reasoning is explained by \citet{Loogen2012} as a demand problem.

Next, we zip the resulting \ensuremath{([\mskip1.5mu \Varid{i}\mskip1.5mu],[\mskip1.5mu \Varid{fut}\;\Varid{r}\mskip1.5mu])} to \ensuremath{[\mskip1.5mu (\Varid{i},\Varid{fut}\;\Varid{r})\mskip1.5mu]} with \ensuremath{\Varid{arr}\;(\Varid{uncurry}\;\Varid{zip})}. We then feed this into our parallel Arrow \ensuremath{\Varid{arr}\;[\mskip1.5mu (\Varid{i},\Varid{fut}\;\Varid{r})\mskip1.5mu]\;[\mskip1.5mu (\Varid{o},\Varid{fut}\;\Varid{r})\mskip1.5mu]} obtained by transforming our input Arrow \ensuremath{\Varid{f}\mathbin{::}\Varid{arr}\;(\Varid{i},\Varid{r})\;(\Varid{o},\Varid{r})} into \ensuremath{\Varid{arr}\;(\Varid{i},\Varid{fut}\;\Varid{r})\;(\Varid{o},\Varid{fut}\;\Varid{r})} before \ensuremath{\Varid{repeat}}ing and lifting it with \ensuremath{\Varid{loopParEvalN}}. Finally we unzip the output list \ensuremath{[\mskip1.5mu (\Varid{o},\Varid{fut}\;\Varid{r})\mskip1.5mu]} list into \ensuremath{([\mskip1.5mu \Varid{o}\mskip1.5mu],[\mskip1.5mu \Varid{fut}\;\Varid{r}\mskip1.5mu])}.

Plugging this Arrow \ensuremath{\Varid{arr}\;([\mskip1.5mu \Varid{i}\mskip1.5mu],[\mskip1.5mu \Varid{fut}\;\Varid{r}\mskip1.5mu])\;([\mskip1.5mu \Varid{o}\mskip1.5mu],\Varid{fut}\;\Varid{r})} into the definition of \ensuremath{\Varid{loop}} from earlier gives us \ensuremath{\Varid{arr}\;[\mskip1.5mu \Varid{i}\mskip1.5mu]\;[\mskip1.5mu \Varid{o}\mskip1.5mu]}, our ring Arrow (Fig.~\ref{fig:ringFinal}). To make sure this algorithm has speedup on shared-memory machines as well, we pass the result of this Arrow to \ensuremath{\Varid{postLoopParEvalN}\;\Varid{conf}\;(\Varid{repeat}\;(\Varid{arr}\;\Varid{id}))}.
This combinator can, for example, be used to calculate the shortest paths in a graph using Warshall's algorithm.

\begin{figure}[tb]
\begin{hscode}\SaveRestoreHook
\column{B}{@{}>{\hspre}l<{\hspost}@{}}%
\column{5}{@{}>{\hspre}l<{\hspost}@{}}%
\column{9}{@{}>{\hspre}l<{\hspost}@{}}%
\column{E}{@{}>{\hspre}l<{\hspost}@{}}%
\>[B]{}\Varid{ring}\mathbin{::}(\Conid{Future}\;\Varid{fut}\;\Varid{r}\;\Varid{conf},{}\<[E]%
\\
\>[B]{}\hsindent{5}{}\<[5]%
\>[5]{}\Conid{ArrowLoop}\;\Varid{arr},{}\<[E]%
\\
\>[B]{}\hsindent{5}{}\<[5]%
\>[5]{}\Conid{ArrowLoopParallel}\;\Varid{arr}\;(\Varid{i},\Varid{fut}\;\Varid{r})\;(\Varid{o},\Varid{fut}\;\Varid{r})\;\Varid{conf},{}\<[E]%
\\
\>[B]{}\hsindent{5}{}\<[5]%
\>[5]{}\Conid{ArrowLoopParallel}\;\Varid{arr}\;\Varid{o}\;\Varid{o}\;\Varid{conf})\Rightarrow {}\<[E]%
\\
\>[B]{}\hsindent{5}{}\<[5]%
\>[5]{}\Varid{conf}\to \Varid{arr}\;(\Varid{i},\Varid{r})\;(\Varid{o},\Varid{r})\to \Varid{arr}\;[\mskip1.5mu \Varid{i}\mskip1.5mu]\;[\mskip1.5mu \Varid{o}\mskip1.5mu]{}\<[E]%
\\
\>[B]{}\Varid{ring}\;\Varid{conf}\;\Varid{f}\mathrel{=}{}\<[E]%
\\
\>[B]{}\hsindent{5}{}\<[5]%
\>[5]{}\Varid{loop}\;(\Varid{second}\;(\Varid{rightRotate}\mathbin{>\!\!>\!\!>}\Varid{lazy})\mathbin{>\!\!>\!\!>}{}\<[E]%
\\
\>[5]{}\hsindent{4}{}\<[9]%
\>[9]{}\Varid{arr}\;(\Varid{uncurry}\;\Varid{zip})\mathbin{>\!\!>\!\!>}{}\<[E]%
\\
\>[5]{}\hsindent{4}{}\<[9]%
\>[9]{}\Varid{loopParEvalN}\;\Varid{conf}\;(\Varid{repeat}\;(\Varid{second}\;(\Varid{get}\;\Varid{conf})\mathbin{>\!\!>\!\!>}\Varid{f}\mathbin{>\!\!>\!\!>}\Varid{second}\;(\Varid{put}\;\Varid{conf})))\mathbin{>\!\!>\!\!>}{}\<[E]%
\\
\>[5]{}\hsindent{4}{}\<[9]%
\>[9]{}\Varid{arr}\;\Varid{unzip})\mathbin{>\!\!>\!\!>}{}\<[E]%
\\
\>[B]{}\hsindent{5}{}\<[5]%
\>[5]{}\Varid{postLoopParEvalN}\;\Varid{conf}\;(\Varid{repeat}\;(\Varid{arr}\;\Varid{id})){}\<[E]%
\ColumnHook
\end{hscode}\resethooks
\caption{\ensuremath{\Varid{ring}} skeleton definition.}
\label{fig:ringFinal}
\end{figure}

\subsubsection{Torus skeleton}\label{sec:torus}
\begin{figure}
	\includegraphics[scale=0.75]{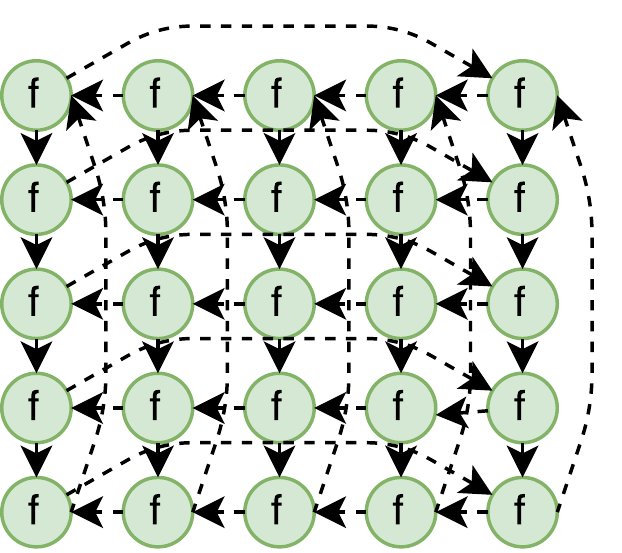}
	\caption{\ensuremath{\Varid{torus}} skeleton depiction.}
	\label{fig:ringTorusImg}
\end{figure}
If we take the concept of a \ensuremath{\Varid{ring}} from Section~\ref{sec:ring} one dimension further, we obtain a \ensuremath{\Varid{torus}} skeleton (Fig.~\ref{fig:ringTorusImg},~\ref{fig:torus}). Every node sends and receives data from horizontal and vertical neighbours in each communication round.
With our Parallel Arrows we re-implement the \ensuremath{\Varid{torus}} combinator\footnote{Available on Hackage: \url{https://hackage.haskell.org/package/edenskel-2.1.0.0/docs/Control-Parallel-Eden-Topology.html}.} from Eden---yet again with the help of the \ensuremath{\Conid{ArrowLoop}} type class.

Similar to the \ensuremath{\Varid{ring}}, we start by rotating the input (Fig.~\ref{fig:edenlazyrightrotate}), but this time not only in one direction, but in two. This means that the intermediary input from the neighbour nodes has to be stored in a tuple \ensuremath{([\mskip1.5mu [\mskip1.5mu \Varid{fut}\;\Varid{a}\mskip1.5mu]\mskip1.5mu],[\mskip1.5mu [\mskip1.5mu \Varid{fut}\;\Varid{b}\mskip1.5mu]\mskip1.5mu])} in the second argument (loop only allows for two arguments) of our looped Arrow \ensuremath{\Varid{arr}\;([\mskip1.5mu [\mskip1.5mu \Varid{c}\mskip1.5mu]\mskip1.5mu],([\mskip1.5mu [\mskip1.5mu \Varid{fut}\;\Varid{a}\mskip1.5mu]\mskip1.5mu],[\mskip1.5mu [\mskip1.5mu \Varid{fut}\;\Varid{b}\mskip1.5mu]\mskip1.5mu]))\;([\mskip1.5mu [\mskip1.5mu \Varid{d}\mskip1.5mu]\mskip1.5mu],([\mskip1.5mu [\mskip1.5mu \Varid{fut}\;\Varid{a}\mskip1.5mu]\mskip1.5mu],[\mskip1.5mu [\mskip1.5mu \Varid{fut}\;\Varid{b}\mskip1.5mu]\mskip1.5mu]))} and our rotation Arrow becomes 
\begin{hscode}\SaveRestoreHook
\column{B}{@{}>{\hspre}l<{\hspost}@{}}%
\column{E}{@{}>{\hspre}l<{\hspost}@{}}%
\>[B]{}\Varid{second}\;((\Varid{mapArr}\;\Varid{rightRotate}\mathbin{>\!\!>\!\!>}\Varid{lazy})\mathbin{*\!*\!*}(\Varid{arr}\;\Varid{rightRotate}\mathbin{>\!\!>\!\!>}\Varid{lazy})){}\<[E]%
\ColumnHook
\end{hscode}\resethooks
instead of the singular rotation in the ring as we rotate \ensuremath{[\mskip1.5mu [\mskip1.5mu \Varid{fut}\;\Varid{a}\mskip1.5mu]\mskip1.5mu]} horizontally and \ensuremath{[\mskip1.5mu [\mskip1.5mu \Varid{fut}\;\Varid{b}\mskip1.5mu]\mskip1.5mu]} vertically. Then, we zip the inputs for the input Arrow with 
\begin{hscode}\SaveRestoreHook
\column{B}{@{}>{\hspre}l<{\hspost}@{}}%
\column{E}{@{}>{\hspre}l<{\hspost}@{}}%
\>[B]{}\Varid{arr}\;(\Varid{uncurry3}\;\Varid{zipWith3}\;\Varid{lazyzip3}){}\<[E]%
\ColumnHook
\end{hscode}\resethooks
from \ensuremath{([\mskip1.5mu [\mskip1.5mu \Varid{c}\mskip1.5mu]\mskip1.5mu],([\mskip1.5mu [\mskip1.5mu \Varid{fut}\;\Varid{a}\mskip1.5mu]\mskip1.5mu],[\mskip1.5mu [\mskip1.5mu \Varid{fut}\;\Varid{b}\mskip1.5mu]\mskip1.5mu]))} to \ensuremath{[\mskip1.5mu [\mskip1.5mu (\Varid{c},\Varid{fut}\;\Varid{a},\Varid{fut}\;\Varid{b})\mskip1.5mu]\mskip1.5mu]}, which we then evaluate in parallel.

This, however, is more complicated than in the ring case as we have one more dimension of inputs that needs to be transformed. We first have to \ensuremath{\Varid{shuffle}} all the inputs to then pass them into \ensuremath{\Varid{loopParEvalN}\;\Varid{conf}\;(\Varid{repeat}\;(\Varid{ptorus}\;\Varid{conf}\;\Varid{f}))} to get an output of \ensuremath{[\mskip1.5mu (\Varid{d},\Varid{fut}\;\Varid{a},\Varid{fut}\;\Varid{b})\mskip1.5mu]}. We then unshuffle this list back to its original ordering by feeding it into \ensuremath{\Varid{arr}\;(\Varid{uncurry}\;\Varid{unshuffle})} which takes the input length we saved one step earlier as additional input to get a result matrix \ensuremath{[\mskip1.5mu [\mskip1.5mu [\mskip1.5mu (\Varid{d},\Varid{fut}\;\Varid{a},\Varid{fut}\;\Varid{b})\mskip1.5mu]\mskip1.5mu]}. Finally, we unpack this matrix  with \ensuremath{\Varid{arr}\;(\Varid{map}\;\Varid{unzip3})\mathbin{>\!\!>\!\!>}\Varid{arr}\;\Varid{unzip3}\mathbin{>\!\!>\!\!>}\Varid{threetotwo}} to get \ensuremath{([\mskip1.5mu [\mskip1.5mu \Varid{d}\mskip1.5mu]\mskip1.5mu],([\mskip1.5mu [\mskip1.5mu \Varid{fut}\;\Varid{a}\mskip1.5mu]\mskip1.5mu],[\mskip1.5mu [\mskip1.5mu \Varid{fut}\;\Varid{b}\mskip1.5mu]\mskip1.5mu]))}.

This internal looping computation is once again fed into \ensuremath{\Varid{loop}} and we also compose a final \ensuremath{\Varid{postLoopParEvalN}\;\Varid{conf}\;(\Varid{repeat}\;(\Varid{arr}\;\Varid{id}))} for the same reasons as explained for the \ensuremath{\Varid{ring}} skeleton. 

\begin{figure}[tb]
\begin{hscode}\SaveRestoreHook
\column{B}{@{}>{\hspre}l<{\hspost}@{}}%
\column{5}{@{}>{\hspre}l<{\hspost}@{}}%
\column{7}{@{}>{\hspre}l<{\hspost}@{}}%
\column{9}{@{}>{\hspre}l<{\hspost}@{}}%
\column{11}{@{}>{\hspre}l<{\hspost}@{}}%
\column{E}{@{}>{\hspre}l<{\hspost}@{}}%
\>[B]{}\Varid{torus}\mathbin{::}(\Conid{Future}\;\Varid{fut}\;\Varid{a}\;\Varid{conf},\Conid{Future}\;\Varid{fut}\;\Varid{b}\;\Varid{conf},{}\<[E]%
\\
\>[B]{}\hsindent{7}{}\<[7]%
\>[7]{}\Conid{ArrowLoop}\;\Varid{arr},\Conid{ArrowChoice}\;\Varid{arr},{}\<[E]%
\\
\>[B]{}\hsindent{7}{}\<[7]%
\>[7]{}\Conid{ArrowLoopParallel}\;\Varid{arr}\;(\Varid{c},\Varid{fut}\;\Varid{a},\Varid{fut}\;\Varid{b})\;(\Varid{d},\Varid{fut}\;\Varid{a},\Varid{fut}\;\Varid{b})\;\Varid{conf},{}\<[E]%
\\
\>[B]{}\hsindent{7}{}\<[7]%
\>[7]{}\Conid{ArrowLoopParallel}\;\Varid{arr}\;[\mskip1.5mu \Varid{d}\mskip1.5mu]\;[\mskip1.5mu \Varid{d}\mskip1.5mu]\;\Varid{conf})\Rightarrow {}\<[E]%
\\
\>[B]{}\hsindent{7}{}\<[7]%
\>[7]{}\Varid{conf}\to \Varid{arr}\;(\Varid{c},\Varid{a},\Varid{b})\;(\Varid{d},\Varid{a},\Varid{b})\to \Varid{arr}\;[\mskip1.5mu [\mskip1.5mu \Varid{c}\mskip1.5mu]\mskip1.5mu]\;[\mskip1.5mu [\mskip1.5mu \Varid{d}\mskip1.5mu]\mskip1.5mu]{}\<[E]%
\\
\>[B]{}\Varid{torus}\;\Varid{conf}\;\Varid{f}\mathrel{=}{}\<[E]%
\\
\>[B]{}\hsindent{5}{}\<[5]%
\>[5]{}\Varid{loop}\;(\Varid{second}\;((\Varid{mapArr}\;\Varid{rightRotate}\mathbin{>\!\!>\!\!>}\Varid{lazy})\mathbin{*\!*\!*}(\Varid{arr}\;\Varid{rightRotate}\mathbin{>\!\!>\!\!>}\Varid{lazy}))\mathbin{>\!\!>\!\!>}{}\<[E]%
\\
\>[5]{}\hsindent{4}{}\<[9]%
\>[9]{}\Varid{arr}\;(\Varid{uncurry3}\;(\Varid{zipWith3}\;\Varid{lazyzip3}))\mathbin{>\!\!>\!\!>}{}\<[E]%
\\
\>[5]{}\hsindent{4}{}\<[9]%
\>[9]{}\Varid{arr}\;\Varid{length}\mathbin{\&\!\&\!\&}(\Varid{shuffle}\mathbin{>\!\!>\!\!>}\Varid{loopParEvalN}\;\Varid{conf}\;(\Varid{repeat}\;(\Varid{ptorus}\;\Varid{conf}\;\Varid{f})))\mathbin{>\!\!>\!\!>}{}\<[E]%
\\
\>[5]{}\hsindent{4}{}\<[9]%
\>[9]{}\Varid{arr}\;(\Varid{uncurry}\;\Varid{unshuffle})\mathbin{>\!\!>\!\!>}{}\<[E]%
\\
\>[5]{}\hsindent{4}{}\<[9]%
\>[9]{}\Varid{arr}\;(\Varid{map}\;\Varid{unzip3})\mathbin{>\!\!>\!\!>}\Varid{arr}\;\Varid{unzip3}\mathbin{>\!\!>\!\!>}\Varid{threetotwo})\mathbin{>\!\!>\!\!>}{}\<[E]%
\\
\>[B]{}\hsindent{5}{}\<[5]%
\>[5]{}\Varid{postLoopParEvalN}\;\Varid{conf}\;(\Varid{repeat}\;(\Varid{arr}\;\Varid{id})){}\<[E]%
\\[\blanklineskip]%
\>[B]{}\Varid{ptorus}\mathbin{::}(\Conid{Arrow}\;\Varid{arr},\Conid{Future}\;\Varid{fut}\;\Varid{a}\;\Varid{conf},\Conid{Future}\;\Varid{fut}\;\Varid{b}\;\Varid{conf})\Rightarrow {}\<[E]%
\\
\>[B]{}\hsindent{11}{}\<[11]%
\>[11]{}\Varid{conf}\to {}\<[E]%
\\
\>[B]{}\hsindent{11}{}\<[11]%
\>[11]{}\Varid{arr}\;(\Varid{c},\Varid{a},\Varid{b})\;(\Varid{d},\Varid{a},\Varid{b})\to {}\<[E]%
\\
\>[B]{}\hsindent{11}{}\<[11]%
\>[11]{}\Varid{arr}\;(\Varid{c},\Varid{fut}\;\Varid{a},\Varid{fut}\;\Varid{b})\;(\Varid{d},\Varid{fut}\;\Varid{a},\Varid{fut}\;\Varid{b}){}\<[E]%
\\
\>[B]{}\Varid{ptorus}\;\Varid{conf}\;\Varid{f}\mathrel{=}{}\<[E]%
\\
\>[B]{}\hsindent{9}{}\<[9]%
\>[9]{}\Varid{arr}\;(\lambda \mathord{\sim}(\Varid{c},\Varid{a},\Varid{b})\to (\Varid{c},\Varid{get}\;\Varid{conf}\;\Varid{a},\Varid{get}\;\Varid{conf}\;\Varid{b}))\mathbin{>\!\!>\!\!>}{}\<[E]%
\\
\>[B]{}\hsindent{9}{}\<[9]%
\>[9]{}\Varid{f}\mathbin{>\!\!>\!\!>}{}\<[E]%
\\
\>[B]{}\hsindent{9}{}\<[9]%
\>[9]{}\Varid{arr}\;(\lambda \mathord{\sim}(\Varid{d},\Varid{a},\Varid{b})\to (\Varid{d},\Varid{put}\;\Varid{conf}\;\Varid{a},\Varid{put}\;\Varid{conf}\;\Varid{b})){}\<[E]%
\ColumnHook
\end{hscode}\resethooks
\caption{\ensuremath{\Varid{torus}} skeleton definition. \ensuremath{\Varid{lazyzip3}}, \ensuremath{\Varid{uncurry3}} and \ensuremath{\Varid{threetotwo}} definitions are in Fig.~\ref{fig:lazyzip3etc}}.
\label{fig:torus}
\end{figure}
\begin{figure}[tb]
\olcomment{more nodes!!!}
	\centering
	\includegraphics[width=0.9\textwidth]{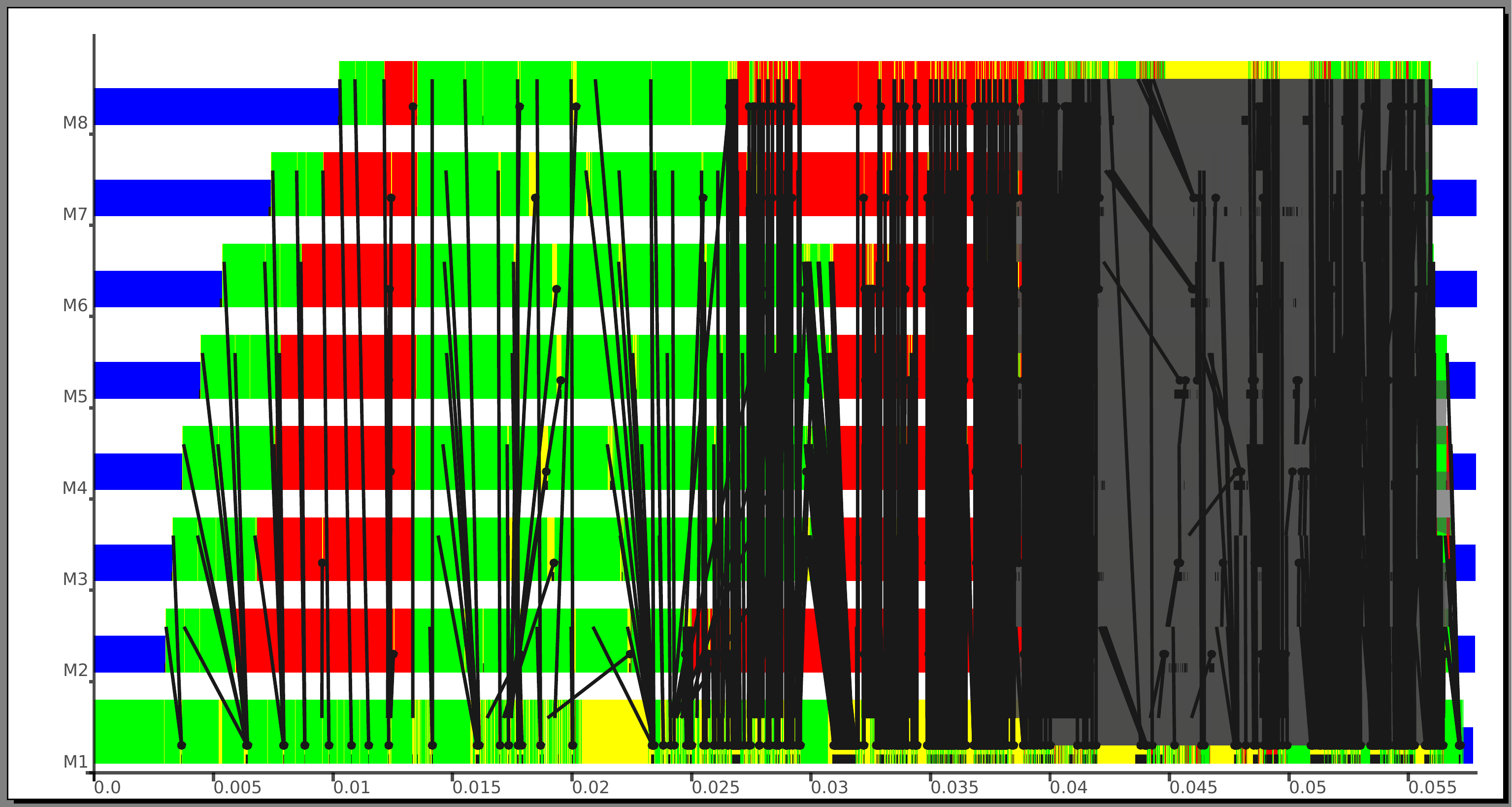}
	\caption[Matrix multiplication with \ensuremath{\Varid{torus}} (PArrows)]{Matrix multiplication with \ensuremath{\Varid{torus}} (PArrows).}
	\label{fig:torus_parrows_trace}
\end{figure}
\begin{figure}[tb]
	\centering
	\includegraphics[width=0.9\textwidth]{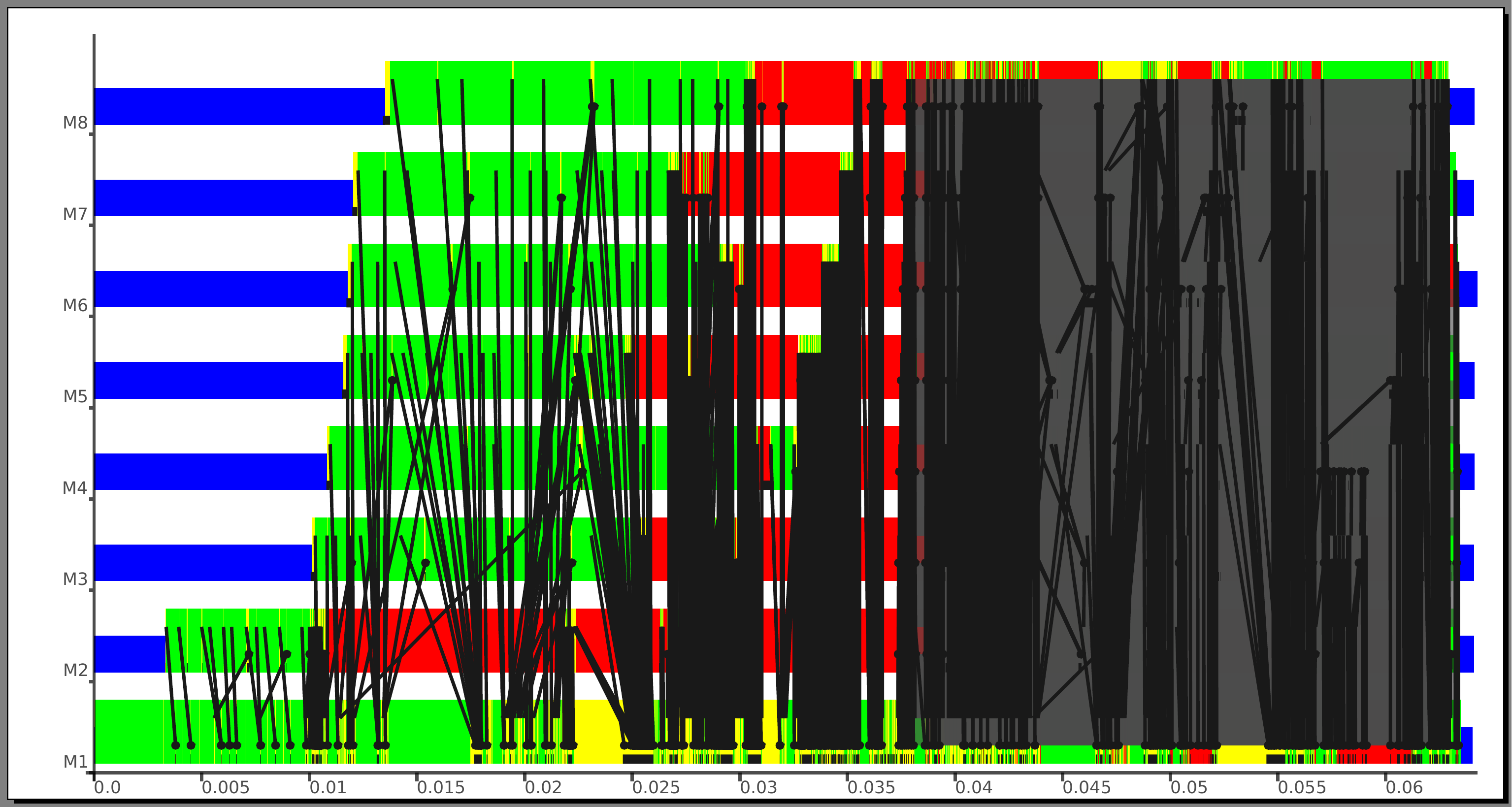}
	\caption[Matrix multiplication with \ensuremath{\Varid{torus}} (Eden)]{Matrix multiplication with \ensuremath{\Varid{torus}} (Eden).}
	\label{fig:torus_eden_trace}
\end{figure}

As an example of using this skeleton, \citet{Eden:SkeletonBookChapter02} showed the matrix multiplication using the Gentleman algorithm (\citeyear{Gentleman1978}). An adapted version can be found in Fig.~\ref{fig:torusMatMult}.
\begin{figure}[tb]
\begin{hscode}\SaveRestoreHook
\column{B}{@{}>{\hspre}l<{\hspost}@{}}%
\column{5}{@{}>{\hspre}l<{\hspost}@{}}%
\column{9}{@{}>{\hspre}l<{\hspost}@{}}%
\column{11}{@{}>{\hspre}l<{\hspost}@{}}%
\column{17}{@{}>{\hspre}l<{\hspost}@{}}%
\column{E}{@{}>{\hspre}l<{\hspost}@{}}%
\>[B]{}\mathbf{type}\;\Conid{Matrix}\mathrel{=}[\mskip1.5mu [\mskip1.5mu \Conid{Int}\mskip1.5mu]\mskip1.5mu]{}\<[E]%
\\[\blanklineskip]%
\>[B]{}\Varid{prMM\char95 torus}\mathbin{::}\Conid{Int}\to \Conid{Int}\to \Conid{Matrix}\to \Conid{Matrix}\to \Conid{Matrix}{}\<[E]%
\\
\>[B]{}\Varid{prMM\char95 torus}\;\Varid{numCores}\;\Varid{problemSizeVal}\;\Varid{m1}\;\Varid{m2}\mathrel{=}{}\<[E]%
\\
\>[B]{}\hsindent{9}{}\<[9]%
\>[9]{}\Varid{combine}\mathbin{\$}\Varid{torus}\;()\;(\Varid{mult}\;\Varid{torusSize})\mathbin{\$}\Varid{zipWith}\;(\Varid{zipWith}\;(,))\;(\Varid{split}\;\Varid{m1})\;(\Varid{split}\;\Varid{m2}){}\<[E]%
\\
\>[B]{}\hsindent{9}{}\<[9]%
\>[9]{}\mathbf{where}\;{}\<[17]%
\>[17]{}\Varid{torusSize}\mathrel{=}(\Varid{floor}\mathbin{\circ}\Varid{sqrt})\mathbin{\$}\Varid{fromIntegral}\;\Varid{numCores}{}\<[E]%
\\
\>[17]{}\Varid{combine}\mathrel{=}\Varid{concat}\mathbin{\circ}(\Varid{map}\;(\Varid{foldr}\;(\Varid{zipWith}\;(\plus ))\;(\Varid{repeat}\;[\mskip1.5mu \mskip1.5mu]))){}\<[E]%
\\
\>[17]{}\Varid{split}\mathrel{=}\Varid{splitMatrix}\;(\Varid{problemSizeVal}\mathbin{\Varid{`div`}}\Varid{torusSize}){}\<[E]%
\\[\blanklineskip]%
\>[B]{}\mbox{\onelinecomment  Function performed by each worker}{}\<[E]%
\\
\>[B]{}\Varid{mult}\mathbin{::}\Conid{Int}\to ((\Conid{Matrix},\Conid{Matrix}),[\mskip1.5mu \Conid{Matrix}\mskip1.5mu],[\mskip1.5mu \Conid{Matrix}\mskip1.5mu])\to (\Conid{Matrix},[\mskip1.5mu \Conid{Matrix}\mskip1.5mu],[\mskip1.5mu \Conid{Matrix}\mskip1.5mu]){}\<[E]%
\\
\>[B]{}\Varid{mult}\;\Varid{size}\;((\Varid{sm1},\Varid{sm2}),\Varid{sm1s},\Varid{sm2s})\mathrel{=}(\Varid{result},\Varid{toRight},\Varid{toBottom}){}\<[E]%
\\
\>[B]{}\hsindent{5}{}\<[5]%
\>[5]{}\mathbf{where}\;\Varid{toRight}\mathrel{=}\Varid{take}\;(\Varid{size}\mathbin{-}\mathrm{1})\;(\Varid{sm1}\mathbin{:}\Varid{sm1s}){}\<[E]%
\\
\>[5]{}\hsindent{6}{}\<[11]%
\>[11]{}\Varid{toBottom}\mathrel{=}\Varid{take}\;(\Varid{size}\mathbin{-}\mathrm{1})\;(\Varid{sm2'}\mathbin{:}\Varid{sm2s}){}\<[E]%
\\
\>[5]{}\hsindent{6}{}\<[11]%
\>[11]{}\Varid{sm2'}\mathrel{=}\Varid{transpose}\;\Varid{sm2}{}\<[E]%
\\
\>[5]{}\hsindent{6}{}\<[11]%
\>[11]{}\Varid{sms}\mathrel{=}\Varid{zipWith}\;\Varid{prMMTr}\;(\Varid{sm1}\mathbin{:}\Varid{sm1s})\;(\Varid{sm2'}\mathbin{:}\Varid{sm2s}){}\<[E]%
\\
\>[5]{}\hsindent{6}{}\<[11]%
\>[11]{}\Varid{result}\mathrel{=}\Varid{foldl1'}\;\Varid{matAdd}\;\Varid{sms}{}\<[E]%
\ColumnHook
\end{hscode}\resethooks
\caption{Adapted matrix multiplication in Eden using a the \ensuremath{\Varid{torus}} skeleton. \ensuremath{\Varid{prMM\char95 torus}} is the parallel matrix multiplication. \ensuremath{\Varid{mult}} is the function performed by each worker. \ensuremath{\Varid{prMMTr}} calculates $AB^T$ and is used for the (sequential) calculation in the chunks. \ensuremath{\Varid{splitMatrix}} splits the Matrix into chunks. \ensuremath{\Varid{matAdd}} calculates $A + B$. Omitted definitions can be found in \ref{fig:torus_example_rest}. }
\label{fig:torusMatMult}
\end{figure}
If we compare the trace from a call using our Arrow definition of the torus (Fig.~\ref{fig:torus_parrows_trace}) with the Eden version (Fig.~\ref{fig:torus_eden_trace}) we can see that the behaviour of the Arrow version and execution times are comparable. We discuss further benchmarks on larger clusters and in a more detail in the next section.

%
\section{Performance results and discussion}
\label{sec:benchmarks}

The preceding section has shown that PArrows are expressive. This section evaluates the performance overhead of this compositional abstraction in comparison to  GpH and the \ensuremath{\Conid{Par}} Monad on shared memory architectures and Eden on  a distributed memory cluster.
We describe our measurement platform, the benchmark results -- the shared-memory variants (GpH, \ensuremath{\Conid{Par}} Monad and Eden~CP) followed by Eden in a distributed memory setting, and conclude that PArrows hold up in terms of performance when compared to the original parallel Haskells.

\newcommand{\rmtest}{Rabin--Miller test\xspace}
\newcommand{\sudokutest}{Sudoku\xspace}
\newcommand{\jacobitest}{Jacobi sum test\xspace}
\newcommand{\torustest}{Gentleman\xspace}
\newlength{\plotwidthSMP}
\setlength{\plotwidthSMP}{0.39\textwidth}
\newlength{\plotwidthDist}
\setlength{\plotwidthDist}{0.6\textwidth}

\newcommand{\benchmarkDir}{benchmarks}

\newcommand{\speedupplot}[8]{
\begin{tikzpicture}
\begin{axis}[title={#1},
title style={align=center},
scale only axis, width=#7,
xlabel=Threads,
xtick distance=#4,
ytick distance=#4,
ylabel=Speedup,
ylabel near ticks,
grid=major,
legend entries={linear, #2},
legend style={at={(0.01,0.99)},anchor=north west},
max space between ticks=50pt,
grid style={line width=.1pt, draw=gray!10},
major grid style={line width=.2pt,draw=gray!50},
ymin=-1,
xmin=-1,
ymax=#8,
xmax=#6]
\addplot [domain=0:#3, no markers,dotted,thick]{x};
#5
\end{axis}
\end{tikzpicture}
}

\subsection{Measurement platform}
We start by explaining the hardware and software stack and outline the benchmark programs and motivation for choosing them. We also shortly address hyper-threading and why we do not use it in our benchmarks.

\subsubsection{Hardware and software}

The benchmarks are executed  both in a shared and in a distributed memory setting using the Glasgow GPG Beowulf cluster, consisting of
16 machines with 2 Intel\SymbReg~Xeon\SymbReg~E5-2640 v2 and 64 GB of DDR3 RAM each. Each processor has 8 cores and 16 (hyper-threaded) threads with a base frequency of 2 GHz and a turbo frequency of 2.50 GHz. This results in a total of 256 cores and 512 threads for the whole cluster. The operating system was Ubuntu 14.04 LTS with Kernel 3.19.0-33. Non-surprisingly, we found that hyper-threaded 32 cores do not behave in the same manner as real 16 cores (numbers here for a single machine). We disregarded the hyper-threading ability in most of the cases.

Apart from Eden, all benchmarks and libraries were compiled with Stack's\footnote{see \url{https://www.haskellstack.org/}} lts-7.1 GHC compiler which is equivalent to a standard GHC 8.0.1 with the base package in version 4.9.0.0. Stack itself was used in version 1.3.2. For GpH in its Multicore variant we used the parallel package in version 3.2.1.0\footnote{see \url{https://hackage.haskell.org/package/parallel-3.2.1.0}}, while for the \ensuremath{\Conid{Par}} Monad we used monad-par in version 0.3.4.8\footnote{see \url{https://hackage.haskell.org/package/monad-par-0.3.4.8}}. For all Eden tests, we used its GHC-Eden compiler in version 7.8.2\footnote{see \url{http://www.mathematik.uni-marburg.de/~eden/?content=build_eden_7_&navi=build}} together with OpenMPI 1.6.5\footnote{see \url{https://www.open-mpi.org/software/ompi/v1.6/}}.

Furthermore, all benchmarks were done with help of the bench\footnote{see \url{https://hackage.haskell.org/package/bench}} tool in version 1.0.2 which uses criterion (>=1.1.1.0 \&\& < 1.2)\footnote{see \url{https://hackage.haskell.org/package/criterion-1.1.1.0}} internally. All runtime data (mean runtime, max stddev, etc.) was collected with this tool.

We used a single node with 16 real cores as a shared memory test-bed
and the whole grid with 256 real cores as a device to test our
distributed memory software.

\subsubsection{Benchmarks}

We measure four benchmarks from different
sources. Most of them are parallel mathematical computations, initially
implemented in Eden. Table~\ref{tab:benches} summarises.

\begin{table}
\centering
\caption{The benchmarks we use in this paper.}
\label{tab:benches}
\renewcommand{\tabcolsep}{0.5em}
\begin{tabular}{lccll}
\toprule
Name & Area & Type & Origin & Source \\
\midrule
\rmtest & Mathematics & \ensuremath{\Varid{parMap}\mathbin{+}\Varid{reduce}} & Eden & \citet{Lobachev2012}\\
\jacobitest & Mathematics & \ensuremath{\Varid{workpool}\mathbin{+}\Varid{reduce}} & Eden & \citet{Lobachev2012}\\
\torustest & Mathematics & \ensuremath{\Varid{torus}} & Eden & \citet{Eden:SkeletonBookChapter02}\\
\sudokutest & Puzzle & \ensuremath{\Varid{parMap}} & \ensuremath{\Conid{Par}} Monad & \citet{par-monad}\tablefootnote{actual code from: \url{http://community.haskell.org/\~simonmar/par-tutorial.pdf} and \url{https://github.com/simonmar/parconc-examples}}\\
\bottomrule
\end{tabular}
\end{table}

\rmtest is a probabilistic primality test that iterates multiple (here: 32--256)
\enquote{subtests}. Should a subtest fail, the input is definitely not a
prime. If all $n$ subtest pass, the input is composite with the
probability of $1/4^{n}$. 

Jacobi sum test or APRCL is also a primality test, that however,
guarantees the correctness of the result. It is probabilistic in the
sense that its run time is not certain. Unlike \rmtest, the subtests
of Jacobi sum test have very different durations. \citet{lobachev-phd}
discusses some optimisations of parallel APRCL. Generic parallel
implementations of \rmtest and APRCL were presented in \citet{Lobachev2012}.

\enquote{Gentleman} is a standard Eden test program, developed
for their \ensuremath{\Varid{torus}} skeleton. It implements a Gentleman's algorithm for parallel matrix
multiplication \citep{Gentleman1978}. We ported an Eden-based version \citep{Eden:SkeletonBookChapter02} to PArrows.

A~parallel Sudoku solver was used by \citet{par-monad} to compare \ensuremath{\Conid{Par}} Monad
to GpH, we ported it to PArrows.

\subsubsection{What parallel Haskells run where}

The \ensuremath{\Conid{Par}} monad and GpH -- in its multicore version \cite{Marlow2009} --  can be executed on shared memory machines only.
Although GpH is available on distributed memory
clusters, and newer distributed memory Haskells such as HdpH exist,
current support of distributed memory in PArrows is limited to
Eden. We used the MPI backend of Eden in a distributed memory
setting. However, for shared memory Eden features a \enquote{CP} backend
that merely copies the memory blocks between disjoint heaps. In
this mode, Eden still operates in the \enquote{nothing shared} setting, but
is adapted better to multicore machines. We call this version of Eden
\enquote{Eden~CP}.

\subsubsection{Effect of hyper-threading}

In preliminary tests, the PArrows version of \rmtest on a single node of the Glasgow cluster
showed almost linear speedup on up to 16 shared-memory cores (as supplementary materials show). The speedup
of 64-task PArrows/Eden at 16 real cores version was 13.65 giving a parallel
efficiency of 85.3\%. However, if we increased the number of
requested cores to 32 -- \ie if we use hyper-threading on 16 real
cores -- the speedup did not increase that well. It was merely 15.99
for 32 tasks with PArrows/Eden. This was worse for other implementations.  As
for 64 tasks, we obtained a speedup of 16.12 with PArrows/Eden at 32
hyper-threaded cores and only 13.55 with PArrows/GpH. 

While this shows that hyper-threading can be of benefit in scenarios similar to the ones presented in the benchmarks, we only use real cores for the performance measurements in Section~\ref{sec:benchmarkResults} as the purpose of this paper is to show the performance of PArrows and not to investigate parallel behaviour with hyper-threading.



\subsection{Benchmark results}\label{sec:benchmarkResults}

We compare the PArrow performance with direct implementations of the benchmarks in Eden, GpH and the \ensuremath{\Conid{Par}} Monad.
We start with the definition of mean overhead to compare both PArrows-enabled and standard benchmark implementations. We continue comparing speedups and overheads for the shared memory implementations and then study OpenMPI variants of the Eden-enabled PArrows as a representative of a distributed memory backend. We plot all speedup curves and all overhead values in the supplementary materials.

\subsubsection{Defining overhead}

We compare the mean overhead, \ie the mean of relative wall-clock run time differences between the PArrow and direct benchmark implementations executed multiple times with the same settings.
The error margins of the time measurements, supplied by criterion package\footnote{\url{https://hackage.haskell.org/package/criterion-1.1.1.0}}, yield the error margin of the mean overhead. 

Quite often the zero value lies in the error margin of the mean overhead. This means that even though we have measured some difference (against or even in favour of PArrows), it could be merely the error margin of the measurement and the difference might not be existent. We are mostly interested in the cases where above issue does not persist, we call them \emph{significant}. We often denote the error margin with $\pm$ after the mean overhead value.

\subsubsection{Shared memory}

\paragraph{Speedup.}
The \rmtest benchmark showed almost linear speedup for both 32 and 64 tasks, the performance is slightly better in the latter case: 13.7 at 16 cores for input $2^{11213}-1$ and 64 tasks in the best case scenario with Eden~CP. The performance of the \sudokutest benchmark merely reaches a speedup of 9.19 (GpH), 8.78 (\ensuremath{\Conid{Par}} Monad), 8.14 (Eden~CP) for 16 cores and 1000 Sudokus. In contrast to Rabin--Miller, here the \ensuremath{\Conid{GpH}} seems to be the best of all, while Rabin--Miller profited most from Eden~CP (\ie Eden with direct memory copy) implementation of PArrows. Gentleman on shared memory has a plummeting speedup curve with GpH and \ensuremath{\Conid{Par}} Monad and logarithmically increasing speedup for the Eden-based version. The latter reached a speedup of 6.56 at 16 cores.

\paragraph{Overhead.}

For the shared memory \rmtest benchmark, implemented with PArrows using Eden~CP, GpH, and \ensuremath{\Conid{Par}} Monad, the overhead values are within single percents range, but also negative overhead (\ie PArrows are better) and larger error margins happen. To give a few examples, the overhead for Eden~CP with input value $2^{11213}-1$, 32 tasks, and 16 cores is $1.5\%$, but the error margin is around $5.2\%$! Same implementation in the same setting with 64 tasks reaches $-0.2\%$ overhead, PArrows apparently fare better than Eden -- but the error margin of $1.9\%$ disallows this interpretation. We focus now on significant overhead values. To name a few: $0.41\%\; \pm 7\cdot 10^{-2}\%$ for Eden~CP and 64 tasks at 4 cores; $4.7\% \; \pm 0.72\%$ for GpH, 32 tasks, 8 cores; $0.34\% \; \pm 0.31\%$ for \ensuremath{\Conid{Par}} Monad at 4 cores with 64 tasks. The worst significant overhead was in case of GpH  with $8\% \; \pm 6.9\%$ at 16 cores with 32 tasks and input value $2^{11213}-1$. In other words, we notice no major slow-down through PArrows here.

For Sudoku the situation is slightly different. There is a minimal significant ($-1.4\% \; \pm 1.2\%$ at 8 cores) speed \emph{improvement} with PArrows Eden~CP version when compared with the base Eden CP benchmark. However, with increasing number of cores the error margin reaches zero again: $-1.6\% \; \pm 5.0\%$ at 16 cores. The \ensuremath{\Conid{Par}} Monad shows a similar development, \eg with $-1.95\% \; \pm 0.64\%$ at 8 cores. The GpH version shows both a significant speed improvement of $-4.2\% \; \pm 0.26\%$ (for 16 cores) with PArrows and a minor overhead of $0.87\% \; \pm 0.70\%$ (4 cores).

The Gentleman multiplication with Eden CP shows a minor significant overhead of $2.6\% \; \pm 1.0\%$ at 8 cores and an insignificant improvement at 16 cores. Summarising, we observe a low (if significant at all) overhead, induced by PArrows in the shared memory setting.

\subsubsection{Distributed memory}

\begin{figure}[ht]
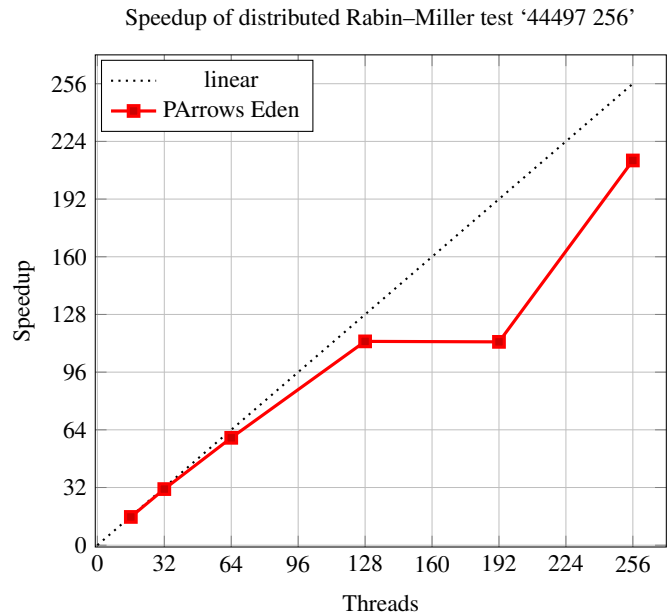

	\centering
	\speedupplot{Speedup of distributed \rmtest \enquote{44497 256}}{PArrows Eden}{256}{32}{
	\addplot+ [very thick] table [scatter, x="nCores", y="speedup", col sep=comma, mark=none,
	smooth]{bench-distributed.bench.skelrm-parrows-44497-256.csv};
	}{272}{\plotwidthDist}{272}
	\caption[Speedup distributed Rabin--Miller]{Speedup of the distributed \rmtest benchmark using PArrows with Eden.}
	\label{fig:rabinMillerDistSpeedup}
\end{figure}

\paragraph{Speedup.}
The speedup of distributed memory Rabin--Miller benchmark with PArrows and Eden showed an almost 
linear speedup excepting around 192 cores where an unfortunate task distribution reduces performance.
As seen in Fig.~\ref{fig:rabinMillerDistSpeedup}, we reached a speedup of 213.4 with PArrrows at 256 cores (vs. 207.7 with pure Eden). Because of memory limitations, the speedup of Jacobi sum test for large inputs (such as $2^{4253}-1$) could be measured only in a massively distributed setting. PArrows improved there from \SI{9193}{\second} (at 128 cores) to \SI{1649}{\second} (at 256 cores). A~scaled-down version with input $2^{3217}-1$ stagnates the speedup at about 11 for both PArrows and Eden for more than 64 cores. There is apparently not enough work for that many cores. The Gentleman test with input 4096 had an almost linear speedup first, then plummeted between 128 and 224 cores, and recovered at 256 cores with speedup of~129.

\paragraph{Overhead.}
We use our mean overhead quality measure and the notion of significance also for distributed memory benchmarks. The mean overhead of Rabin-Miller test in the distributed memory setting ranges from $0.29\%$ to $-2.8\%$ (last value in favour of PArrows), but these values are not significant with error margins $\pm 0.8\%$ and $\pm 2.9\%$ correspondingly. A sole significant (by a very low margin) overhead is $0.35\% \; \pm 0.33\%$ at 64 cores.
We measured the mean overhead for Jacobi benchmark for an input of $2^{3217}-1$ for up to 256 cores.
We reach the flattering value $-3.8\% \; \pm 0.93\%$ at 16 cores in favour of PArrows, it was the sole significant overhead value. The  value for 256 cores was $0.31\% \; \pm 0.39\%$.
Mean overhead for distributed Gentleman multiplication was also low. Significant values include $1.23\% \; \pm 1.20\%$ at 64 cores and $2.4\% \; \pm 0.97\%$ at 256 cores. It took PArrows 64.2 seconds at 256 cores to complete the benchmark.

Similar to the shared memory setting, PArrows only imply a very low penalty with distributed memory that lies in lower single-percent digits at most.

\subsection{Discussion}


\begin{table}[]
\centering
\caption{Overhead in the shared memory benchmarks. Bold marks values in favour of PArrows.}
\label{tab:meanOverheadSharedMemory}
\centering
\begingroup\catcode`"=9
\sisetup{round-mode=places, round-precision=3}
\begin{tabular}{C{4cm} C{2cm} C{1.7cm} C{2.3cm} C{1.7cm}}
	\thead{Benchmark} & \thead{Base}             & \thead{Mean of \\ mean \\overheads} & \thead{Maximum \\ normalised \\ stdDev} & \thead{Runtime for \\ 16 cores (s)} \\ \hline \\
	\multirow{3}{*}{\sudokutest 1000}
	\begin{filecontents*}{bestAndWorstSudoku-1000.csv}
"benchmark","vs","params","displayName","","worstName","worstTime","worstNCores","worstSpeedup","worstMaxstddev","worstFactor","worstFactorStdDev","worstOverhead","worstStdDevForOverhead","worstRuntimeX","worstRuntimeY","","bestName","bestTime","bestNCores","bestSpeedup","bestMaxstddev","bestFactor","bestFactorStdDev","bestOverhead","bestStdDevForOverhead","bestRuntimeX","bestRuntimeY","meanFactor","maxStdDevFactor","meanOverhead","maxStdDevForOverhead","runtimeMaxCores"
"Sudoku (Shared-Memory)","Eden CP","1000","Eden CP vs. PArrows 1000","4","bench/./eden-sudoku sudoku17.1000.txt-bench/./parrows-sudoku-parmap-eden sudoku17.1000.txt",1.637621762024022e-2,8,-7.986537640347269e-2,1.2917229656086132e-2,0.986,1.1038863805406956e-2,-1.42e-2,1.1195543888032955e-2,1.1701593464500515,1.1537831288298113,"2","bench/./eden-sudoku sudoku17.1000.txt-bench/./parrows-sudoku-parmap-eden sudoku17.1000.txt",0.10213296612103795,2,-6.216963687210475e-2,7.712255704127203e-3,0.969,2.308806659243841e-3,-3.15e-2,2.381625811588361e-3,3.3403644576514906,3.2382314915304526,0.979,0.0498,"\textbf{-2.1\%}",5.1\%,1.17
"Sudoku (Shared-Memory)","GpH","1000","GpH vs. PArrows 1000","3","bench/./multicore-sudoku sudoku17.1000.txt-bench/./parrows-sudoku-parmap-mult sudoku17.1000.txt",-1.7040419081846725e-2,4,2.929276395028424e-2,1.38547393824998e-2,1.01,7.1100393963584784e-3,8.67e-3,7.048401958053585e-3,1.9486164014218725,1.9656568205037193,"5","bench/./multicore-sudoku sudoku17.1000.txt-bench/./parrows-sudoku-parmap-mult sudoku17.1000.txt",3.0452317247788185e-2,16,-0.374411033652299,1.8554577054043255e-3,0.959,2.483262765942491e-3,-4.25e-2,2.5887708085167715e-3,0.7471854089915894,0.7167330917438012,0.992,0.00711,"\textbf{-0.82\%}",0.7\%,1.11
"Sudoku (Shared-Memory)","Par Monad ","1000","Par Monad vs. PArrows 1000","5","bench/./parmonad-sudoku sudoku17.1000.txt-bench/./parrows-sudoku-parmap-par sudoku17.1000.txt",-4.968114507695076e-3,16,5.854298541700231e-2,1.611224415196458e-2,1.01,2.1626379165803972e-2,6.62e-3,2.1483121863732923e-2,0.7450273588766795,0.7499954733843746,"4","bench/./parmonad-sudoku sudoku17.1000.txt-bench/./parrows-sudoku-parmap-par sudoku17.1000.txt",2.1793455713325027e-2,8,-0.11286356897026284,7.0987795985597455e-3,0.981,6.23510180089946e-3,-1.95e-2,6.356782960550528e-3,1.1385186361409037,1.1167251804275786,0.988,0.0216,"\textbf{-1.3\%}",2.1\%,1.14
\end{filecontents*}
\csvreader[head to column names]{bestAndWorstSudoku-1000.csv}{}{& \vs & \meanOverhead & \maxStdDevForOverhead & \runtimeMaxCores \\}
	\\ \hline \\
	\multirow{1}{*}{\torustest 512}
	\begin{filecontents*}{bestAndWorstTorusSM-512.csv}
"benchmark","vs","params","displayName","","worstName","worstTime","worstNCores","worstSpeedup","worstMaxstddev","worstFactor","worstFactorStdDev","worstOverhead","worstStdDevForOverhead","worstRuntimeX","worstRuntimeY","","bestName","bestTime","bestNCores","bestSpeedup","bestMaxstddev","bestFactor","bestFactorStdDev","bestOverhead","bestStdDevForOverhead","bestRuntimeX","bestRuntimeY","meanFactor","maxStdDevFactor","meanOverhead","maxStdDevForOverhead","runtimeMaxCores"
"Torus (Shared-Memory)","Eden CP","512","Eden vs. PArrows 512","4","bench/./torus_matrix_eden_sm 512-bench/./torus_matrix_parrows_eden_sm 512",-4.422902189091671e-2,8,0.1297177337063422,1.7913418837691958e-2,1.03,1.080812152000734e-2,2.6e-2,1.052719562871802e-2,1.6574035371948515,1.7016325590857682,"5","bench/./torus_matrix_eden_sm 512-bench/./torus_matrix_parrows_eden_sm 512",7.168860122975573e-3,16,-3.707054701625889e-2,8.547446046412084e-2,0.994,6.739100657135076e-2,-5.68e-3,6.777407759057727e-2,1.2683363079556331,1.2611674478326576,1.01,0.0674,"\textit{0.81\%}",6.8\%,1.66
\end{filecontents*}
\csvreader[head to column names]{bestAndWorstTorusSM-512.csv}{}{& \vs & \meanOverhead & \maxStdDevForOverhead & \runtimeMaxCores \\}
	\\ \hline \\
	\multirow{3}{*}{\rmtest 11213 32}
	\begin{filecontents*}{bestAndWorstRMSM-11213-32.csv}
"benchmark","vs","params","displayName","","worstName","worstTime","worstNCores","worstSpeedup","worstMaxstddev","worstFactor","worstFactorStdDev","worstOverhead","worstStdDevForOverhead","worstRuntimeX","worstRuntimeY","","bestName","bestTime","bestNCores","bestSpeedup","bestMaxstddev","bestFactor","bestFactorStdDev","bestOverhead","bestStdDevForOverhead","bestRuntimeX","bestRuntimeY","meanFactor","maxStdDevFactor","meanOverhead","maxStdDevForOverhead","runtimeMaxCores"
"Rabin-Miller (Shared-Memory)","Eden CP","11213 32","Eden CP vs. PArrows 11213 32","5","bench/./skelrm-eden-cp 11213 32-bench/./skelrm-parr-eden-cp 11213 32",-4.269474227395342e-2,16,0.20225719175084578,0.1465317786293577,1.02,5.234126303721241e-2,1.5e-2,5.155501835834275e-2,2.7995460966461554,2.842240838920109,"2","bench/./skelrm-eden-cp 11213 32-bench/./skelrm-parr-eden-cp 11213 32",-8.875875247435872e-2,2,9.143105109885141e-3,1.8744278260221795e-2,1.0,9.821515322365121e-4,4.63e-3,9.776049572283482e-4,19.084914745831686,19.173673498306044,1.01,0.0523,"\textit{0.79\%}",5.2\%,5.16
"Rabin-Miller (Shared-Memory)","GpH","11213 32","GpH vs. PArrows 11213 32","5","bench/./skelrm-mult 11213 32-bench/./skelrm-parr-mult 11213 32",-0.25967859766549584,16,1.020098622917022,0.22301263698737295,1.09,7.507435200045001e-2,8.04e-2,6.903912569882896e-2,2.9705569351572447,3.2302355328227406,"2","bench/./skelrm-mult 11213 32-bench/./skelrm-parr-mult 11213 32",-5.445944890379906e-2,2,5.4778328142575106e-3,2.0278329686933546e-2,1.0,1.0489910624942255e-3,2.81e-3,1.046044179599149e-3,19.331270219516455,19.385729668420254,1.04,0.0751,"\textit{3.5\%}",6.9\%,5.28
"Rabin-Miller (Shared-Memory)","Par Monad ","11213 32","Par Monad vs. PArrows 11213 32","5","bench/./skelrm-par 11213 32-bench/./skelrm-parr-par 11213 32",-1.5569804960655187e-2,16,6.762528606934914e-2,0.15029939404345763,1.01,5.115390793893668e-2,5.27e-3,5.088426548551566e-2,2.9381800941361638,2.953749899096819,"4","bench/./skelrm-par 11213 32-bench/./skelrm-parr-par 11213 32",0.5582982600770059,8,-0.6818679418683526,0.9838505435373271,0.904,0.16842015507263824,-0.106,0.18621734058802458,5.841643733869028,5.283345473792022,0.978,0.168,"\textbf{-2.5\%}",19.0\%,5.84
\end{filecontents*}
\csvreader[head to column names]{bestAndWorstRMSM-11213-32.csv}{}{& \vs & \meanOverhead & \maxStdDevForOverhead & \runtimeMaxCores \\}
	\\ \hline \\
	\multirow{3}{*}{\rmtest 11213 64}
	\begin{filecontents*}{bestAndWorstRMSM-11213-64.csv}
"benchmark","vs","params","displayName","","worstName","worstTime","worstNCores","worstSpeedup","worstMaxstddev","worstFactor","worstFactorStdDev","worstOverhead","worstStdDevForOverhead","worstRuntimeX","worstRuntimeY","","bestName","bestTime","bestNCores","bestSpeedup","bestMaxstddev","bestFactor","bestFactorStdDev","bestOverhead","bestStdDevForOverhead","bestRuntimeX","bestRuntimeY","meanFactor","maxStdDevFactor","meanOverhead","maxStdDevForOverhead","runtimeMaxCores"
"Rabin-Miller (Shared-Memory)","Eden CP","11213 64","Eden CP vs. PArrows 11213 64","2","bench/./skelrm-eden-cp 11213 64-bench/./skelrm-parr-eden-cp 11213 64",-0.19811010816031427,2,1.0202687069202465e-2,1.609078624767623e-2,1.01,4.215516565790552e-4,5.16e-3,4.1937503593220354e-4,38.1703784021512,38.368488510311515,"5","bench/./skelrm-eden-cp 11213 64-bench/./skelrm-parr-eden-cp 11213 64",1.1685287269453504e-2,16,-2.897939565695573e-2,0.1060196023842691,0.998,1.920421149635597e-2,-2.12e-3,1.924494637322122e-2,5.520643344528177,5.508958057258724,1.0,0.0192,"\textit{0.21\%}",1.9\%,10.3
"Rabin-Miller (Shared-Memory)","GpH","11213 64","GpH vs. PArrows 11213 64","5","bench/./skelrm-mult 11213 64-bench/./skelrm-parr-mult 11213 64",-0.2226828384316626,16,0.4995938552502963,7.574642022807072e-2,1.04,1.3317170214233849e-2,3.77e-2,1.2815440163615218e-2,5.687876554067796,5.910559392499459,"2","bench/./skelrm-mult 11213 64-bench/./skelrm-parr-mult 11213 64",-4.243913634369534e-2,2,2.138593354588547e-3,3.344836082497214e-2,1.0,8.650462674241591e-4,1.1e-3,8.640978621403383e-4,38.66655702078346,38.70899615712715,1.02,0.0133,"\textit{1.6\%}",1.3\%,10.6
"Rabin-Miller (Shared-Memory)","Par Monad ","11213 64","Par Monad vs. PArrows 11213 64","3","bench/./skelrm-par 11213 64-bench/./skelrm-parr-par 11213 64",-6.882146093994379e-2,4,1.2620334333781624e-2,6.309519134219096e-2,1.0,3.1163920267642115e-3,3.39e-3,3.1058346010475308e-3,20.246230512822702,20.315051973762646,"5","bench/./skelrm-par 11213 64-bench/./skelrm-parr-par 11213 64",0.4702964991124139,16,-1.0450499941127287,0.9239802887252248,0.922,0.15232342585814976,-8.4e-2,0.16512580504059476,6.065910634032586,5.595614134920172,0.963,0.152,"\textbf{-4.0\%}",17.0\%,11.4
\end{filecontents*}
\csvreader[head to column names]{bestAndWorstRMSM-11213-64.csv}{}{& \vs & \meanOverhead & \maxStdDevForOverhead & \runtimeMaxCores \\}
\end{tabular}
\endgroup
\end{table}

\begin{table}[]
\centering
\caption{Overhead in the distributed memory benchmarks. Bold marks values
  in favour of PArrows.}
\label{tab:meanOverHeadDistributedMemory}
\centering
\begingroup\catcode`"=9
\begin{tabular}{C{4cm} C{2cm} C{1.7cm} C{2.3cm} C{1.7cm}}
	\thead{Benchmark} & \thead{Base}             & \thead{Mean of \\ mean \\overheads} & \thead{Maximum \\ normalised \\ stdDev} & \thead{Runtime for \\ 256 cores (s)} \\ \hline \\
	\multirow{1}{*}{\torustest 4096}
	\begin{filecontents*}{bestAndWorstTorus-4096.csv}
"benchmark","vs","params","displayName","","worstName","worstTime","worstNCores","worstSpeedup","worstMaxstddev","worstFactor","worstFactorStdDev","worstOverhead","worstStdDevForOverhead","worstRuntimeX","worstRuntimeY","","bestName","bestTime","bestNCores","bestSpeedup","bestMaxstddev","bestFactor","bestFactorStdDev","bestOverhead","bestStdDevForOverhead","bestRuntimeX","bestRuntimeY","meanFactor","maxStdDevFactor","meanOverhead","maxStdDevForOverhead","runtimeMaxCores"
"Torus (Distributed)","Eden","4096","Eden vs. PArrows 4096","10","bench/./torus_matrix_eden 4096-bench/./torus_matrix_parrows_eden 4096",-1.5685654832050275,256,3.238873878214548,0.625426299322997,1.03,9.979498607854406e-3,2.44e-2,9.73582527270879e-3,62.67111444163664,64.23967992484167,"8","bench/./torus_matrix_eden 4096-bench/./torus_matrix_parrows_eden 4096",0.18758766384173953,192,-0.22133814814162633,0.6507934302436146,0.998,7.744667587139149e-3,-2.24e-3,7.76199514123013e-3,84.03116375508831,83.84357609124658,1.01,0.0147,"\textit{0.67\%}",1.5\%,110.0
\end{filecontents*}
\csvreader[head to column names]{bestAndWorstTorus-4096.csv}{}{& \vs & \meanOverhead & \maxStdDevForOverhead & \runtimeMaxCores \\}
	\\ \hline \\
  	\multirow{1}{*}{\rmtest 44497 256}
	\begin{filecontents*}{bestAndWorstRM-44497-256.csv}
"benchmark","vs","params","displayName","","worstName","worstTime","worstNCores","worstSpeedup","worstMaxstddev","worstFactor","worstFactorStdDev","worstOverhead","worstStdDevForOverhead","worstRuntimeX","worstRuntimeY","","bestName","bestTime","bestNCores","bestSpeedup","bestMaxstddev","bestFactor","bestFactorStdDev","bestOverhead","bestStdDevForOverhead","bestRuntimeX","bestRuntimeY","meanFactor","maxStdDevFactor","meanOverhead","maxStdDevForOverhead","runtimeMaxCores"
"Rabin-Miller (Distributed)","Eden","44497 256","Eden vs. PArrows 44497 256","6","bench/./skelrm-eden 44497 256-bench/./skelrm-parrows 44497 256",-0.2543358287463491,192,0.3315338425749843,0.6898809179690073,1.0,7.972864276632251e-3,2.93e-3,7.949498108723297e-3,86.52861682231143,86.78295265105778,"7","bench/./skelrm-eden 44497 256-bench/./skelrm-parrows 44497 256",1.2624170565977693,256,-5.716918551932167,1.3433314464267978,0.973,2.8504234741791114e-2,-2.75e-2,2.9288802902585495e-2,47.12743417234387,45.8650171157461,0.995,0.0285,"\textbf{-0.5\%}",2.9\%,165.0
\end{filecontents*}
\csvreader[head to column names]{bestAndWorstRM-44497-256.csv}{}{& \vs & \meanOverhead & \maxStdDevForOverhead & \runtimeMaxCores \\}
	\\ \hline \\
	\multirow{1}{*}{\jacobitest 3217}
	\begin{filecontents*}{bestAndWorstJacobi-3217.csv}
"benchmark","vs","params","displayName","","worstName","worstTime","worstNCores","worstSpeedup","worstMaxstddev","worstFactor","worstFactorStdDev","worstOverhead","worstStdDevForOverhead","worstRuntimeX","worstRuntimeY","","bestName","bestTime","bestNCores","bestSpeedup","bestMaxstddev","bestFactor","bestFactorStdDev","bestOverhead","bestStdDevForOverhead","bestRuntimeX","bestRuntimeY","meanFactor","maxStdDevFactor","meanOverhead","maxStdDevForOverhead","runtimeMaxCores"
"Jacobi (Distributed)","Eden","3217","Eden vs. PArrows 3217","9","bench/./jacobi-eden 3 3217-bench/./jacobi-parr 3 3217",-1.0052940241164379,256,-4.509725619081095e-2,1.259171740268168,1.0,3.95599612113595e-3,3.15e-3,3.943540931543817e-3,318.2944830356915,319.2997770598079,"5","bench/./jacobi-eden 3 3217-bench/./jacobi-parr 3 3217",16.688426074054576,16,-0.35475384677816635,4.039609740413705,0.963,8.945898630310279e-3,-3.84e-2,9.289202231452924e-3,451.5599726031766,434.87154652912204,0.993,0.0157,"\textbf{-0.74\%}",1.6\%,635.0
\end{filecontents*}
\csvreader[head to column names]{bestAndWorstJacobi-3217.csv}{}{& \vs & \meanOverhead & \maxStdDevForOverhead & \runtimeMaxCores \\}
\end{tabular}
\endgroup
\end{table}

PArrows performed in our benchmarks with little to no overhead. Tables~\ref{tab:meanOverheadSharedMemory} and \ref{tab:meanOverHeadDistributedMemory} clarify this once more: The PArrows-enabled versions trade blows with their vanilla counterparts when comparing the means over all cores of the mean overheads. If we combine these findings with the usability of our DSL,
the minor overhead induced by PArrows is outweighed by their convenience and usefulness to the user.

PArrows is an extendable formalism, they can be easily ported to further parallel Haskells while still maintaining interchangeability. It is straightforward to provide further implementations of algorithmic skeletons in PArrows.
	
\section{Conclusion}
\label{sec:conclusion}
Arrows are a generic concept that allows for powerful composition
combinators. To our knowledge we are first to represent
\emph{parallel} computation with
Arrows, and hence to show their usefulness for
composing parallel
programs. We have shown that for a generic and extensible parallel Haskell, we do not have to restrict ourselves to a monadic interface. 
We argue that Arrows are better suited to  parallelise
pure functions than Monads,  as the functions are already Arrows and can be used
directly in our DSL.
Arrows are a better fit to parallelise pure code than a monadic solution as regular functions are already Arrows and can be used with our DSL in a more natural way. 
We use a non-monadic interface (similar to Eden or GpH) and retain composability.
The DSL allows for a direct parallelisation of monadic code via the Kleisli type and additionally allows to parallelise any Arrow type that has an instance for \ensuremath{\Conid{ArrowChoice}}. (Some skeletons require an additional \ensuremath{\Conid{ArrowLoop}} instance.)

We have demonstrated the generality of the approach by exhibiting PArrow implementations for Eden, GpH, and the \ensuremath{\Conid{Par}} Monad. Hence, parallel programs can be ported between task parallel Haskell implementations with little or no effort. We are confident that it will be straightforward to add other task-parallel Haskells.
%
In other words, PArrows greatly increase portability of parallel Haskell programs.
Our measurements of four benchmarks on both shared and distributed memory platforms shows that the generality and portability of PArrows has very low performance overheads, \ie never more than $8\% \; \pm 6.9\%$ and typically under $2\%$.


\subsection{Future work}
\label{sec:future-work}

Our PArrows DSL can be expanded 
to other task parallel Haskells, and a specific target is HdpH \cite{Maier:2014:HDS:2775050.2633363}.
Further Future-aware versions of Arrow combinators can be defined. Existing combinators could also be improved, for example a more special versions of \ensuremath{\mathbin{>\!\!>\!\!>}} and \ensuremath{\mathbin{*\!*\!*}} combinators are viable.

In ongoing work we are expanding 
both our skeleton library and the number of skeleton-based parallel programs that use our DSL.
It would also be interesting to see a hybrid of PArrows and Accelerate \cite{McDonell:2015:TRC:2887747.2804313}.
Ports of our approach to other languages such as Frege, Eta, or Java directly are at an early development stage.
        \providecommand{\noopsort}[1]{}

        \appendix
	\section{Utility Arrows}\label{utilfns}
Following are definitions of some utility Arrows used in this paper that have been left out for brevity.
We start with the \ensuremath{\Varid{second}} combinator from \citet{HughesArrows}, which is a mirrored version of \ensuremath{\Varid{first}}, which is for example used in the definition of \ensuremath{\mathbin{*\!*\!*}}: 
\begin{hscode}\SaveRestoreHook
\column{B}{@{}>{\hspre}l<{\hspost}@{}}%
\column{9}{@{}>{\hspre}l<{\hspost}@{}}%
\column{E}{@{}>{\hspre}l<{\hspost}@{}}%
\>[B]{}\Varid{second}\mathbin{::}\Conid{Arrow}\;\Varid{arr}\Rightarrow \Varid{arr}\;\Varid{a}\;\Varid{b}\to \Varid{arr}\;(\Varid{c},\Varid{a})\;(\Varid{c},\Varid{b}){}\<[E]%
\\
\>[B]{}\Varid{second}\;\Varid{f}\mathrel{=}\Varid{arr}\;\Varid{swap}\mathbin{>\!\!>\!\!>}\Varid{first}\;\Varid{f}\mathbin{>\!\!>\!\!>}\Varid{arr}\;\Varid{swap}{}\<[E]%
\\
\>[B]{}\hsindent{9}{}\<[9]%
\>[9]{}\mathbf{where}\;\Varid{swap}\;(\Varid{x},\Varid{y})\mathrel{=}(\Varid{y},\Varid{x}){}\<[E]%
\ColumnHook
\end{hscode}\resethooks

Next, we give the definition of \ensuremath{\Varid{evalN}} which also helps us to define \ensuremath{\Varid{map}}, and \ensuremath{\Varid{zipWith}} on Arrows.
The \ensuremath{\Varid{evalN}} combinator in Fig.~\ref{fig:evalN} converts a list of Arrows \ensuremath{[\mskip1.5mu \Varid{arr}\;\Varid{a}\;\Varid{b}\mskip1.5mu]} into an Arrow \ensuremath{\Varid{arr}\;[\mskip1.5mu \Varid{a}\mskip1.5mu]\;[\mskip1.5mu \Varid{b}\mskip1.5mu]}.

\begin{figure}[h]
\begin{hscode}\SaveRestoreHook
\column{B}{@{}>{\hspre}l<{\hspost}@{}}%
\column{10}{@{}>{\hspre}l<{\hspost}@{}}%
\column{16}{@{}>{\hspre}l<{\hspost}@{}}%
\column{32}{@{}>{\hspre}l<{\hspost}@{}}%
\column{E}{@{}>{\hspre}l<{\hspost}@{}}%
\>[B]{}\Varid{evalN}\mathbin{::}(\Conid{ArrowChoice}\;\Varid{arr})\Rightarrow [\mskip1.5mu \Varid{arr}\;\Varid{a}\;\Varid{b}\mskip1.5mu]\to \Varid{arr}\;[\mskip1.5mu \Varid{a}\mskip1.5mu]\;[\mskip1.5mu \Varid{b}\mskip1.5mu]{}\<[E]%
\\
\>[B]{}\Varid{evalN}\;(\Varid{f}\mathbin{:}\Varid{fs})\mathrel{=}\Varid{arr}\;\Varid{listcase}\mathbin{>\!\!>\!\!>}{}\<[E]%
\\
\>[B]{}\hsindent{10}{}\<[10]%
\>[10]{}\Varid{arr}\;(\Varid{const}\;[\mskip1.5mu \mskip1.5mu])\mathbin{\mid\!\mid\!\mid}(\Varid{f}\mathbin{*\!*\!*}\Varid{evalN}\;\Varid{fs}\mathbin{>\!\!>\!\!>}\Varid{arr}\;(\Varid{uncurry}\;(\mathbin{:}))){}\<[E]%
\\
\>[B]{}\hsindent{10}{}\<[10]%
\>[10]{}\mathbf{where}\;\Varid{listcase}\;[\mskip1.5mu \mskip1.5mu]{}\<[32]%
\>[32]{}\mathrel{=}\Conid{Left}\;(){}\<[E]%
\\
\>[10]{}\hsindent{6}{}\<[16]%
\>[16]{}\Varid{listcase}\;(\Varid{x}\mathbin{:}\Varid{xs})\mathrel{=}\Conid{Right}\;(\Varid{x},\Varid{xs}){}\<[E]%
\\
\>[B]{}\Varid{evalN}\;[\mskip1.5mu \mskip1.5mu]\mathrel{=}\Varid{arr}\;(\Varid{const}\;[\mskip1.5mu \mskip1.5mu]){}\<[E]%
\ColumnHook
\end{hscode}\resethooks
\caption{The definition of \ensuremath{\Varid{evalN}}}
\label{fig:evalN}
\end{figure}

The \ensuremath{\Varid{mapArr}} combinator (Fig.~\ref{fig:mapArr}) lifts any Arrow \ensuremath{\Varid{arr}\;\Varid{a}\;\Varid{b}} to an Arrow \ensuremath{\Varid{arr}\;[\mskip1.5mu \Varid{a}\mskip1.5mu]\;[\mskip1.5mu \Varid{b}\mskip1.5mu]}. The original inspiration was from \citet{Hughes2005},
but the definition as then unified with \ensuremath{\Varid{evalN}}. 

\begin{figure}[h]
\begin{hscode}\SaveRestoreHook
\column{B}{@{}>{\hspre}l<{\hspost}@{}}%
\column{E}{@{}>{\hspre}l<{\hspost}@{}}%
\>[B]{}\Varid{mapArr}\mathbin{::}\Conid{ArrowChoice}\;\Varid{arr}\Rightarrow \Varid{arr}\;\Varid{a}\;\Varid{b}\to \Varid{arr}\;[\mskip1.5mu \Varid{a}\mskip1.5mu]\;[\mskip1.5mu \Varid{b}\mskip1.5mu]{}\<[E]%
\\
\>[B]{}\Varid{mapArr}\mathrel{=}\Varid{evalN}\mathbin{\circ}\Varid{repeat}{}\<[E]%
\ColumnHook
\end{hscode}\resethooks
\caption{The definition of \ensuremath{\Varid{map}} over Arrows.}
\label{fig:mapArr}
\end{figure}

Finally, with the help of \ensuremath{\Varid{mapArr}} (Fig.~\ref{fig:mapArr}), we can define \ensuremath{\Varid{zipWithArr}}  (Fig.~\ref{fig:zipWithArr}) that lifts any Arrow \ensuremath{\Varid{arr}\;(\Varid{a},\Varid{b})\;\Varid{c}} to an Arrow \ensuremath{\Varid{arr}\;([\mskip1.5mu \Varid{a}\mskip1.5mu],[\mskip1.5mu \Varid{b}\mskip1.5mu])\;[\mskip1.5mu \Varid{c}\mskip1.5mu]}.
\begin{figure}[h]
\begin{hscode}\SaveRestoreHook
\column{B}{@{}>{\hspre}l<{\hspost}@{}}%
\column{E}{@{}>{\hspre}l<{\hspost}@{}}%
\>[B]{}\Varid{zipWithArr}\mathbin{::}\Conid{ArrowChoice}\;\Varid{arr}\Rightarrow \Varid{arr}\;(\Varid{a},\Varid{b})\;\Varid{c}\to \Varid{arr}\;([\mskip1.5mu \Varid{a}\mskip1.5mu],[\mskip1.5mu \Varid{b}\mskip1.5mu])\;[\mskip1.5mu \Varid{c}\mskip1.5mu]{}\<[E]%
\\
\>[B]{}\Varid{zipWithArr}\;\Varid{f}\mathrel{=}(\Varid{arr}\;(\lambda (\Varid{as},\Varid{bs})\to \Varid{zipWith}\;(,)\;\Varid{as}\;\Varid{bs}))\mathbin{>\!\!>\!\!>}\Varid{mapArr}\;\Varid{f}{}\<[E]%
\ColumnHook
\end{hscode}\resethooks
\caption{\ensuremath{\Varid{zipWith}} over Arrows.} 
\label{fig:zipWithArr}
\end{figure}

These combinators make use of the \ensuremath{\Conid{ArrowChoice}} type class which provides the \ensuremath{\mathbin{\mid\!\mid\!\mid}} combinator. It takes two Arrows \ensuremath{\Varid{arr}\;\Varid{a}\;\Varid{c}} and \ensuremath{\Varid{arr}\;\Varid{b}\;\Varid{c}} and combines them into a new Arrow \ensuremath{\Varid{arr}\;(\Conid{Either}\;\Varid{a}\;\Varid{b})\;\Varid{c}} which pipes all \ensuremath{\Conid{Left}\;\Varid{a}}'s to the first Arrow and all \ensuremath{\Conid{Right}\;\Varid{b}}'s to the second Arrow:
\begin{hscode}\SaveRestoreHook
\column{B}{@{}>{\hspre}l<{\hspost}@{}}%
\column{E}{@{}>{\hspre}l<{\hspost}@{}}%
\>[B]{}(\mathbin{\mid\!\mid\!\mid})\mathbin{::}\Conid{ArrowChoice}\;\Varid{arr}\;\Varid{a}\;\Varid{c}\to \Varid{arr}\;\Varid{b}\;\Varid{c}\to \Varid{arr}\;(\Conid{Either}\;\Varid{a}\;\Varid{b})\;\Varid{c}{}\<[E]%
\ColumnHook
\end{hscode}\resethooks

\section{Profunctor Arrows}
\label{app:profunctorArrows}

In Fig.~\ref{fig:profunctorArrow} we show how specific Profunctors can be turned into Arrows. This works because Arrows are strong Monads in the bicategory \ensuremath{\Conid{Prof}} of Profunctors as shown by \citet{Asada:2010:ASM:1863597.1863607}. In Standard GHC \ensuremath{(\mathbin{>\!\!>\!\!>})} has the type \ensuremath{(\mathbin{>\!\!>\!\!>})\mathbin{::}\Conid{Category}\;\Varid{cat}\Rightarrow \Varid{cat}\;\Varid{a}\;\Varid{b}\to \Varid{cat}\;\Varid{b}\;\Varid{c}\to \Varid{cat}\;\Varid{a}\;\Varid{c}} and is therefore not part of the \ensuremath{\Conid{Arrow}} type class like presented in this paper.\footnote{For additional information on the type classes used, see: \url{https://hackage.haskell.org/package/profunctors-5.2.1/docs/Data-Profunctor.html} and \url{https://hackage.haskell.org/package/base-4.9.1.0/docs/Control-Category.html}.}

\begin{figure}[h]
\begin{hscode}\SaveRestoreHook
\column{B}{@{}>{\hspre}l<{\hspost}@{}}%
\column{3}{@{}>{\hspre}l<{\hspost}@{}}%
\column{E}{@{}>{\hspre}l<{\hspost}@{}}%
\>[B]{}\mathbf{instance}\;(\Conid{Category}\;\Varid{p},\Conid{Strong}\;\Varid{p})\Rightarrow \Conid{Arrow}\;\Varid{p}\;\mathbf{where}{}\<[E]%
\\
\>[B]{}\hsindent{3}{}\<[3]%
\>[3]{}\Varid{arr}\;\Varid{f}\mathrel{=}\Varid{dimap}\;\Varid{id}\;\Varid{f}\;\Varid{id}{}\<[E]%
\\
\>[B]{}\hsindent{3}{}\<[3]%
\>[3]{}\Varid{first}\mathrel{=}\Varid{first'}{}\<[E]%
\\[\blanklineskip]%
\>[B]{}\mathbf{instance}\;(\Conid{Category}\;\Varid{p},\Conid{Strong}\;\Varid{p},\Conid{Costrong}\;\Varid{p})\Rightarrow \Conid{ArrowLoop}\;\Varid{p}\;\mathbf{where}{}\<[E]%
\\
\>[B]{}\hsindent{3}{}\<[3]%
\>[3]{}\Varid{loop}\mathrel{=}\Varid{loop'}{}\<[E]%
\\[\blanklineskip]%
\>[B]{}\mathbf{instance}\;(\Conid{Category}\;\Varid{p},\Conid{Strong}\;\Varid{p},\Conid{Choice}\;\Varid{p})\Rightarrow \Conid{ArrowChoice}\;\Varid{p}\;\mathbf{where}{}\<[E]%
\\
\>[B]{}\hsindent{3}{}\<[3]%
\>[3]{}\Varid{left}\mathrel{=}\Varid{left'}{}\<[E]%
\ColumnHook
\end{hscode}\resethooks
\caption{Profunctors as Arrows.}
\label{fig:profunctorArrow}
\end{figure}

\section{Additional function definitions}
\label{app:omitted}
We have omitted some function definitions in the main text for
brevity, and redeem this here.
We begin with warping Eden's build-in \ensuremath{\Conid{RemoteData}} to \ensuremath{\Conid{Future}} in
Figure~\ref{fig:RDFuture}

\begin{figure}[h]
\begin{hscode}\SaveRestoreHook
\column{B}{@{}>{\hspre}l<{\hspost}@{}}%
\column{5}{@{}>{\hspre}l<{\hspost}@{}}%
\column{E}{@{}>{\hspre}l<{\hspost}@{}}%
\>[B]{}\mathbf{data}\;\Conid{RemoteData}\;\Varid{a}\mathrel{=}\Conid{RD}\;\{\mskip1.5mu \Varid{rd}\mathbin{::}\Conid{RD}\;\Varid{a}\mskip1.5mu\}{}\<[E]%
\\[\blanklineskip]%
\>[B]{}\Varid{put'}\mathbin{::}(\Conid{Arrow}\;\Varid{arr})\Rightarrow \Varid{arr}\;\Varid{a}\;(\Conid{BasicFuture}\;\Varid{a}){}\<[E]%
\\
\>[B]{}\Varid{put'}\mathrel{=}\Varid{arr}\;\Conid{BF}{}\<[E]%
\\[\blanklineskip]%
\>[B]{}\Varid{get'}\mathbin{::}(\Conid{Arrow}\;\Varid{arr})\Rightarrow \Varid{arr}\;(\Conid{BasicFuture}\;\Varid{a})\;\Varid{a}{}\<[E]%
\\
\>[B]{}\Varid{get'}\mathrel{=}\Varid{arr}\;(\lambda (\mathord{\sim}(\Conid{BF}\;\Varid{a}))\to \Varid{a}){}\<[E]%
\\[\blanklineskip]%
\>[B]{}\mathbf{instance}\;\Conid{NFData}\;(\Conid{RemoteData}\;\Varid{a})\;\mathbf{where}{}\<[E]%
\\
\>[B]{}\hsindent{5}{}\<[5]%
\>[5]{}\Varid{rnf}\mathrel{=}\Varid{rnf}\mathbin{\circ}\Varid{rd}{}\<[E]%
\\
\>[B]{}\mathbf{instance}\;\Conid{Trans}\;(\Conid{RemoteData}\;\Varid{a}){}\<[E]%
\\[\blanklineskip]%
\>[B]{}\mathbf{instance}\;(\Conid{Trans}\;\Varid{a})\Rightarrow \Conid{Future}\;\Conid{RemoteData}\;\Varid{a}\;\Conid{Conf}\;\mathbf{where}{}\<[E]%
\\
\>[B]{}\hsindent{5}{}\<[5]%
\>[5]{}\Varid{put}\;\anonymous \mathrel{=}\Varid{put'}{}\<[E]%
\\
\>[B]{}\hsindent{5}{}\<[5]%
\>[5]{}\Varid{get}\;\anonymous \mathrel{=}\Varid{get'}{}\<[E]%
\\[\blanklineskip]%
\>[B]{}\mathbf{instance}\;(\Conid{Trans}\;\Varid{a})\Rightarrow \Conid{Future}\;\Conid{RemoteData}\;\Varid{a}\;()\;\mathbf{where}{}\<[E]%
\\
\>[B]{}\hsindent{5}{}\<[5]%
\>[5]{}\Varid{put}\;\anonymous \mathrel{=}\Varid{put'}{}\<[E]%
\\
\>[B]{}\hsindent{5}{}\<[5]%
\>[5]{}\Varid{get}\;\anonymous \mathrel{=}\Varid{get'}{}\<[E]%
\ColumnHook
\end{hscode}\resethooks
\caption{\ensuremath{\Conid{RD}}-based \ensuremath{\Conid{RemoteData}} version of \ensuremath{\Conid{Future}} for the Eden backend.}
\label{fig:RDFuture}
\end{figure}

Next, we have the definition of \ensuremath{\Conid{BasicFuture}} in Fig.~\ref{fig:BasicFuture} and the corresponding \ensuremath{\Conid{Future}} instances.

\begin{figure}[h]
\begin{hscode}\SaveRestoreHook
\column{B}{@{}>{\hspre}l<{\hspost}@{}}%
\column{5}{@{}>{\hspre}l<{\hspost}@{}}%
\column{E}{@{}>{\hspre}l<{\hspost}@{}}%
\>[B]{}\mathbf{data}\;\Conid{BasicFuture}\;\Varid{a}\mathrel{=}\Conid{BF}\;\Varid{a}{}\<[E]%
\\[\blanklineskip]%
\>[B]{}\Varid{put'}\mathbin{::}(\Conid{Arrow}\;\Varid{arr})\Rightarrow \Varid{arr}\;\Varid{a}\;(\Conid{BasicFuture}\;\Varid{a}){}\<[E]%
\\
\>[B]{}\Varid{put'}\mathrel{=}\Varid{arr}\;\Conid{BF}{}\<[E]%
\\[\blanklineskip]%
\>[B]{}\Varid{get'}\mathbin{::}(\Conid{Arrow}\;\Varid{arr})\Rightarrow \Varid{arr}\;(\Conid{BasicFuture}\;\Varid{a})\;\Varid{a}{}\<[E]%
\\
\>[B]{}\Varid{get'}\mathrel{=}\Varid{arr}\;(\lambda (\mathord{\sim}(\Conid{BF}\;\Varid{a}))\to \Varid{a}){}\<[E]%
\\[\blanklineskip]%
\>[B]{}\mathbf{instance}\;\Conid{NFData}\;\Varid{a}\Rightarrow \Conid{NFData}\;(\Conid{BasicFuture}\;\Varid{a})\;\mathbf{where}{}\<[E]%
\\
\>[B]{}\hsindent{5}{}\<[5]%
\>[5]{}\Varid{rnf}\;(\Conid{BF}\;\Varid{a})\mathrel{=}\Varid{rnf}\;\Varid{a}{}\<[E]%
\\[\blanklineskip]%
\>[B]{}\mathbf{instance}\;\Conid{Future}\;\Conid{BasicFuture}\;\Varid{a}\;(\Conid{Conf}\;\Varid{a})\;\mathbf{where}{}\<[E]%
\\
\>[B]{}\hsindent{5}{}\<[5]%
\>[5]{}\Varid{put}\;\anonymous \mathrel{=}\Varid{put'}{}\<[E]%
\\
\>[B]{}\hsindent{5}{}\<[5]%
\>[5]{}\Varid{get}\;\anonymous \mathrel{=}\Varid{get'}{}\<[E]%
\\[\blanklineskip]%
\>[B]{}\mathbf{instance}\;\Conid{Future}\;\Conid{BasicFuture}\;\Varid{a}\;()\;\mathbf{where}{}\<[E]%
\\
\>[B]{}\hsindent{5}{}\<[5]%
\>[5]{}\Varid{put}\;\anonymous \mathrel{=}\Varid{put'}{}\<[E]%
\\
\>[B]{}\hsindent{5}{}\<[5]%
\>[5]{}\Varid{get}\;\anonymous \mathrel{=}\Varid{get'}{}\<[E]%
\ColumnHook
\end{hscode}\resethooks
\caption{\ensuremath{\Conid{BasicFuture}} type and its \ensuremath{\Conid{Future}} instance for the \ensuremath{\Conid{Par}} Monad and GpH.}
\label{fig:BasicFuture}
\end{figure}

\begin{figure}[thb]
\includegraphics[scale=0.7]{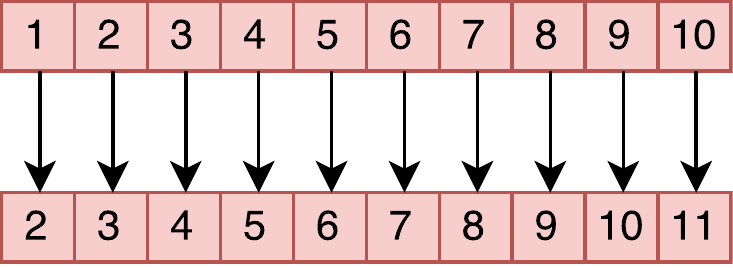}
\caption{\ensuremath{\Varid{parMap}} depiction.}
\label{fig:parMapImg}

\begin{hscode}\SaveRestoreHook
\column{B}{@{}>{\hspre}l<{\hspost}@{}}%
\column{E}{@{}>{\hspre}l<{\hspost}@{}}%
\>[B]{}\Varid{parMap}\mathbin{::}(\Conid{ArrowParallel}\;\Varid{arr}\;\Varid{a}\;\Varid{b}\;\Varid{conf})\Rightarrow \Varid{conf}\to (\Varid{arr}\;\Varid{a}\;\Varid{b})\to (\Varid{arr}\;[\mskip1.5mu \Varid{a}\mskip1.5mu]\;[\mskip1.5mu \Varid{b}\mskip1.5mu]){}\<[E]%
\\
\>[B]{}\Varid{parMap}\;\Varid{conf}\;\Varid{f}\mathrel{=}\Varid{parEvalN}\;\Varid{conf}\;(\Varid{repeat}\;\Varid{f}){}\<[E]%
\ColumnHook
\end{hscode}\resethooks
\caption{Definition of parMap.}
\label{fig:parMap}

\includegraphics[scale=0.7]{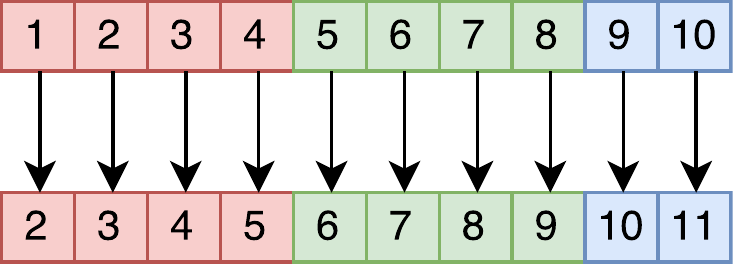}
\caption{\ensuremath{\Varid{parMapStream}} depiction.}
\label{fig:parMapStreamImg}

\begin{hscode}\SaveRestoreHook
\column{B}{@{}>{\hspre}l<{\hspost}@{}}%
\column{9}{@{}>{\hspre}l<{\hspost}@{}}%
\column{E}{@{}>{\hspre}l<{\hspost}@{}}%
\>[B]{}\Varid{parMapStream}\mathbin{::}(\Conid{ArrowParallel}\;\Varid{arr}\;\Varid{a}\;\Varid{b}\;\Varid{conf},\Conid{ArrowChoice}\;\Varid{arr},\Conid{ArrowApply}\;\Varid{arr})\Rightarrow {}\<[E]%
\\
\>[B]{}\hsindent{9}{}\<[9]%
\>[9]{}\Varid{conf}\to \Conid{ChunkSize}\to \Varid{arr}\;\Varid{a}\;\Varid{b}\to \Varid{arr}\;[\mskip1.5mu \Varid{a}\mskip1.5mu]\;[\mskip1.5mu \Varid{b}\mskip1.5mu]{}\<[E]%
\\
\>[B]{}\Varid{parMapStream}\;\Varid{conf}\;\Varid{chunkSize}\;\Varid{f}\mathrel{=}\Varid{parEvalNLazy}\;\Varid{conf}\;\Varid{chunkSize}\;(\Varid{repeat}\;\Varid{f}){}\<[E]%
\ColumnHook
\end{hscode}\resethooks
\caption{\ensuremath{\Varid{parMapStream}} definition.}
\label{fig:parMapStream}
\end{figure}

Figures~\ref{fig:parMapImg}--\ref{fig:parMapStream} show the definitions and a visualisations of two parallel \ensuremath{\Varid{map}} variants, defined using \ensuremath{\Varid{parEvalN}} and its lazy counterpart.

Arrow versions of Eden's \ensuremath{\Varid{shuffle}}, \ensuremath{\Varid{unshuffle}} and the definition of \ensuremath{\Varid{takeEach}} are in Figure~\ref{fig:edenshuffleetc}. Similarly, Figure~\ref{fig:edenlazyrightrotate} contains the definition of Arrow versions of Eden's \ensuremath{\Varid{lazy}} and \ensuremath{\Varid{rightRotate}} utility functions. Fig.~\ref{fig:lazyzip3etc} contains Eden's definition of \ensuremath{\Varid{lazyzip3}} together with the utility functions \ensuremath{\Varid{uncurry3}} and \ensuremath{\Varid{threetotwo}}.
The full definition of \ensuremath{\Varid{farmChunk}} is in Figure~\ref{fig:farmChunk}.
Eden definition of \ensuremath{\Varid{ring}} skeleton is in Figure~\ref{fig:ringEden}. It
follows \citet{Loogen2012}.

\begin{figure}[h]
\begin{hscode}\SaveRestoreHook
\column{B}{@{}>{\hspre}l<{\hspost}@{}}%
\column{E}{@{}>{\hspre}l<{\hspost}@{}}%
\>[B]{}\Varid{shuffle}\mathbin{::}(\Conid{Arrow}\;\Varid{arr})\Rightarrow \Varid{arr}\;[\mskip1.5mu [\mskip1.5mu \Varid{a}\mskip1.5mu]\mskip1.5mu]\;[\mskip1.5mu \Varid{a}\mskip1.5mu]{}\<[E]%
\\
\>[B]{}\Varid{shuffle}\mathrel{=}\Varid{arr}\;(\Varid{concat}\mathbin{\circ}\Varid{transpose}){}\<[E]%
\\[\blanklineskip]%
\>[B]{}\Varid{unshuffle}\mathbin{::}(\Conid{Arrow}\;\Varid{arr})\Rightarrow \Conid{Int}\to \Varid{arr}\;[\mskip1.5mu \Varid{a}\mskip1.5mu]\;[\mskip1.5mu [\mskip1.5mu \Varid{a}\mskip1.5mu]\mskip1.5mu]{}\<[E]%
\\
\>[B]{}\Varid{unshuffle}\;\Varid{n}\mathrel{=}\Varid{arr}\;(\lambda \Varid{xs}\to [\mskip1.5mu \Varid{takeEach}\;\Varid{n}\;(\Varid{drop}\;\Varid{i}\;\Varid{xs})\mid \Varid{i}\leftarrow [\mskip1.5mu \mathrm{0}\mathinner{\ldotp\ldotp}\Varid{n}\mathbin{-}\mathrm{1}\mskip1.5mu]\mskip1.5mu]){}\<[E]%
\\[\blanklineskip]%
\>[B]{}\Varid{takeEach}\mathbin{::}\Conid{Int}\to [\mskip1.5mu \Varid{a}\mskip1.5mu]\to [\mskip1.5mu \Varid{a}\mskip1.5mu]{}\<[E]%
\\
\>[B]{}\Varid{takeEach}\;\Varid{n}\;[\mskip1.5mu \mskip1.5mu]\mathrel{=}[\mskip1.5mu \mskip1.5mu]{}\<[E]%
\\
\>[B]{}\Varid{takeEach}\;\Varid{n}\;(\Varid{x}\mathbin{:}\Varid{xs})\mathrel{=}\Varid{x}\mathbin{:}\Varid{takeEach}\;\Varid{n}\;(\Varid{drop}\;(\Varid{n}\mathbin{-}\mathrm{1})\;\Varid{xs}){}\<[E]%
\ColumnHook
\end{hscode}\resethooks
\caption{\ensuremath{\Varid{shuffle}}, \ensuremath{\Varid{unshuffle}}, \ensuremath{\Varid{takeEach}} definition.}
\label{fig:edenshuffleetc}
\end{figure}

\begin{figure}[h]
\begin{hscode}\SaveRestoreHook
\column{B}{@{}>{\hspre}l<{\hspost}@{}}%
\column{3}{@{}>{\hspre}l<{\hspost}@{}}%
\column{E}{@{}>{\hspre}l<{\hspost}@{}}%
\>[B]{}\Varid{lazy}\mathbin{::}(\Conid{Arrow}\;\Varid{arr})\Rightarrow \Varid{arr}\;[\mskip1.5mu \Varid{a}\mskip1.5mu]\;[\mskip1.5mu \Varid{a}\mskip1.5mu]{}\<[E]%
\\
\>[B]{}\Varid{lazy}\mathrel{=}\Varid{arr}\;(\lambda \mathord{\sim}(\Varid{x}\mathbin{:}\Varid{xs})\to \Varid{x}\mathbin{:}\Varid{lazy}\;\Varid{xs}){}\<[E]%
\\[\blanklineskip]%
\>[B]{}\Varid{rightRotate}\mathbin{::}(\Conid{Arrow}\;\Varid{arr})\Rightarrow \Varid{arr}\;[\mskip1.5mu \Varid{a}\mskip1.5mu]\;[\mskip1.5mu \Varid{a}\mskip1.5mu]{}\<[E]%
\\
\>[B]{}\Varid{rightRotate}\mathrel{=}\Varid{arr}\mathbin{\$}\lambda \Varid{list}\to \mathbf{case}\;\Varid{list}\;\mathbf{of}{}\<[E]%
\\
\>[B]{}\hsindent{3}{}\<[3]%
\>[3]{}[\mskip1.5mu \mskip1.5mu]\to [\mskip1.5mu \mskip1.5mu]{}\<[E]%
\\
\>[B]{}\hsindent{3}{}\<[3]%
\>[3]{}\Varid{xs}\to \Varid{last}\;\Varid{xs}\mathbin{:}\Varid{init}\;\Varid{xs}{}\<[E]%
\ColumnHook
\end{hscode}\resethooks
\caption{\ensuremath{\Varid{lazy}} and \ensuremath{\Varid{rightRotate}} definitions.}
\label{fig:edenlazyrightrotate}
\end{figure}

\begin{figure}[h]
\begin{hscode}\SaveRestoreHook
\column{B}{@{}>{\hspre}l<{\hspost}@{}}%
\column{E}{@{}>{\hspre}l<{\hspost}@{}}%
\>[B]{}\Varid{lazyzip3}\mathbin{::}[\mskip1.5mu \Varid{a}\mskip1.5mu]\to [\mskip1.5mu \Varid{b}\mskip1.5mu]\to [\mskip1.5mu \Varid{c}\mskip1.5mu]\to [\mskip1.5mu (\Varid{a},\Varid{b},\Varid{c})\mskip1.5mu]{}\<[E]%
\\
\>[B]{}\Varid{lazyzip3}\;\Varid{as}\;\Varid{bs}\;\Varid{cs}\mathrel{=}\Varid{zip3}\;\Varid{as}\;(\Varid{lazy}\;\Varid{bs})\;(\Varid{lazy}\;\Varid{cs}){}\<[E]%
\\[\blanklineskip]%
\>[B]{}\Varid{uncurry3}\mathbin{::}(\Varid{a}\to \Varid{b}\to \Varid{c}\to \Varid{d})\to (\Varid{a},(\Varid{b},\Varid{c}))\to \Varid{d}{}\<[E]%
\\
\>[B]{}\Varid{uncurry3}\;\Varid{f}\;(\Varid{a},(\Varid{b},\Varid{c}))\mathrel{=}\Varid{f}\;\Varid{a}\;\Varid{b}\;\Varid{c}{}\<[E]%
\\[\blanklineskip]%
\>[B]{}\Varid{threetotwo}\mathbin{::}(\Conid{Arrow}\;\Varid{arr})\Rightarrow \Varid{arr}\;(\Varid{a},\Varid{b},\Varid{c})\;(\Varid{a},(\Varid{b},\Varid{c})){}\<[E]%
\\
\>[B]{}\Varid{threetotwo}\mathrel{=}\Varid{arr}\mathbin{\$}\lambda \mathord{\sim}(\Varid{a},\Varid{b},\Varid{c})\to (\Varid{a},(\Varid{b},\Varid{c})){}\<[E]%
\ColumnHook
\end{hscode}\resethooks
\caption{\ensuremath{\Varid{lazyzip3}}, \ensuremath{\Varid{uncurry3}} and \ensuremath{\Varid{threetotwo}} definitions.}
\label{fig:lazyzip3etc}
\end{figure}

\begin{figure}[h]
\begin{hscode}\SaveRestoreHook
\column{B}{@{}>{\hspre}l<{\hspost}@{}}%
\column{9}{@{}>{\hspre}l<{\hspost}@{}}%
\column{14}{@{}>{\hspre}l<{\hspost}@{}}%
\column{E}{@{}>{\hspre}l<{\hspost}@{}}%
\>[B]{}\Varid{farmChunk}\mathbin{::}(\Conid{ArrowParallel}\;\Varid{arr}\;\Varid{a}\;\Varid{b}\;\Varid{conf},\Conid{ArrowParallel}\;\Varid{arr}\;[\mskip1.5mu \Varid{a}\mskip1.5mu]\;[\mskip1.5mu \Varid{b}\mskip1.5mu]\;\Varid{conf},{}\<[E]%
\\
\>[B]{}\hsindent{14}{}\<[14]%
\>[14]{}\Conid{ArrowChoice}\;\Varid{arr},\Conid{ArrowApply}\;\Varid{arr})\Rightarrow {}\<[E]%
\\
\>[B]{}\hsindent{9}{}\<[9]%
\>[9]{}\Varid{conf}\to \Conid{ChunkSize}\to \Conid{NumCores}\to \Varid{arr}\;\Varid{a}\;\Varid{b}\to \Varid{arr}\;[\mskip1.5mu \Varid{a}\mskip1.5mu]\;[\mskip1.5mu \Varid{b}\mskip1.5mu]{}\<[E]%
\\
\>[B]{}\Varid{farmChunk}\;\Varid{conf}\;\Varid{chunkSize}\;\Varid{numCores}\;\Varid{f}\mathrel{=}{}\<[E]%
\\
\>[B]{}\hsindent{9}{}\<[9]%
\>[9]{}\Varid{unshuffle}\;\Varid{numCores}\mathbin{>\!\!>\!\!>}{}\<[E]%
\\
\>[B]{}\hsindent{9}{}\<[9]%
\>[9]{}\Varid{parEvalNLazy}\;\Varid{conf}\;\Varid{chunkSize}\;(\Varid{repeat}\;(\Varid{mapArr}\;\Varid{f}))\mathbin{>\!\!>\!\!>}{}\<[E]%
\\
\>[B]{}\hsindent{9}{}\<[9]%
\>[9]{}\Varid{shuffle}{}\<[E]%
\ColumnHook
\end{hscode}\resethooks
\caption{\ensuremath{\Varid{farmChunk}} definition.}
\label{fig:farmChunk}
\end{figure}

\begin{figure}[h]
\begin{hscode}\SaveRestoreHook
\column{B}{@{}>{\hspre}l<{\hspost}@{}}%
\column{3}{@{}>{\hspre}l<{\hspost}@{}}%
\column{7}{@{}>{\hspre}l<{\hspost}@{}}%
\column{9}{@{}>{\hspre}l<{\hspost}@{}}%
\column{16}{@{}>{\hspre}l<{\hspost}@{}}%
\column{19}{@{}>{\hspre}l<{\hspost}@{}}%
\column{20}{@{}>{\hspre}l<{\hspost}@{}}%
\column{22}{@{}>{\hspre}l<{\hspost}@{}}%
\column{E}{@{}>{\hspre}l<{\hspost}@{}}%
\>[B]{}\Varid{ringSimple}\mathbin{::}(\Conid{Trans}\;\Varid{i},\Conid{Trans}\;\Varid{o},\Conid{Trans}\;\Varid{r})\Rightarrow (\Varid{i}\to \Varid{r}\to (\Varid{o},\Varid{r}))\to [\mskip1.5mu \Varid{i}\mskip1.5mu]\to [\mskip1.5mu \Varid{o}\mskip1.5mu]{}\<[E]%
\\
\>[B]{}\Varid{ringSimple}\;\Varid{f}\;\Varid{is}\mathrel{=}{}\<[20]%
\>[20]{}\Varid{os}{}\<[E]%
\\
\>[B]{}\hsindent{3}{}\<[3]%
\>[3]{}\mathbf{where}\;(\Varid{os},\Varid{ringOuts})\mathrel{=}\Varid{unzip}\;(\Varid{parMap}\;(\Varid{toRD}\mathbin{\$}\Varid{uncurry}\;\Varid{f})\;(\Varid{zip}\;\Varid{is}\mathbin{\$}\Varid{lazy}\;\Varid{ringIns})){}\<[E]%
\\
\>[3]{}\hsindent{6}{}\<[9]%
\>[9]{}\Varid{ringIns}\mathrel{=}\Varid{rightRotate}\;\Varid{ringOuts}{}\<[E]%
\\[\blanklineskip]%
\>[B]{}\Varid{toRD}\mathbin{::}(\Conid{Trans}\;\Varid{i},\Conid{Trans}\;\Varid{o},\Conid{Trans}\;\Varid{r})\Rightarrow ((\Varid{i},\Varid{r})\to (\Varid{o},\Varid{r}))\to ((\Varid{i},\Conid{RD}\;\Varid{r})\to (\Varid{o},\Conid{RD}\;\Varid{r})){}\<[E]%
\\
\>[B]{}\Varid{toRD}\;{}\<[7]%
\>[7]{}\Varid{f}\;(\Varid{i},\Varid{ringIn}){}\<[22]%
\>[22]{}\mathrel{=}(\Varid{o},\Varid{release}\;\Varid{ringOut}){}\<[E]%
\\
\>[B]{}\hsindent{3}{}\<[3]%
\>[3]{}\mathbf{where}\;(\Varid{o},\Varid{ringOut})\mathrel{=}\Varid{f}\;(\Varid{i},\Varid{fetch}\;\Varid{ringIn}){}\<[E]%
\\[\blanklineskip]%
\>[B]{}\Varid{rightRotate}{}\<[16]%
\>[16]{}\mathbin{::}[\mskip1.5mu \Varid{a}\mskip1.5mu]\to [\mskip1.5mu \Varid{a}\mskip1.5mu]{}\<[E]%
\\
\>[B]{}\Varid{rightRotate}\;[\mskip1.5mu \mskip1.5mu]\mathrel{=}{}\<[19]%
\>[19]{}[\mskip1.5mu \mskip1.5mu]{}\<[E]%
\\
\>[B]{}\Varid{rightRotate}\;\Varid{xs}\mathrel{=}{}\<[19]%
\>[19]{}\Varid{last}\;\Varid{xs}\mathbin{:}\Varid{init}\;\Varid{xs}{}\<[E]%
\\[\blanklineskip]%
\>[B]{}\Varid{lazy}\mathbin{::}[\mskip1.5mu \Varid{a}\mskip1.5mu]\to [\mskip1.5mu \Varid{a}\mskip1.5mu]{}\<[E]%
\\
\>[B]{}\Varid{lazy}\mathord{\sim}(\Varid{x}\mathbin{:}\Varid{xs})\mathrel{=}\Varid{x}\mathbin{:}\Varid{lazy}\;\Varid{xs}{}\<[E]%
\ColumnHook
\end{hscode}\resethooks
\caption{Eden's definition of the \ensuremath{\Varid{ring}} skeleton.}
\label{fig:ringEden}
\end{figure}

The \ensuremath{\Varid{parEval2}} skeleton is defined in Figure~\ref{fig:parEval2}. 
We start by transforming the \ensuremath{(\Varid{a},\Varid{c})} input into a two-element list \ensuremath{[\mskip1.5mu \Conid{Either}\;\Varid{a}\;\Varid{c}\mskip1.5mu]} by first tagging the two inputs with \ensuremath{\Conid{Left}} and \ensuremath{\Conid{Right}} and wrapping the right element in a singleton list with \ensuremath{\Varid{return}} so that we can combine them with \ensuremath{\Varid{arr}\;(\Varid{uncurry}\;(\mathbin{:}))}. Next, we feed this list into a parallel Arrow running on two instances of \ensuremath{\Varid{f}\mathbin{+\!\!+\!\!+}\Varid{g}} as described in the paper. After the calculation is finished, we convert the resulting \ensuremath{[\mskip1.5mu \Conid{Either}\;\Varid{b}\;\Varid{d}\mskip1.5mu]} into \ensuremath{([\mskip1.5mu \Varid{b}\mskip1.5mu],[\mskip1.5mu \Varid{d}\mskip1.5mu])} with \ensuremath{\Varid{arr}\;\Varid{partitionEithers}}. The two lists in this tuple contain only one element each by construction, so we can finally just convert the tuple to \ensuremath{(\Varid{b},\Varid{d})} in the last step.
\begin{figure}[h]
\begin{hscode}\SaveRestoreHook
\column{B}{@{}>{\hspre}l<{\hspost}@{}}%
\column{9}{@{}>{\hspre}l<{\hspost}@{}}%
\column{E}{@{}>{\hspre}l<{\hspost}@{}}%
\>[B]{}\Varid{parEval2}\mathbin{::}(\Conid{ArrowChoice}\;\Varid{arr},{}\<[E]%
\\
\>[B]{}\hsindent{9}{}\<[9]%
\>[9]{}\Conid{ArrowParallel}\;\Varid{arr}\;(\Conid{Either}\;\Varid{a}\;\Varid{c})\;(\Conid{Either}\;\Varid{b}\;\Varid{d})\;\Varid{conf})\Rightarrow {}\<[E]%
\\
\>[B]{}\hsindent{9}{}\<[9]%
\>[9]{}\Varid{conf}\to \Varid{arr}\;\Varid{a}\;\Varid{b}\to \Varid{arr}\;\Varid{c}\;\Varid{d}\to \Varid{arr}\;(\Varid{a},\Varid{c})\;(\Varid{b},\Varid{d}){}\<[E]%
\\
\>[B]{}\Varid{parEval2}\;\Varid{conf}\;\Varid{f}\;\Varid{g}\mathrel{=}{}\<[E]%
\\
\>[B]{}\hsindent{9}{}\<[9]%
\>[9]{}\Varid{arr}\;\Conid{Left}\mathbin{*\!*\!*}(\Varid{arr}\;\Conid{Right}\mathbin{>\!\!>\!\!>}\Varid{arr}\;\Varid{return})\mathbin{>\!\!>\!\!>}{}\<[E]%
\\
\>[B]{}\hsindent{9}{}\<[9]%
\>[9]{}\Varid{arr}\;(\Varid{uncurry}\;(\mathbin{:}))\mathbin{>\!\!>\!\!>}{}\<[E]%
\\
\>[B]{}\hsindent{9}{}\<[9]%
\>[9]{}\Varid{parEvalN}\;\Varid{conf}\;(\Varid{replicate}\;\mathrm{2}\;(\Varid{f}\mathbin{+\!\!+\!\!+}\Varid{g}))\mathbin{>\!\!>\!\!>}{}\<[E]%
\\
\>[B]{}\hsindent{9}{}\<[9]%
\>[9]{}\Varid{arr}\;\Varid{partitionEithers}\mathbin{>\!\!>\!\!>}{}\<[E]%
\\
\>[B]{}\hsindent{9}{}\<[9]%
\>[9]{}\Varid{arr}\;\Varid{head}\mathbin{*\!*\!*}\Varid{arr}\;\Varid{head}{}\<[E]%
\ColumnHook
\end{hscode}\resethooks
	\caption{\ensuremath{\Varid{parEval2}} definition.}
	\label{fig:parEval2}
\end{figure}
Furthermore, Fig.~\ref{fig:torus_example_rest} contains the omitted definitions of \ensuremath{\Varid{prMMTr}} (which calculates $AB^T$ for two matrices $A$ and $B$), \ensuremath{\Varid{splitMatrix}} (which splits the a matrix into chunks), and lastly \ensuremath{\Varid{matAdd}}, that calculates $A + B$ for two matrices $A$ and $B$.
\begin{figure}[h]
\begin{hscode}\SaveRestoreHook
\column{B}{@{}>{\hspre}l<{\hspost}@{}}%
\column{E}{@{}>{\hspre}l<{\hspost}@{}}%
\>[B]{}\Varid{prMMTr}\;\Varid{m1}\;\Varid{m2}\mathrel{=}[\mskip1.5mu [\mskip1.5mu \Varid{sum}\;(\Varid{zipWith}\;(\mathbin{*})\;\Varid{row}\;\Varid{col})\mid \Varid{col}\leftarrow \Varid{m2}\mskip1.5mu]\mid \Varid{row}\leftarrow \Varid{m1}\mskip1.5mu]{}\<[E]%
\\[\blanklineskip]%
\>[B]{}\Varid{splitMatrix}\mathbin{::}\Conid{Int}\to \Conid{Matrix}\to [\mskip1.5mu [\mskip1.5mu \Conid{Matrix}\mskip1.5mu]\mskip1.5mu]{}\<[E]%
\\
\>[B]{}\Varid{splitMatrix}\;\Varid{size}\;\Varid{matrix}\mathrel{=}\Varid{map}\;(\Varid{transpose}\mathbin{\circ}\Varid{map}\;(\Varid{chunksOf}\;\Varid{size}))\mathbin{\$}\Varid{chunksOf}\;\Varid{size}\mathbin{\$}\Varid{matrix}{}\<[E]%
\\[\blanklineskip]%
\>[B]{}\Varid{matAdd}\mathrel{=}\Varid{chunksOf}\;(\Varid{dimX}\;\Varid{x})\mathbin{\$}\Varid{zipWith}\;(\mathbin{+})\;(\Varid{concat}\;\Varid{x})\;(\Varid{concat}\;\Varid{y}){}\<[E]%
\ColumnHook
\end{hscode}\resethooks
	\caption{\ensuremath{\Varid{prMMTr}}, \ensuremath{\Varid{splitMatrix}} and \ensuremath{\Varid{matAdd}} definition.}
	\label{fig:torus_example_rest}
\end{figure} 

\section{Syntactic sugar} \label{syntacticSugar}
Finally, we also give the definitions for some syntactic sugar for PArrows, namely \ensuremath{\mathbin{\mid\!\!*\!*\!*\!\!\mid}} and \ensuremath{\mathbin{\mid\!\!\&\!\&\!\&\!\!\mid}}.
For basic Arrows, we have the \ensuremath{\mathbin{*\!*\!*}} combinator (Fig.~\ref{fig:syntacticSugarArrows}) which allows us to combine two Arrows \ensuremath{\Varid{arr}\;\Varid{a}\;\Varid{b}} and \ensuremath{\Varid{arr}\;\Varid{c}\;\Varid{d}} into an Arrow \ensuremath{\Varid{arr}\;(\Varid{a},\Varid{c})\;(\Varid{b},\Varid{d})} which does both computations at once. This can easily be translated into a parallel version \ensuremath{\mathbin{\mid\!\!*\!*\!*\!\!\mid}} with the use of \ensuremath{\Varid{parEval2}}, but for this we require a backend which has an implementation that does not require any configuration (hence the \ensuremath{()} as the \ensuremath{\Varid{conf}} parameter
):
\begin{hscode}\SaveRestoreHook
\column{B}{@{}>{\hspre}l<{\hspost}@{}}%
\column{9}{@{}>{\hspre}l<{\hspost}@{}}%
\column{E}{@{}>{\hspre}l<{\hspost}@{}}%
\>[B]{}(\mathbin{\mid\!\!*\!*\!*\!\!\mid})\mathbin{::}(\Conid{ArrowChoice}\;\Varid{arr},\Conid{ArrowParallel}\;\Varid{arr}\;(\Conid{Either}\;\Varid{a}\;\Varid{c})\;(\Conid{Either}\;\Varid{b}\;\Varid{d})\;()))\Rightarrow {}\<[E]%
\\
\>[B]{}\hsindent{9}{}\<[9]%
\>[9]{}\Varid{arr}\;\Varid{a}\;\Varid{b}\to \Varid{arr}\;\Varid{c}\;\Varid{d}\to \Varid{arr}\;(\Varid{a},\Varid{c})\;(\Varid{b},\Varid{d}){}\<[E]%
\\
\>[B]{}(\mathbin{\mid\!\!*\!*\!*\!\!\mid})\mathrel{=}\Varid{parEval2}\;(){}\<[E]%
\ColumnHook
\end{hscode}\resethooks
We define the parallel \ensuremath{\mathbin{\mid\!\!\&\!\&\!\&\!\!\mid}} 
in a similar manner to its sequential pendant \ensuremath{\mathbin{\&\!\&\!\&}} (Fig.~\ref{fig:syntacticSugarArrows}):
\begin{hscode}\SaveRestoreHook
\column{B}{@{}>{\hspre}l<{\hspost}@{}}%
\column{9}{@{}>{\hspre}l<{\hspost}@{}}%
\column{E}{@{}>{\hspre}l<{\hspost}@{}}%
\>[B]{}(\mathbin{\mid\!\!\&\!\&\!\&\!\!\mid})\mathbin{::}(\Conid{ArrowChoice}\;\Varid{arr},\Conid{ArrowParallel}\;\Varid{arr}\;(\Conid{Either}\;\Varid{a}\;\Varid{a})\;(\Conid{Either}\;\Varid{b}\;\Varid{c})\;())\Rightarrow {}\<[E]%
\\
\>[B]{}\hsindent{9}{}\<[9]%
\>[9]{}\Varid{arr}\;\Varid{a}\;\Varid{b}\to \Varid{arr}\;\Varid{a}\;\Varid{c}\to \Varid{arr}\;\Varid{a}\;(\Varid{b},\Varid{c}){}\<[E]%
\\
\>[B]{}(\mathbin{\mid\!\!\&\!\&\!\&\!\!\mid})\;\Varid{f}\;\Varid{g}\mathrel{=}(\Varid{arr}\mathbin{\$}\lambda \Varid{a}\to (\Varid{a},\Varid{a}))\mathbin{>\!\!>\!\!>}\Varid{f}\mathbin{\mid\!\!*\!*\!*\!\!\mid}\Varid{g}{}\<[E]%
\ColumnHook
\end{hscode}\resethooks

\end{document}